\documentclass{mn2e}
\usepackage{times,graphicx,astrobib,amssymb,epsfig,subfigure,epstopdf}
\pdfoutput=1
\newcommand{\de}{{\rm d}}
\newcommand{\bea}{\begin{eqnarray}}
\newcommand{\eea}{\end{eqnarray}}
\newcommand{\f}{\frac}
\title[Galactic feedback]
{Constrained semi-analytical models of Galactic outflows}
\author[Samui, Subramanian \& Srianand] 
{Saumyadip Samui\thanks{E-mail: samui@iucaa.ernet.in},
	Kandaswamy Subramanian\thanks{E-mail: kandu@iucaa.ernet.in},
        Raghunathan Srianand\thanks{E-mail: anand@iucaa.ernet.in}\\
        IUCAA, Post Bag 4, Ganeshkhind, Pune 411 007, India.}

\begin{document}

\maketitle 
\begin{abstract}

We present semi-analytic models of galactic outflows that are constrained by
available observations on high redshift star formation and reionization.
Galactic outflows are modeled in a manner akin to models of stellar wind
blown bubbles. Large scale outflows can generically escape from low mass halos
($M\lesssim 10^9~M_\odot$) for a wide range of model parameters while this is
not the case in high mass halos ($M\gtrsim 10^{11}~M_\odot$). The flow
generically accelerates within the halo virial radius, then starts to
decelerate, and traverses well into the intergalactic medium (IGM),
before freezing to the Hubble flow. The acceleration phase can result in
shell fragmentation due to the Rayleigh-Taylor instability, although the final
outflow radius is not significantly altered. The gas phase metallicity of
the outflow and within the galaxy are computed assuming uniform instantaneous
mixing. Ionization states of different metal species are calculated and used
to examine the detectability of metal lines from the outflows.
The global influence of galactic outflows is also investigated using porosity
weighted averages and probability density functions of various physical
quantities. Models with only atomic cooled halos significantly fill the IGM
at $z\sim3$ with metals (with $-2.5\gtrsim [Z/Z_\odot]\gtrsim -3.7 $), the
actual extent depending on the efficiency of winds, the initial mass function
(IMF) and the fractional mass that goes through star formation.
The reionization history has a significant effect on the volume
filling factor, due to radiative feedback. In these models, a large fraction
of outflows at $z\sim3$ are supersonic, hot ($T\ge 10^5$K)
and have low density, making metal lines difficult to detect.
They may also result in significant perturbations in the IGM gas on scales
probed by the Lyman-$\alpha$ forest. On the contrary, models including 
molecular cooled halos with a normal mode of star formation can potentially
volume fill the universe at $z\ge 8$ without drastic dynamic effects
on the IGM, thereby setting up a possible metallicity floor
($-4.0\le [Z/Z_\odot]\le-3.6$). The order unity fluctuations at $z\sim8$
that becomes the mildly non-linear fluctuations traced by Lyman-$\alpha$
forest at $z<4$ will then have this metallicity. 
Interestingly, molecular cooled halos with a ``top-heavy'' mode of star
formation are not very successful in  establishing the metallicity floor 
because of the additional radiative feedback, that they induce. 
\end{abstract}

 
\section{Introduction} 

The rapid growth of observations of the high redshift universe has
raised several intriguing questions regarding physics of galaxy formation
and the physical state of the intergalactic medium (IGM).
Some of the important issues are:  how and when the dark ages ended with
the reionization of the IGM, the origin of the metals, and temperature of the
low density  IGM traced by Lyman-$\alpha$ forest at $z\simeq2.5$.
In this work we concentrate on the issue of IGM metal enrichment.

Presence of metals in the Lyman-$\alpha$ forest (with $\tau\ge1$) is
now well established though C~{\sc iv}  (Tytler et al. 1995;
Songaila \& Cowie, 1996) and O~{\sc vi} (Carswell et al. 2002;
Simcoe et al. 2002;  Bergeron et al. 2002)
absorption lines  detected in the echelle spectra of high redshift QSOs.
The measured N(C~{\sc iv})/N(H~{\sc i}) at $z\sim3$ are consistent with
[C/H]$\sim-2.5$ with large errors (see Rauch et al. 1997). Songaila (2001)
has shown that the C~{\sc iv} column density distribution function
is consistent with being invariant  between $z=1.5$ and $z=5$ and
a minimum metallicity  of [C/H]$\sim-3.5$ is already in place at $z \sim 5.5$
(see also Ryan-Weber et al. 2006). As under-dense space occupy most of the volume,
measuring metallicity in the low density region is very important. 
Given the  expected low metallicities and large ionization corrections,
direct detection of metals from such low 
neutral hydrogen optical depth (i.e $13\le$~log~[N(H~{\sc i})~{cm}$^{-2}$]~$\le 14$)
is currently impossible and usually the metallicity is estimated using 
pixel statistics \cite{ellison00,schaye03,aracil04}.  Schaye et al. (2003) 
obtained [C/H]$\sim-3.5$ for $\log \delta >-0.5$ at $z = 1.8-4.1$. Here,
$\delta$ being the over density defined as $\delta =\rho/\bar{\rho}\ge
10 (N(H~{\sc i})/10^{15} {\rm cm}^{-2})^{2/3} [(1+z)/4]^{-3}$ (Schaye 2001)
and $\bar{\rho}$ is the mean IGM density. However, Aracil et al. (2004)
have not detected C~{\sc iv} from $\tau < 1$ Lyman-$\alpha$
systems in their UVES data. Thus,  what fraction of the  IGM is filled with
metals is a subject of ongoing debates and investigations (Schaye et al. 2003;
Aracil et al. 2004; Scannapieco et al. 2006). In any case, in the standard
framework of LCDM models the Lyman-$\alpha$ forest 
absorption lines originate from density fluctuations that are
either in the linear or in the quasi-linear regime (Bi \& Davidsen, 1997). 
Thus, metals that may be present in these regions need to be transported from
the neighboring star-forming regions. The amount and  distribution of metals
in the Lyman-$\alpha$ forest provides information on different feedbacks 
from star forming galaxies.

Observations of high-$z$ Lyman break galaxies (LBGs) frequently show
galactic scale superwinds (Pettini et al. 2001), high value for the
escape fraction of the UV photons (Steidel et al. 2001) and 
a strong correlation between the C~{\sc iv} absorption systems and 
the LBGs at $z\sim3$ within an impact parameter of $\sim 50$ kpc
(Adelberger et al. 2005).  Profiles of C~{\sc iv} and O~{\sc vi} absorption
lines seen in the high redshift damped Lyman-$\alpha$ systems (DLAs)
are also consistent with them originating from outflows in DLA protogalaxies
(Fox et al. 2007, 2007a).
Adelberger et al. (2005) have argued that
considerable fraction of high  column density C~{\sc iv} systems may
originate from the large scale galactic outflows. Clustering properties
of strong C~{\sc iv} absorption lines (Scannapieco et al. 2006) and the
velocity profiles C~{\sc iv} absorption lines (Songaila et al. 2006) are consistent
with a good fraction of the high column density C~{\sc iv}
systems originating from the region near massive galaxies. However,
the important question is whether these metals are freshly emitted by
the bright galaxies or due to biased clustering of gas ejected by
low mass galaxies from early (say $z >6$) epochs (Porciani \& Madau, 2005).
Aguirre et al. (2005) by comparing spectra predicted by various simulations 
that includes winds found that the predicted metal distribution in the models are 
highly inhomogeneous and can not reproduce the observed probability distribution of 
C~{\sc iv} absorption. They suggested that strong winds from galaxies at $z \le6$ 
cannot fully explain the observed enrichment and additional pre-enrichment from
higher redshift galaxies are needed.

Note that star formation is the key element in controlling the
outflows from galaxies. Star formation also has an important effect
on the reionization of the universe which in turn affects the star formation
in low mass halos through radiative feedback. There exists a growing body 
of data constraining the star formation rate (SFR) density in the high
redshift universe \cite{bouwens05,richard06,hopkins06}.
Also observational constraints on the reionization are provided by the spectra
of highest redshift QSOs (Fan et al. 2006) and ongoing WMAP satellite observations
of Cosmic Microwave Background (CMB) polarization (Spergel et al. 2007).
Therefore one must deal with reionization and galactic outflows in a manner
consistent with the observed star formation history of the universe.
In our previous paper Samui, Srianand \& Subramanian (2007; here after Paper I)
we have described a set of models which correctly produce the observed luminosity
functions of LBGs and hence the SFR density at different redshifts.
Here we take those constrained models of star formation and reionization
from Paper I and used them to predict the various properties galactic outflows
and their impact on the IGM such as volume filling factor of wind/metals,
mean metallicity, temperature of these polluted region etc.  

There are several attempts to model the galactic outflow using semi-analytic
calculations (for example, Madau, Ferrara \& Rees 2001; Scannapieco,
Ferrara \& Madau 2002; Furlanetto \& Loeb 2003; Scannapieco 2005). The model
presented here will be broadly in line with
these above studies with significant improvements. In particular, the 
star formation is continuous and duration of the star formation and the amount
of gas going into stars are constrained by the observed high-$z$
luminosity function (see Paper I and section 2 below). This is very different
from the previous attempts where the star formation is usually in the form
of bursts. Our model takes into account the reionization and the radiative
feedback to the star formation in a self-consistent way. This is very
important, in particular, to estimate the metal pollution due to low mass
halos and constrain the initial mass function (IMF) at high redshifts. We
discuss various other improvements and their effects in detail.

The paper is organized in following manner. In the next
section we first briefly outline the model of star formation.
In section 3 we introduce our model of galactic outflows in detail.
The structural properties of the outflow and  sensitivity of our results to
the adopted initial conditions are discussed in detail in 
section 4.  In section 5 we discuss the dependence of outflow properties
on the model parameters. In section 6 we investigate the growth of
Rayleigh-Tayler instabilities in the accelerating phase of the outflow and
study its consequences. The ionization correction and the 
detectability of the winds in different stages of its evolution are
discussed in section 7. The global effects of the outflow are 
studied in section 8 and the summary and conclusions are provided 
in section 9.
In most of this work we use the cosmological parameters consistent with the
recent WMAP data ($\Omega=1$, $\Omega_m = 0.26$, $\Omega_\Lambda = 0.74$,
$\Omega_b=0.044$, $h = 0.71$, $\sigma_8=0.75$ and $n_s =0.95$).

\section{Star formation rate}
\label{sfr}

We follow the same prescription as in Samui, Srianand \& Subramanian (2007) 
to get the star formation rate (SFR) density in a semi-analytical fashion.
Here, we only briefly outline the model.
We use the modified Press-Schechter (PS) formalism of Sasaki (1994) to
calculate the number density of collapsed objects having
mass between $M$ and $M + \de M$, which are formed at the redshift
interval $(z_c, z_c + \de z_c)$ and survive till redshift
$z$. This is given by (Chiu \& Ostriker 2000; Choudhury \&
Srianand 2002)
\bea
N(M,z,z_c)~ \de M~ \de z_c &=& N_M(z_c) \left(\f{\delta_c}{D(z_c) 
\sigma(M)}\right)^2 \f{\dot{D}(z_c)}{D(z_c)}\;\nonumber \\
& \times & \f{D(z_c)}{D(z)} \f{\de z_c}{H(z_c) (1 + z_c)}~ \de M.
\label{eqnmPS}
\eea
Here, $N_M(z_c)~ \de M$ is the PS mass function \cite{ps74},
$\delta_c$ is the critical over density for collapse, usually taken to be
equal to $1.686$. Further, $H(z)$ is the
Hubble parameter, $D(z)$ the growth factor for linear perturbations and
$\sigma(M)$ the rms mass fluctuation at a mass scale~$M$. 

We model the star formation in a given halo of mass $M$ collapsed at
$z_c$ and observed at $z$ as,

\bea
\dot{M}_{\rm SF}(M,z,z_c) &=& f_{*} \left(\f{\Omega_b}{\Omega_m} M \right) 
\f{t(z)-t(z_c)}{\kappa ^2~ t_{\rm dyn}^2(z_c)} \nonumber \\
& &  \times \exp\left[-\f{t(z)-t(z_c)}{ \kappa ~t_{\rm dyn}(z_c)}\right].
\label{eqnsf}
\eea
Here, $f_{*}$ is the fraction  of total baryonic mass in a halo that will be
converted to stars. The function $t(z)$ gives the age of the universe at
redshift $z$; thus, $t(z)-t(z_c)$ is the age of the collapsed halo at $z$ and
$t_{\rm dyn}$ is the dynamical time-scale \cite{chiu,bl01}.
As mentioned in Paper I the duration of star formation activity in a
halo depends on the value of $\kappa $.
Note that our prescription of SFR as a function of time
(i.e. Eq.~\ref{eqnsf}) is purely empirical and not obtained by
taking into account all possible physical processes in the
interstellar medium (ISM). However, as the model parameters ($f_*$ \&
$\kappa$) are constrained by the observed luminosity functions,
we can treat it as the net SFR as a function of time resulting
from various competing physical processes that govern the
star formation.

Having modeled the formation rate of halos and the star formation in
an individual halo, we can calculate the comoving star formation rate
density as,
\begin{equation}
\dot{\rho}_{\rm SF}(z) = \int\limits_z^{\infty} \de z_c 
\int\limits_{M_{\rm low}}^{\infty} \de M' \dot{M}_{\rm SF}(M',z,z_c) 
\times N(M',z,z_c).
\label{eqnsfr}
\end{equation}
The lower mass cutoff ($M_{\rm low}$) at a given epoch is decided by the
cooling efficiency of the gas and different feedback processes (see Paper I
for detail discussions). We consider models with  $M_{\rm low}$
corresponding to a virial temperature, $T_{\rm vir} = 10^4$~K ( as
``atomic cooling model'') and $300$~K (``molecular cooling model'')
for the neutral gas. For ionized regions of the universe,
our models assume complete
suppression of star formation in halos below circular velocity
$v_c=35$~km~s$^{-1}$, no suppression above circular velocity of
$95~$km s$^{-1}$ and a linear fit  from $1$
to $0$ for the intermediate masses (as in Bromm \& Loeb 2002).
The SFR in the high mass halos are reduced by a suppression 
factor $[1+(M/10^{12} M_\odot)^3]^{-1}$. However, our models
at present do not incorporate the chemical and SNe feedback.

The reionization history of the universe is also calculated
as outlined in Paper I. We assume all the Lyman-continuum photon which
escape the star forming region (with escape fraction $f_{esc}=0.1$)
are used in ionization. We only consider the ionization of
hydrogen and take case B recombination (with a recombination coefficient
$\alpha_B $) to calculate the recombination rate. The redshift evolution
of ionized hydrogen fraction ($f_{HII}$) is given by,
\bea
\f {\de f_{HII}}{\de z } & =& \f {\dot{N}_{\gamma}}{n_H(z)}~\f{\de t}{\de z}
 ~-~\alpha_B n_H(z)f_{HII} C~\f{\de t }{\de z}. 
\eea
Here, $\dot{N}_{\gamma}$ is the rate of UV photons escaping into the IGM and
obtained from $\dot{\rho}_{\rm SF}(z)$. Further, $n_H(z)$ is the proper
number density of the hydrogen atoms and $C$ is the clumping factor
of the IGM. The redshift evolution of $C$ is assumed to be in the form
$C(z) = 1+ 9\times[7/(1+z)]^2$  for $z\ge6$ and  $C=10$ for $z<6$.
(Haiman \& Bryan 2006) $\dot{N}_{\gamma}$ for a given IMF is calculated
as explained in Paper I (see Eq.~(13) and Table 1 there).

In Paper I, the observed UV luminosity function of galaxies at $3<z<6$ are
well fitted by our models with $f_*=0.5$ and $\kappa = 1$ for the cosmological
parameters constrained by WMAP 3rd year data. Models with 
$f_*$ less than $0.5$ require $\kappa<1$. However, the measured ages of
high-$z$ star formation galaxies favor $\kappa\simeq 1$. It has also been
pointed out in Paper I, that models with 
slightly higher values of $\sigma_8$ and $n_s$ than that derived from
the 3rd year WMAP data require lower values of $f_*$ even for $\kappa = 1$. 
Thus, in our global models discussed in the following sections we will use
three sets of parameters for the star formation models 
(i) $\kappa= 1$ and $f_*=0.5$; (ii) $\kappa=0.5$ and $f_*=0.25$ with
$\sigma_8 = 0.75$ \& $n_s = 0.95$ and 
(iii) $\kappa =1 $ and $f_*=0.25$ with $\sigma_8 =0.85$ and $n_s =1.00 $.  
All these models reproduce the observed 
high-$z$ luminosity functions reasonably well.
Models with low values of $\kappa$ and $f_*$ correspond to cases where the
fraction of cold gas that can be used for forming stars
decreases more rapidly as a function of time.

In Paper I, it has been shown that various ongoing deep imaging surveys
will not be able to directly detect the  molecular cooled halos that may be 
present prior to reionization. However,  presence of 
these objects influences the reionization history and hence the extent of 
radiative feedback at $z\ge 6$
(see section~6 in Paper I). As the gravitational potential of these sources are
expected to be small it is usually believed that even a small amount of 
star formation in these objects can drive outflows. Therefore, presence of 
star formation in molecular cooled halos can have an important role to play
in the pre-enrichment of the IGM to some metallicity floor. We discuss this 
issue by considering the models with molecular cooled halos that 
consistently produce reionization history constrained by the WMAP
3rd year data (see the list in Table~3 of Paper I).

\section {Outflows from galaxies into the IGM}

The metals detected in the IGM can only have been synthesized by stars 
in galaxies, and galactic outflows are the primary means by which they 
can be transported from galaxies into the IGM. The mechanical energy 
that drives such outflows may arise either from an active galactic
nuclei (AGN) in the galaxy or from the supernovae (SNe) explosions
associated with the star formation activities in the galaxy.
Here we concentrate on the effects of the star formation activity and the 
resulting SNe in high redshift protogalaxies.

\subsection{The general outflow scenario}

When a single SNe explodes, it creates a bubble of shock heated 
interstellar medium (ISM) around itself which expands supersonically into 
surrounding ISM.  The occurrence of clustered and coherent explosions of 
several SNe can lead to the merger of the associated supernova remnants
to form a super bubble.  This super bubble  expands as it is fed by consequent 
SNe explosions, and decelerates as it sweeps up the ISM of the galaxy.
In the case of a disk galaxy, as the super bubble radius approaches the 
disk scale height,  the decreasing density of the ISM generally causes a 
reacceleration of the swept up ISM shell, which then begins to fragment 
through growing Rayleigh-Taylor (RT) instabilities (see the discussion in 
Veilleux, Cecil \& Bland-Hawthorn 2005). These fragments and the shock 
heated hot gas are then vented out into the galaxy halo.
Further, in a protogalaxy with high enough rates of star formation,
supernova remnants, even if randomly distributed, could 
fill a significant volume of the whole galaxy to create a 
galaxy wide super bubble. This will also feed hot gas into
the galactic halo. 

In the absence of any gas in the halo of the galaxy,  the hot gas being fed into 
the halo by super bubbles would escape as a thermally driven wind into the 
intergalactic space. However, it is more likely that the halo of  a forming galaxy 
itself has residual gas which has not yet collapsed to the centre or gas which is 
continuing to fall in from the intergalactic medium.  For example, if accretion of gas
is bimodal, with both an early dominant cold mode along filaments and a latter
hot mode via an accretion shock (cf. Dekel \& Birnboim 2006; Keres et al. 2005), 
then significant star formation can occur before the hot gas in the halo 
completely accretes onto the galaxy. In this case, the bubble of SNe heated 
hot gas ejected from the galaxy, will be initially confined by the outside medium
and will only escape as its increasing internal pressure drives out the
external medium.

\begin{figure}
\subfigure[]{
\centerline{
\includegraphics[width=0.4\textwidth]{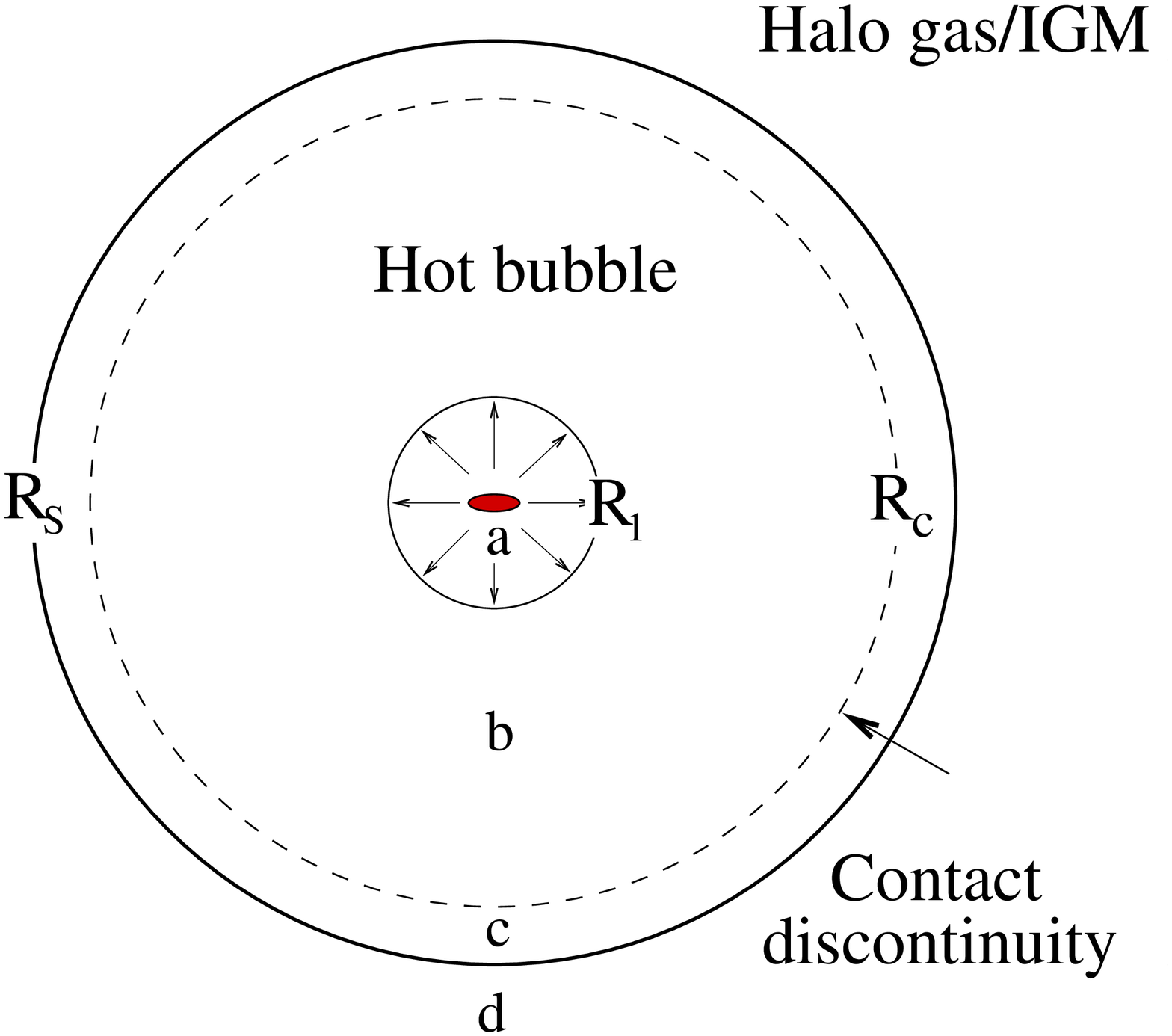}
}}
\subfigure[]{\centerline{
\includegraphics[width=0.4\textwidth]{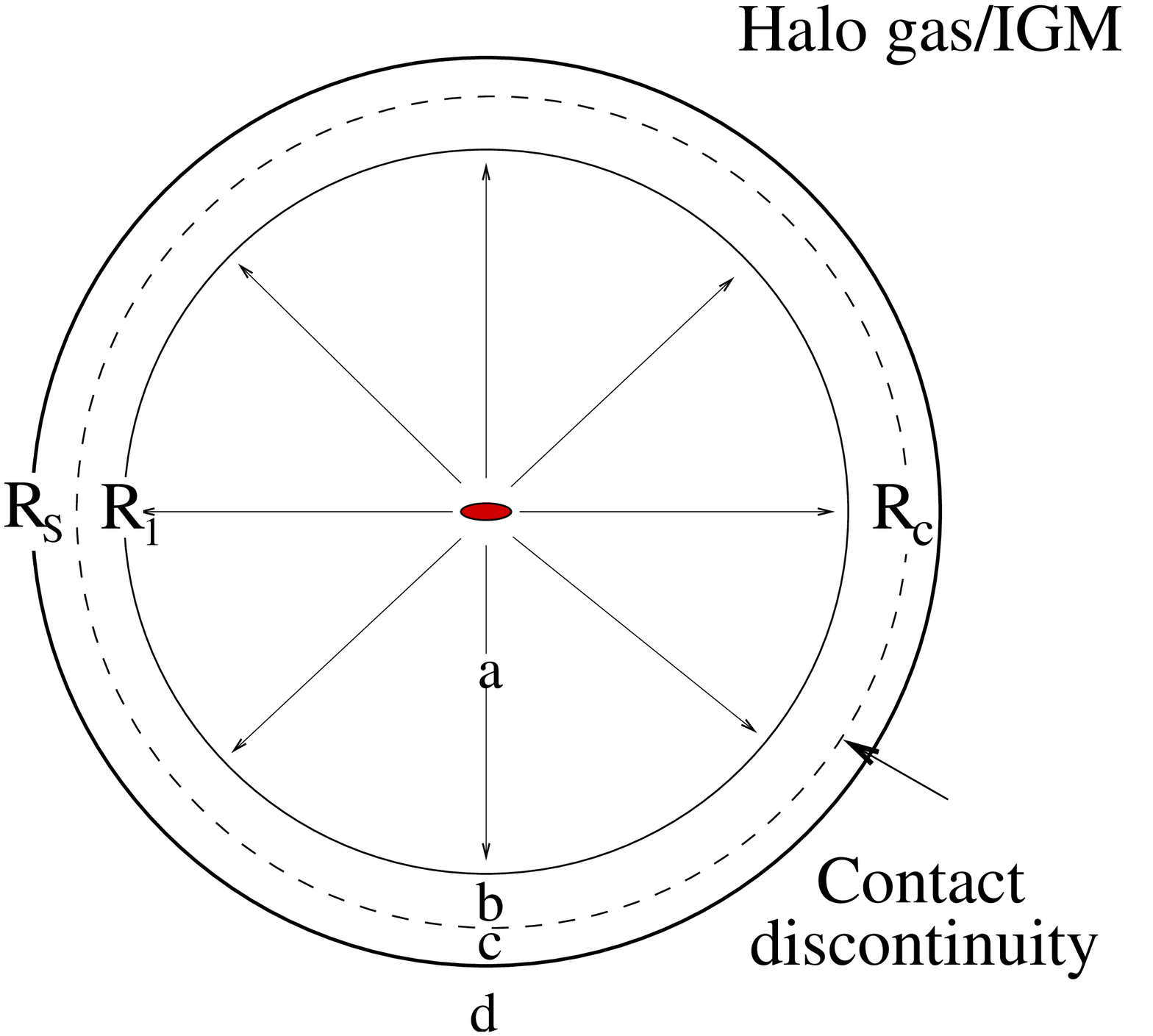}
}}
\caption[]{Schematic diagram of the structure of the galactic outflow in our
 models. Panel (a) is for the `pressure driven flow' and panel (b) is for the 
`momentum driven flow'. 
}
\label{wind_profile}
\end{figure}

Our subsequent treatment of the dynamics of the galactic outflow assumes a 
thin shell approximation, analogous to the treatment of
interstellar bubbles driven by stellar winds (cf. Castor et al. 1975;
Weaver et al. 1977; Ostriker \& McKee 1988; Tegmark et al. 1993).
In this picture (see panel (a) in Fig.~\ref{wind_profile}) the wind blown bubble, 
at some stage, has an onion-like structure
with 4 concentric zones: (a) an innermost region consisting of 
the galactic wind blowing out (called ``free wind'') (b) a hot bubble of
shocked wind gas (the galactic wind entering the surrounding halo/IGM 
gets shocked at an inner shock at radius say $R_1$)
(c) a thin dense shell of shocked IGM/halo gas separated from
the shocked galactic wind by a contact discontinuity at $R_c$ and (d)
undisturbed halo/IGM gas outside an outer shock at radius $R_s$.
In what follows we shall describe an outflow with the above
structure as  a ``pressure driven outflow". 

While the above picture is valid for most of the time,
in the initial stages, the hot gas could be a filled bubble of radius $R_s$, 
without the innermost shock at $R_1$. In this case, 
the SNe in the galaxy would just feed this bubble directly with its hot ejecta. 
But subsequently, as this bubble expands, its pressure decreases, and then 
the fresh supply of SNe heated gas from the galaxy would be
at a higher pressure. A sufficiently large pressure difference between
the SNe heated hot gas and the bubble gas would lead to  
a thermally driven wind being driven out of the galaxy, where the
thermal energy of the SNe heated gas is converted into directed
kinetic energy of the wind (cf. Chevalier \& Clegg 1985). Such a wind
will be again stopped at a distance where its ram pressure equals
the previously created hot bubble pressure. This is the origin of
the inner shock at radius $R_1$. The above evolutionary sequence is
different from the case of a stellar wind blown bubble, where the
wind from the star has already a high velocity and low temperature that
it leads to the ``reverse shock" at $R_1$ right from the early
stages. Nevertheless the subsequent evolution can be very similar
to the stellar case.

In the case when the hot bubble material in region (b) between $R_1$ 
and $R_c$ cools efficiently, $R_1$ can approach $R_c$ with zone (b) 
becoming of negligible thickness. The wind material from the galaxy
then directly impinges on the shell (as shown in panel (b) in 
Fig.~\ref{wind_profile}). The wind momentum is directly transferred
to the shell and the outflow then becomes momentum driven. We shall refer
to the outflow then as a ``momentum driven outflow".

Note, the thin shell approximation we will be using
is valid if the cooling time in post-shocked halo/IGM gas is small 
compared to the expansion time of the bubble. For interstellar bubbles,
being blown into a constant density ISM, Weaver et al. (1977) show that, 
even during the adiabatic evolution of the system, when the fluid in 
region (c) does not cool efficiently, the thickness of region (c) is only 
about $0.14 R_s$.  
Also the self-similar solutions of Ryu \& Vishniac (1991) and 
Koo \& McKee (1992), for a range of 
power law density profiles, $\rho_B(R) \propto R^{-\alpha}$,
and wind luminosity $L(t) \propto t^{\beta}$,
the thickness of region (c) is only $\sim 0.09-0.16 R_s$.
We verify this in the context of our model parameters
in Appendix~\ref{sec_shell_struc}. Further, explosions in an expanding universe 
generically have the swept matter concentrated in a thin shell 
(Bertschinger 1985; Voit 1996).
These results encourage the use of the thin shell approximation even
when the shocked halo/IGM gas cannot cool efficiently. 
We generally find that the shocked halo gas can
indeed cool efficiently, but the shocked IGM gas
need not do so, especially if it cannot mix efficiently with
the metal enriched galactic wind material in region (b).

\subsection{Modeling the outflow dynamics}

The evolution of the spherically symmetric outflows is 
governed by the following equations in the pressure driven
case as in Fig. 1a 
(cf. Weaver et al. 1977; Ostriker \& McKee 1988; Tegmark et al. 1993),
\bea
\f{\de ^2 R_s}{\de t^2} &=& \f{4\pi R_s^2 \left(P_b -P_0\right)}{m_s(R_s)}
- \f {\dot m_s(R_s)(\dot{R}_s - v_0(R_s)) }{m_s(R_s)}
\nonumber \\
 & & \:\:\:\:\:\:\:- \f {G M(R_s)}{R_s^2}, 
 \label {eqnradius} 
\eea
\begin{equation}
\f{\de m_s}{\de t}(R_s) =
 \epsilon 4 \pi R_s^2 \rho_B(R_s) (\dot{R}_s - v_0(R_s)).
 \label {eqnmaass} 
\end{equation}
Here, the subscript $s$ represents the thin shell variables while
a subscript $b$ represents the bubble variables.
In particular, $R_s$ is the physical radius of the shell and 
$P_b$ the pressure inside the bubble. Further, $m_s(R_s)$ is 
the baryonic mass in the thin shell, $M(R_s)$ the total gravitational 
mass within $R_s$ and $\rho_B(R_s)$ is the baryonic density of the
ambient medium into which the supper bubble is expanding. 

The first term in the right hand side of Eq.~(\ref{eqnradius})
represents the outward force due to the difference
between the hot bubble pressure ($P_b$) and the outside medium
pressure ($P_0$). The second term takes into account both
the drag on the outflow due to mass
swept up from the surrounding medium and the addition of momentum
of this medium to the shell. The last term is deceleration of the
shell due to gravity. Further, $v_0$ is the velocity of the fluid 
outside $R_s$.
As the wind propagates it sweeps up mass from the surrounding medium.
We assume that a fraction $\epsilon$ of this swept up mass
is added to the shell and rest of it is incorporated into the
hot bubble. The resulting evolution of the shell mass
$m_s$ is described by Eq.~(\ref{eqnmaass}).

The pressure $P_b$ and the thermal energy of the bubble, $E_b$,  
are related by, 
\begin{equation}
P_b = \frac{E_b}{ 2\pi (R_s^3-R_1^3)},
\label{pressure_energy}
\end{equation}
with the assumption that the adiabatic index of the gas $\gamma = 5/3$. 
The evolution of the energy in the bubble is  in turn given by
\begin{equation}
\f {\de E_b}{\de t} = L(t) - \Lambda (t,T,Z) 
   - 4\pi (R_s^2 \dot{R}_s - R_1^2 \dot{R}_1) P_b.
    \label {eqnenergy}
\end{equation}
Energy is fed into the bubble from the supernovae explosions 
in the host halo and is lost due to cooling and the $P dV$ work done during
the expansion of the bubble. In Eq.~(\ref{pressure_energy}),
we have taken into account of the fact that the wind may 
enter the hot bubble through an inner shock at $R_1$.  This
leads to a smaller hot bubble volume compared to the models that
do not consider the inner shock. Further since it takes a finite 
time, ($R_1/v_w$), for the free wind to propagate to $R_1$, 
the effective energy input to the bubble in a time $dt$ is, 
$L(t) dt = L_0(t_e)dt_e$, where $t_e = t- R_1/v_w$ is the time at which 
the wind leaves the galaxy and $L_0(t_e)$ is its mechanical luminosity
generated from the SNe explosions in the galaxy.
Here $v_w$ is the asymptotic velocity of the wind from the galaxy
before it encounters the shock at $R_1$. As noted earlier,
such a thermally driven wind can arise when the bubble pressure
drops sufficiently below the ram pressure of the
gas outflowing from the galaxy. The luminosity output
in this situation can also be written as $L_0(t_e) = \dot{M}_w(t_e)v_w^2/2$,
where $\dot{M}_w$ is the rate at which mass is ejected by the galaxy.
The evolution equation for $R_1$ is obtained simply from the jump
condition across the inner shock, assuming it to be strong:
\begin{equation}
P_b =  \frac{3}{4} \f{\dot{M}_w(t_e) }{4 \pi R_1^2 v_w}\left[ v_w - \dot{R}_1\right]^2.
\label{R1eqn}
\end{equation}
Earlier semi-analytical studies of SNe driven outflows from galaxies, 
do not take into account the evolution of the inner shock at $R_1$.
Indeed for decelerating self-similar flows, $R_1$ grows with time slower
than $R_s$ (cf. Weaver et al. 1977), and so its evolution changes 
$P_b$ negligibly, and hence the evolution of $R_s$. However, we find
that the outflows accelerate in the early stages of evolution, basically because
of an increasing $L(t)$ and a steeply decreasing $\rho_B(R)$. In this case,
we will see below from naive scaling arguments that $R_1$ can approach $R_s$.
In addition, as discussed above when there is efficient cooling of the bubble
$R_1$ approaches $R_s$ and the  `pressure driven flow' transits to a
`momentum driven flow'. Therefore, it is important to take account of
the dynamics of $R_1$ to determine the correct evolution of the outflow.

The cooling of the bubble (represented by $\Lambda (t,T,Z)$ in Eq.~(\ref{eqnenergy}))
is due to Compton drag against  the CMBR, bremsstrahlung and recombination line 
cooling. The cooling rate depends on the hot bubble density, temperature and 
metallicity ($Z$). The metallicity of the outflowing gas and interstellar medium
of the galaxy are computed in a self-consistent way assuming instantaneous
uniform mixing (see Appendix~\ref{metal_evolution}). As expected the metallicity
grows with time. However, to keep our computation simple, we estimate bubble cooling
rate as a function of time for an assumed constant metallicity (say $Z = 0.01 
Z_\odot$). We note that the maximum metallicity achieved in a typical
bubble is of this order.  Usually the bubble material reaches the maximum
metallicity when it is outside the halo where the adiabatic cooling dominates
over the radiative cooling.  Thus assuming this constant metallicity slightly
over estimates the cooling in the early stages of the wind and have no 
effect in the latter stages.  We adopt here the cooling rates $\Lambda$,
given by Sutherland \& Dopita (1993). The temperature and bubble density
(required also for estimating $\Lambda$) are computed as follows.
The mass of the hot bubble evolves as,
\bea
M_{\rm b}&=&
\int \dot{M}_w~ \de t \nonumber \\
&&+ \int (1-\epsilon) 4 \pi R_s^2 \rho_B(R_s)(\dot{R}_s-v_{0})~\de t.
\label{eqnwindmass}
\eea
Here, the mass outflow rate from the galaxy is $\dot{M}_w$
which we assume 
to be proportional to the SFR ; i.e  $\dot{M}_w = \eta \dot{M}_{SF}$,
where $\eta$ is the mass loading factor. Most outflow models 
that are discussed in  the literature use $\eta\ge$1 (see for example,
Furlanetto \& Loeb 2003, Oppenheimer \& Dave 2006; 
Bertone, De Lucia \& Thomas 2007).
The value of $\eta$ is found to be greater than 1 for local star forming
dwarf galaxies \cite{Martin99}, and in the range $0.1 \le \eta \le 0.7$
for ultra luminous infrared galaxies at $ 0.04 \le z \le 0.27$ \cite{rupke02}.
Martin (2005) finds a median values of $\eta$ of 0.19 and 0.09 for high and
low redshift ultra luminous infrared galaxies respectively.
 Even though there are predictions of the mass dependence of $\eta$
(see Murray, Quataert \& Thompson 2005)
available observations do not strongly support such a dependence \cite{rupke02}.
In our model calculations we assume $\eta$ to be independent of
mass and use $\eta=0.3$ in most of the models below.

The second term in Eq.~(\ref{eqnwindmass})
takes into account the mass loading of the
hot bubble due to processes like, evaporation of the shell, and 
we have assumed simply that the fraction ($1-\epsilon$) of the 
halo/IGM gas that has not being swept by the shell
is added to the hot bubble (cf. Furlanetto \& Loeb 2003). 
The hot bubble is taken to be of nearly uniform density given
 by $\rho_b=3M_b/(4\pi (R_s^3 -R_1^3))$; the self-similar
solution derived by Weaver et al. (1977) for example, supports such a view,
with only the region near the 
contact discontinuity between regions (b) and (c), being at a higher density.
Knowing the bubble density, we can also derive its temperature $T_b$,
from the total thermal energy, using $3k (M_b/m_p) T_b = E_b$
where $m_p$ is the proton mass.
Here we have assumed the gas is mostly ionized hydrogen. 

Given the temperature, density and metallicity of the bubble gas, we can calculate
its cooling rate. We show below that for
certain range of parameters, the bubble can cool efficiently enough that
$R_1$ becomes very close to $R_s$. The outflow then
transits to a `momentum driven flow'. The wind then directly deposits
momentum at the shell
and the evolution of $R_s$ is governed
by a modified equation (Ostriker \& McKee 1988; Bertone, Stoehr \& White 2005) 
\bea
\f{\de ^2 R_s}{\de t^2} &=& 
\f{\dot{M}_w(t) v_w ( 1 - \dot{R}_s/v_w )}{m_s(R_s)}
-\f{4\pi R_s^2 P_0}{m_s(R_s)}
\nonumber \\
& & \:\:\:\:\:\:\: 
- \f {\dot m_s(R_s)(\dot{R}_s - v_0(R_s)) }{m_s(R_s)}
 - \f {G M(R_s)}{R_s^2},
  \label {eqnradius1}
  \eea
where, the shell mass $m_s$ now evolves as,
\begin{equation}
\f{\de m_s}{\de t}(R_s) = \dot{M}_w +
4 \pi R_s^2 \rho_B(R_s) (\dot{R}_s - v_0(R_s)).
\label {eqnmaass1}
\end{equation}
We always begin the evolution of outflows using Eq.~(\ref{eqnradius}) and
Eq.~(\ref{eqnmaass}) and 
switch over to Eqs.~(\ref{eqnradius1}) and (\ref{eqnmaass1}), 
if and when $R_1$ becomes very close to $R_s$.

We also have to specify the distribution of the dark matter and baryons 
in the halo through which the outflow propagates.
The dark matter distribution within the virial radius of the halo is assumed to be
a NFW density profile \cite{nfw} 
and smooth outside the halo with mean cosmological density.
The baryonic density $\rho_B$ is estimated in the following manner.
Within the virial radius, a fraction $f_h \sim 0.1$ of the total baryonic mass
is taken to be still in the halo in hydrostatic equilibrium with
the dark matter potential at the virial temperature $T_{\rm vir}$.
This can represent for example, the gas which is being accreted into
the halo in the hot-mode of accretion (cf. Keres et al. 2005; Dekel \& Birnboim 2006)
and which has not yet cooled and fallen into the galaxy. We note in passing that
earlier work on outflows have adopted values ranging from
$f_h=1$ (Madau, Ferrara \& Rees 2001) to $f_h=0$ (Furlanetto \& Loeb 2003).
Kobayashi, Springel \& White (2006) have used $f_h=0.1$ with an NFW profile
in their isolated disk models.
A gas in hydrostatic equilibrium in an NFW dark halo with virial
radius $R_{{ \rm vir}}$ and concentration $c$, has a density profile that
is well fitted by a beta model (Makino, Sasaki \& Suto 1998):
\bea
\rho_B(R) = \f{\rho_c}{ [ 1 + (R/R_c)^2]^{1.4}}
\label{gasprof}
\eea
where, for a gas at the virial temperature, the core radius is
$R_c = (0.22/c)R_{{ \rm vir}}$ and $\rho_c$ is the central gas density, determined
from normalizing the gas mass in this profile to the total mass.
We have taken a typical concentration parameter of 4.8 for all the halos
(Madau, Ferrara \& Rees 2001).
Hence $\rho_B$ is equal to the left over gas density in the halo for
$R_s<R_{\rm vir}$ and outside $R_{\rm vir}$, it is simply the background IGM
gas density. Note that at $R_{\rm vir}$ the gas density
is still larger than the background IGM density. To avoid any unphysical
jump we assume an exponential decay of density of width $0.2R_{\rm vir}$ 
(Madau, Ferrara \& Rees 2001).
In what follows we will examine the sensitivity of our results to $f_h$.

The outside pressure $P_0$ is therefore fixed to be the pressure of the
halo gas at the virial temperature ($T_{\rm vir}$) within the virial radius. 
Outside this radius $P_0$ is calculated assuming that the gas in the IGM
is at $10^4~$K. The latter assumption is justified as the ionization front
from the galaxy moves faster than the outflow and hence
the wind always passes through an ionized medium which has also 
been photoheated to a temperature of $10^4~$K. 

Further, in Eq.~(\ref{eqnmaass}), $v_0$ is the velocity of the surrounding
medium.  We adopt the following form for $v_0$  (Furlanetto and Loeb 2001),
\bea
v_0(R) &=& 0.0 \;\; {\rm for} \;\; R \le R_{{\rm vir}} \\ \nonumber
		&=& \f{\sigma}{3}\left(\f{R}{R_{{\rm vir}}} - 4 \right) \;\;
		{\rm for} \;\;R_{{\rm vir}} <R \le 4 R_{{\rm vir}} \\ \nonumber
		&=& \f{3}{2}H(R - 4 R_{{\rm vir}})\;\; 
		{\rm for} \;\;4R_{{\rm vir}} <R \le 12 R_{{\rm vir}} \\ \nonumber
		&=& HR \;\;\;{\rm for} \;\;12R_{{\rm vir}} > R.  \nonumber
\label{vinfall}
\eea
Thus the medium is assumed to be at rest within the halo.

The mechanical luminosity, $L_0(t)$, fed into the wind
coming from the SNe produced by a continuous star formation
(as in Eq.~(\ref{eqnsfr})) is given by,
\bea
L_0(t) &=& 10^{51}\times\epsilon_{\rm w}~\nu~f_* 
\left(\f{\Omega_b}{\Omega_m} M \right) 
\f{t(z)-t(z_c)}{\kappa^2 t_{\rm dyn}^2(z_c)} \nonumber \\
& & \exp\left[-\f{t(z)-t(z_c)}{\kappa t_{\rm dyn}(z_c)}\right]{\rm erg ~s}^{-1}.
\eea
Here, we have assume that each SNe produces $10^{51}$ ergs
of energy and a fraction $\epsilon_{\rm w}$ of this energy goes to power the
wind.
For most of our work we take $\epsilon_{\rm w} = 0.1$. Note that,
Mori et al. (2002) from numerical simulation find an efficiency of 20-30\%
of converting supernova energy into kinetic energy of the outflowing gas.
This is factor 2-3 higher than the maximum $\epsilon_{\rm w}$ we
use in this work. Further, $\nu$ is the number of SNe per unit solar mass
of stars formed.
For a Salpeter IMF with mass range $1-100~M_\odot$, one SNe occurs every
$\nu ^{-1} ~\approx~50 M_{\odot}$ of stars formed. However, if we assume
Salpeter IMF with mass range $0.1-100~M_\odot$
then $\nu ^{-1} ~\approx~130 M_{\odot}$. 

For the major part of the present work we adopt the continuous star formation
model above,  with the parameters constrained by fitting the observed 
high redshift luminosity functions (see Paper I).
However, much of the earlier 
semi-analytic models of outflows assume that the star formation 
in a halo occurs in a single instantaneous burst 
\cite{scannapieco,madau_ferrara_rees}.  
To compare our results with previous work we will also show the results
obtained with such a model, where we assume that  a fraction 
$\epsilon_{\rm SF}$ of total baryonic mass goes into stars instantaneously. In such
a model, the last SNe explosion will occur at a time $\sim 33$ Myr 
(assuming that lower mass limit of a star to explore as SNe
is $\sim 8 M_{\odot}$) after the burst of star formation. This gives rise to
\bea
L_0(t) &=& 10^{51} \times \epsilon_{\rm w}~\nu~\epsilon_{\rm SF}
\f{\Omega_b}{\Omega_m} M \f{\theta ( t_{\rm OB} - t )}{t_{\rm OB}}~{\rm erg ~s}^{-1}
\eea
where $t_{\rm OB}$ is the maximum life time of an OB star which is assumed to be
33 Myr. 

Finally to solve the above listed equations one has to specify the
initial conditions.
\begin{itemize}
\item 
The initial radius of the hot bubble is taken to be $R_i = R_{{ \rm vir}}/15$.
This is much smaller than halo size but typically larger than the 
radius of a disk galaxy forming in the halo (Efstathiou 2000).

\item We wait until the bubble pressure determined from $L_0(t)$ is equal to
the outside thermal pressure of the halo gas. This time is taken to be
the initial time $t_i$ for the integration of the outflow equations.
Typically $t_i \ll t_{\rm dyn}$ and so most of the star formation and 
energy input into the hot bubble will take place at later times.

\item The initial mass loaded into the bubble is taken to be $\eta$ times
the mass of formed stars upto $t_i$, motivated by the proportionality of
the wind mass loss rate and the SFR noted above. 

\item The initial mass of the shell is taken to be equal to that of the super
bubble at $t_i$.
However, we find that the subsequent swept up
mass by the shell form the ambient medium greatly exceeds
this initial mass (even after the first few time-steps), and
so the evolution of the outflow is almost
independent of this initial assumption. 

\item Although the shock at $R_1$ arises perhaps after an initial
period of filled bubble evolution, we shall solve the equations above as if it was
always present. This only changes the initial period of the evolution.

\item We start the evolution with a initial zero velocity for the shell.
Again, we will show that starting from a supersonic shell velocity makes very
little difference to the final evolution (mainly because the bubble
pressure at $t_i$ is not large enough to maintain this velocity). 
In passing we note that even for subsonic expansion of buoyant bubbles into uniform
fluid, one gets an evolution equation very similar to the equations we follow here
(see Batchelor, 1997, p. 479), including an effective mass loading due
to the displaced mass. We also set $R_1 = 0$ initially.

\end{itemize}

We follow the evolution equations upto when the peculiar velocity of the shock
reaches to the local sound speed and afterward allow 
it expand with the Hubble flow. Note that if the velocity of an outflow
becomes zero within the virial radius of the halo and the elapsed time is less
than $t_{\rm dyn}$, we start the outflow again from $R_i$. The reason is that
the $L_0(t)$ increases upto $t_{\rm dyn}$ in our model and hence there is still
possibility of creating a new supper bubble which can escape the halo potential.

Having drawn the basic framework of our calculations, in the following section
we discuss various generic properties of the outflow solutions in our models.
We also test our model equations, and its starting conditions by comparing to
the known self-similar evolution when both $L_0(t)$ and $\rho_B(R)$ have a  
scale-free power law form.
We show that the outflow dynamics is robust to fairly large variations
of $R_i$, $v_i$ and $t_i$ around the above fiducial values. 

\section{Structural properties of the outflow}

\label {sec_wind_profile}
\begin{figure}
\centerline{
\includegraphics[width=0.45\textwidth]{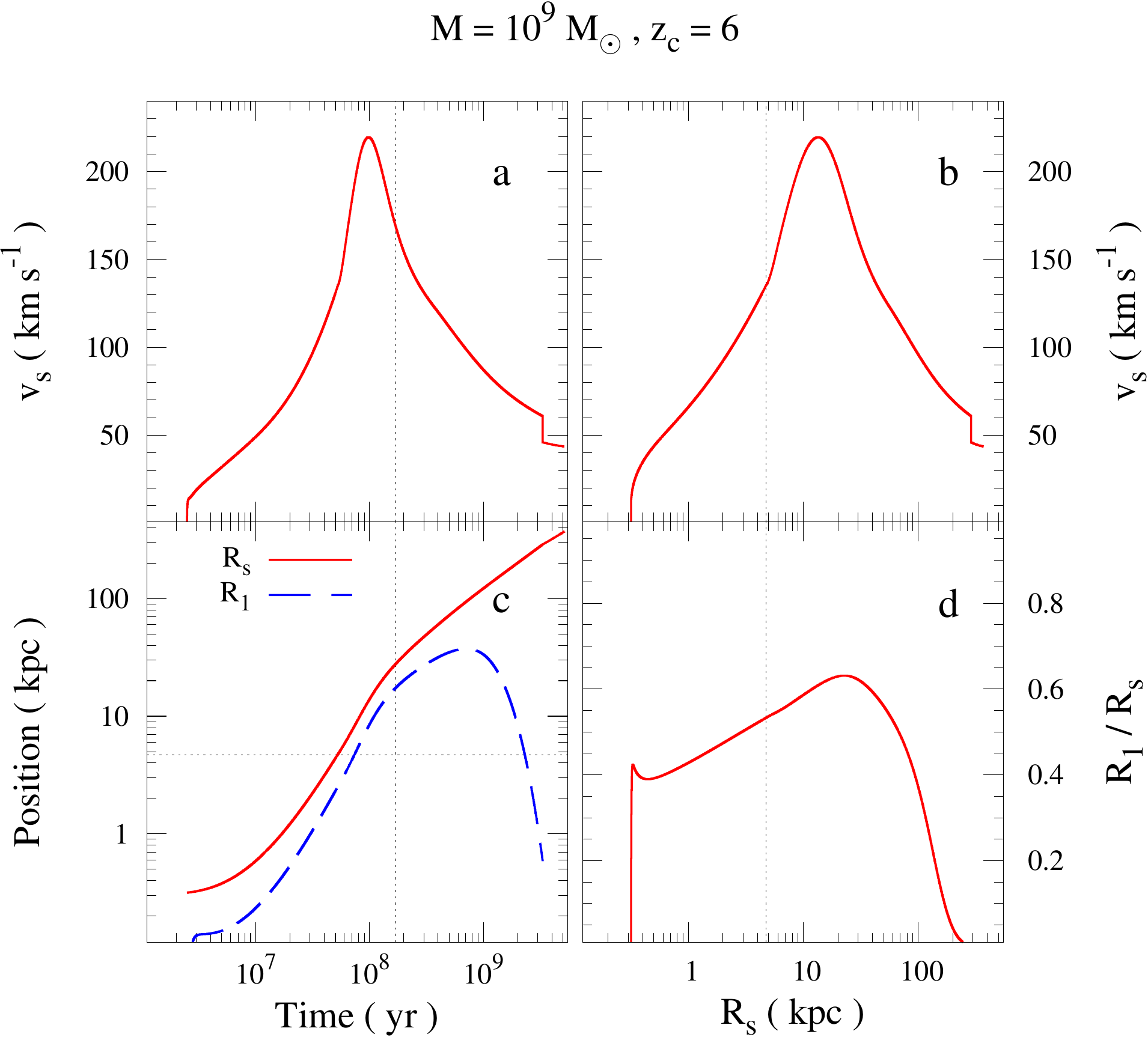}
}
\caption[]{ 
Properties of an outflow originating from a
galaxy of dark matter halo of mass $10^9M_{\odot}$ that has collapsed
at $z_c = 6$. Panel (a) shows the time evolution of the shock velocity ($v_s$),
while panel (c) shows the outer shock location ($R_s$) and the 
inner shock location ($R_1$) as a function of time.
The velocity $v_s$ and the ratio $R_1/R_s$ as a function of the 
outer shock location are shown in panel (b) and (d) respectively.
The vertical lines in panel (a) and (c) mark the dynamical time-scale
for this halo. The vertical lines in panels (b) and (d) and
the horizontal line in panel (c) mark the virial radius.
}
\label{fig_profile}
\end{figure}
\begin{figure}
\centerline{
\includegraphics[width=0.45\textwidth]{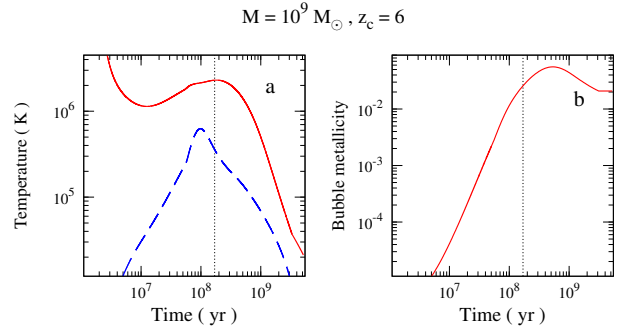}
}
\caption[]{ The time evolution of hot bubble temperature (solid line)
and shell temperature (dashed line) are shown in panel (a) for
the outflow originating from $10^9~M_\odot$ halo that has
collapsed at $z_c = 6$}. Panel (b) shows the metallicity evolution
of the bubble for the same outflow. The vertical dotted lines in both the panels
represent the dynamical time-scale for the halo.

\label{fig_T_Z}
\end{figure}
In this section, we illustrate typical structural properties of galactic
outflows, by focusing on the case of 
a $10^9 M_\odot$ halo collapsing at $z_c=6$. We will find that halos having 
masses around this value, dominate in determining the global consequences of
outflows.
For this halo the virial radius is $R_{{ \rm vir}} = 4.7$ kpc.
The evolution of various physical quantities are
shown in Fig.~\ref{fig_profile} and Fig.~\ref{fig_T_Z}. 
We have taken $f_* = 0.5$, $\kappa = 1$, $f_h = 0.1$,
$\epsilon=0.9$, $\epsilon_{\rm w} = 0.1$, $\eta = 0.3$, 
$\nu^{-1}=50~M_\odot$ and the initial 
conditions are set as described above, with $R_i=R_{{ \rm vir}}/15$, $v_i=0$.
We call it as our fiducial model.

Panel (a) of Fig.~\ref{fig_profile} shows the time evolution of the shell velocity ($v_s$).
The outflow gets accelerated initially, due to the increasing energy input 
from the galaxy and decreasing halo density profile. This phase 
lasts typically for a timescale $\sim t_{\rm dyn}$ when $L_0(t)$ increases,
after which the outflow starts to decelerate.
Further, the acceleration period is mostly when the shock is 
within the virial radius of the halo or very close to the virial radius. 
To make this more clear we have plotted the velocity of the outflow
as a function of location of the outer shock in panel (b)
of Fig.~\ref{fig_profile}. The virial radius
is shown here with the dotted vertical line. 
At the virial radius the density of outside medium changes sharply, 
which results in steepening  in the slope of the velocity profile.
The outflow subsequently decelerates till the 
peculiar velocity decreases to the sound speed of the IGM,
at which stage we assume it freezes into the Hubble flow.
In the above example this happens at $3\times 10^9$
years when the radius of the outflow is $\sim 290$~kpc.
By $z=1$ the outflowing material has spread
to a proper distance of $\sim 370$~kpc from the galaxy.

In panel (c) of Fig.~\ref{fig_profile} we show the time
evolution of $R_1$ (dashed line) and $R_s$ (solid line).
In this model $R_1$ becomes $\sim 0.6 R_S$ 
over a period of few dynamical time scale, as seen from
panel (d). When the injection of mechanical luminosity stops $R_1$ becomes
zero as there is no ram pressure to support the bubble pressure.

The evolution of the bubble temperature, $T_b$, (solid line in panel (a)
of Fig.~\ref{fig_T_Z}) is governed by competing effects.
Adiabatic expansion and radiative cooling lead to a decrease of $T_b$, 
while energy input from the galaxy can result in its increase.
The radiative cooling depends on the metallicity of the gas. 
In panel (b) of  Fig.~\ref{fig_T_Z} we show our computed bubble metallicity
as a function of time. The details of our metallicity calculations are
given in Appendix~\ref{metal_evolution}.
In most of our models we use the Carbon as the metallicity indicator with
an yield of $0.1~M_\odot$ per solar mass of SNe. Initially 
the metallicity increases  with time as most of the bubble 
material comes from the galaxy. The maximum metallicity of the bubble gas is
$\sim0.05~Z_\odot$. In the later stages the metallicity begins to decrease 
when the star formation in the halo stops while more and more
primordial  gas gets added to the bubble.
In the calculations presented above, 
the radiative cooling of the bubble gas is calculated assuming the 
average metallicity of the gas to be $0.01~Z_\odot$. 
In Fig.~\ref{fig_T_Z},  $T_b$ starts from an initial value
$\sim 5 \times 10^6$~K (determined by the mass loading factor $\eta$), 
decreases for $t \sim 10^7$ yrs, before increasing again due to the
growing SFR of the galaxy till $t\sim t_{\rm dyn}$ (Eq.~\ref{eqnsf}).
After $t\sim t_{\rm dyn}$ the SFR starts to decrease and 
the effect of adiabatic expansion dominates over the energy
input to the bubble, leading to decrease of $T_b$. We notice that
the results do not change even when we use the average metallicity
to be $0.1~Z_\odot$. This is mainly because at $T_b>10^6$ K the 
adiabatic cooling is much faster than the radiative cooling.

The dashed curve in panel (a) of Fig.~\ref{fig_T_Z} is the post shock
temperature of the shell assuming it to be an adiabatic shock. As this temperature
depends on the velocity of the shell it follows the time evolution of the $v_s$ 
shown in panel (a) of Fig.~\ref{fig_profile}.  In the case of thin shell 
approximation we expect the swept up shell material to cool very efficiently.
We check this in the following subsection.

We note that the pressure of the bubble $P_b$ is always
a monotonically decreasing function of time. This arises due to the
fact that the density inside the bubble always decreases and this decrease
is faster than the increase of temperature at any time in the evolution of the
outflow. 

\subsection{Cooling of the shell gas}

\begin{figure}
\centerline{
\includegraphics[width=0.45\textwidth]{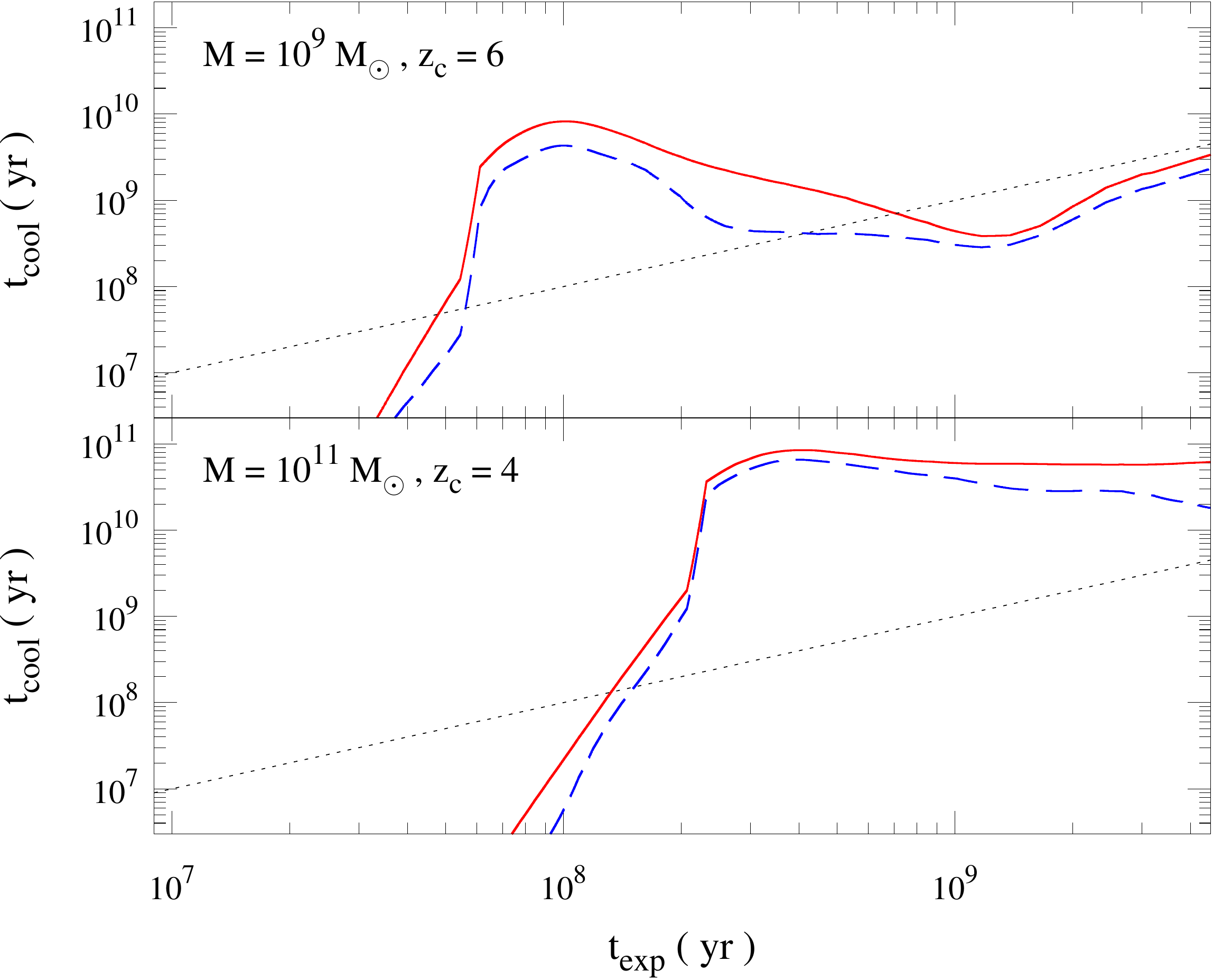}
}
\caption[]{Top and bottom panels show the cooling time for the shell 
originating from halos of mass $10^{9}~M_\odot$  and
$10^{11}~M_\odot$ respectively. The solid and dashed curves
are for primordial abundance and $Z = 10^{-2}~Z_\odot$ respectively. 
The dotted lines in both the panels correspond to
cooling time equals to the expansion time. The collapse
redshifts, $z_c$, are indicated in each panel.
}
\label{fig_cooling}
\end{figure}

It is also of interest to examine whether the swept up gas in the
shell can cool efficiently. In Fig.~\ref{fig_cooling} we show the cooling time
($t_{\rm cool}$) of the shell gas as a function of the expansion time
($t_{\rm exp}$) of the outflow, assuming the shell density is simply $4$
times the pre-shock density,
(as would obtain for a strong adiabatic shock of a $\gamma=5/3$ gas)
and its temperature is the post-shock temperature. The cooling rate
has been taken from Sutherland \& Dopita (1993) assuming primordial
abundance (solid lines) or $Z = 10^{-2}~Z_\odot$ (dashed lines).
The latter case is to illustrate the effect of having some metals in the halo gas, 
perhaps due to enrichment from earlier generation of outflows,
or from partial mixing with the metal rich bubble. We have also 
plotted as a dotted line the relation $t_{\rm cool} = t_{\rm exp}$.
The top panel is for a $10^9M_\odot$ halo that has been discussed in 
detail above. The bottom panel is for a $10^{11}M_\odot$ halo
collapsed at $z_c = 4$ with all other parameters same as that of the
fiducial model used for the illustration above. 
\begin{figure*}
\centerline{
\includegraphics[width=0.95\textwidth]{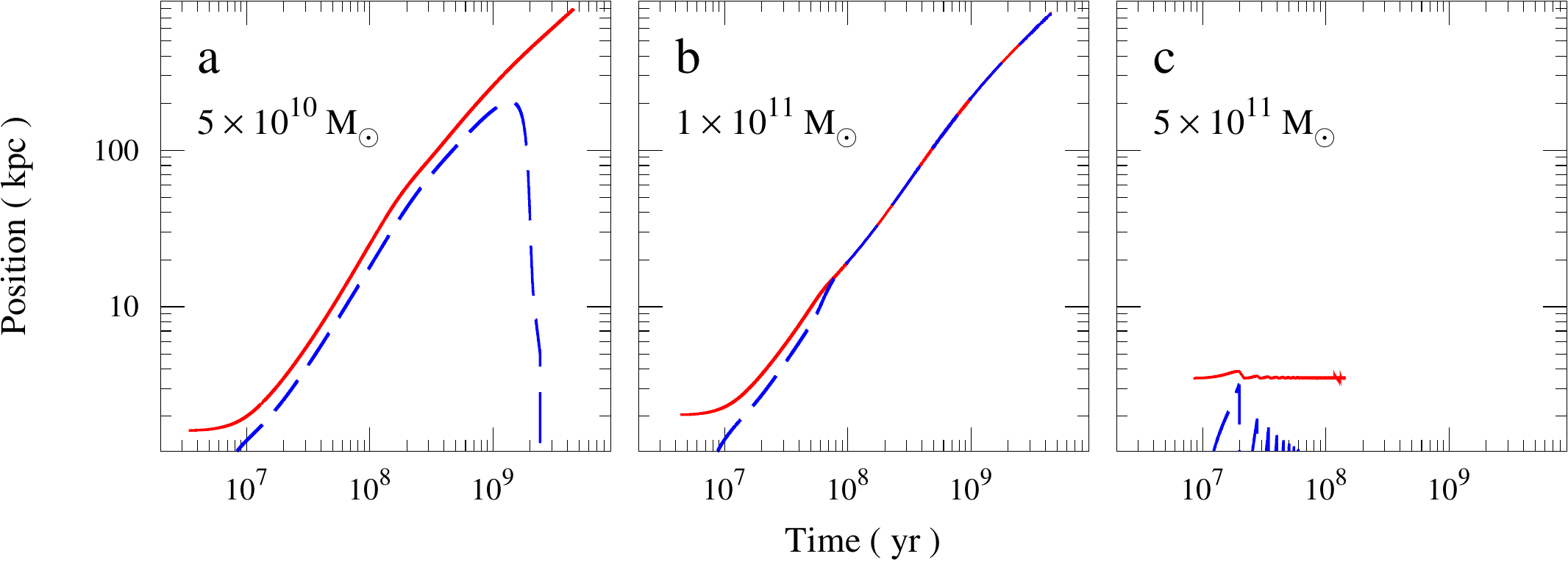}
}
\caption[]{
Time evolution of $R_s$ (solid lines) and $R_1$ (dashed lines) for
three different halo masses.  All the curves are drawn for $f_*=0.25,~\kappa
= 0.5,~\eta = 2.5,~ f_h = 0.02,~\epsilon_w = 0.15$, bubble metallicity
$Z=0.03~Z_\odot$ and $z_c=4$ but different halo masses as indicated in
the corresponding panels. In the case of $M = 5\times 10^{10}~M_\odot$
the flow is a pressure driven flow through out (panel (a)).
However for $M = 10^{11}~M_\odot$
the flow transits from a pressure driven flow to the momentum 
driven flow (panel (b)).  In the case of  $M = 5\times 10^{11}~M_\odot$
the outflow is not set as the gas cools much faster.
}
\label{fig_transit}
\end{figure*}

One sees that in the absence of radiative heating by the
UV flux from galaxy the shell can indeed cool efficiently with 
$t_{\rm cool} < t_{\rm exp}$ while the shock traverses much of the halo.
The epoch when the shock leaves the halo is the time when 
$t_{\rm cool}$ rises abruptly. Just around this epoch or just before,
one gets $t_{\rm cool} = t_{\rm exp}$. One can also see that the
cooling becomes inefficient in the IGM ($t_{\rm cool} > t_{\rm exp}$). 
Note that the density in the shell is not expected to be
uniform while the outflow traverses the declining density distribution in the
halo. Indeed self-similar solutions for the adiabatic
evolution of the shell in a medium with declining density profiles,
by Ryu and Vishniac (1991) show that the
density (temperature) in the shell rises (falls) steeply
as one goes from the shock at $R_s$ to the contact discontinuity $R_c$.
For example, for the case $L(t) = constant$, $\rho_B(R) \propto R^{-2.5}$ 
(see Fig. 1e in Ryu and Vishniac 1991), the density (temperature) is a factor of
2 larger (smaller) than the value at the shock, for the inner half of the shell.
We have also obtained the shell structure assuming an adiabatic
strong shock for our model parameters in a self-similar outflow
(see appendix~\ref{sec_shell_struc}).
When the outflow is traversing the halo, we also get a similar steep density 
enhancement in the shell compared to its density at the shock.
Assuming $\Lambda \propto T^{-1/2}$ in the relevant temperature range,
the density and pressure profile together make the $t_{\rm cool}$
smaller by a factor $\sim 5.7$ for the inner half of the shell.
So our estimates of the shell cooling times, given in Fig.~\ref{fig_cooling}, 
are more upper limits while the outflow traverses the halo. In addition, since 
the shell/hot bubble interface is likely to be unstable, leading to mixing 
(see below), the metallicity of the shell gas could also be much higher
than we have assumed. Overall it appears reasonable to assume
that the shell can cool efficiently while it is traversing the halo,
but not in the IGM. Later when the outflow has
gathered up mass from the IGM and slowed down sufficiently, 
$t_{\rm cool}$ can again become smaller than $t_{\rm exp}$.
In case of efficient cooling, the thin shell density will be enhanced
to $\rho_s \sim \rho_B M^2$, where $M$ is the Mach number of the shock.
These results will be useful in our discussions below.

\subsection{ Transition from pressure to momentum driven flow}

We note that the basic trend seen in the illustrative example 
discussed till now is the case with most of our models.  
That is throughout the evolution, the flow is pressure driven.  
However, in few cases, $R_1$ can become close to 
$R_s$ and the outflow can transit to a ``momentum driven case''.
This arises when the radiative cooling of the hot
bubble gas is efficient.  When the bubble is well within the halo
its density and metallicity depends mostly on the wind material.
As we have already mentioned, the recombination 
line cooling depends on metallicity, temperature and 
density of the hot bubble.  Higher density and lower temperature
that favors cooling can be achieved by increasing the mass 
loading of the bubble (i.e higher value of $\eta$).

To illustrate these points we have chosen three examples and show
the time evolution of $R_1$ and $R_s$ in Fig.~\ref{fig_transit}.
For all the three examples, we have assumed $f_*=0.25,~\kappa
= 0.5,~\eta = 2.5,~ f_h = 0.02,~\epsilon_w = 0.15, $
the metallicity of the bubble gas  $Z=0.03~Z_\odot$ and
$z_c=4$.  The halo masses are indicated in the corresponding
panel. 
The initial temperature of the gas is $\sim10^6$ K in all the
models as $\eta$ is same in all cases. However, bubble 
density is higher in the case of higher mass halo.
In the first case, for $M=5\times10^{10}~M_\odot$,
the flow is completely pressure driven (panel (a) of Fig.~\ref{fig_transit}). 
Here $R_1$ is $\sim 0.7 R_s$. The time evolution of various physical
quantities follow the example discussed before.
In the second example, for $M=10^{11}~M_\odot$,
$R_1\simeq R_s$ at $t = 10^8$ yr (panel (b) of Fig.~\ref{fig_transit}).
Afterward the flow transits to a momentum driven wind case. 
In our models the mass loading into the bubble is
proportional to the halo mass.
The main difference in this model compared to the model discussed in panel (a)
is that the bubble density is a factor 4 higher in the second
case. This makes the radiative cooling rate a factor 4 times
faster. This is sufficient to cool the bubble gas to $T = 10^4$~K
in $\sim 8\times 10^7$ yrs. As the shell+bubble gas is already 
been accelerated to high velocity and mechanical luminosity is 
still raising (as $t<t_{\rm dyn}$) the gas continuous to move outwards.
When we consider $M=5\times10^{11}~M_\odot$ (panel (c) of Fig.~\ref{fig_transit})
the initial gas density is higher by a factor 7. At this density the gas
cools to $10^4$~K within $1.5\times10^7$ yrs.  This time is very
short and gas does not have sufficient energy to initiate an
outflow. So the bubble gas remain confined to the galaxy
itself.

In principle one can also get pressure driven flow transiting into a momentum
driven flow in cases where there is high mass loading with higher metallicity.
However, we find that for a wide range of model parameters the pressure
driven flow is more generic than a flow which transits to momentum driven case
(especially because we take $\eta=0.3$ for most models).
In any case, as shown in the above example, 
our calculations switch to correct set of equations as and 
when the outflow transits to a momentum driven case.

\subsection { Comparison with scale-free solution}
\label{sec_comp}
The thin shell equations allow for a power-law solution when
$L_0(t)$ as well as $\rho_B(R)$ are scale-free, provided we ignore 
gravity, outside pressure and the effect the inner shock.
Suppose the halo gas density goes as $\rho_B \propto R^{-\alpha}$
and $L_0(t) \propto t^{\beta}$, then simple scaling argument suggests that 
the outflow radius will scale as
$R_s(t) \propto (L_0t^3/\rho_B)^{1/5} \propto t^n$, where $n = (3+\beta)/(5 -\alpha)$.
Acceleration of the outflow obtains for $n>1$ or $\alpha + \beta > 2$, while
deceleration obtains in the opposite limit. 
For example, if $\alpha = 2.8$ and $\beta =1$ (as one expects
within the virial radius of the halo and at early times in our model), one
expects $R_s(t) \propto t^{2/1.1}$, $v_s(t) =\dot{R}_s(t)\propto t^{-0.9/1.1}$ and
$P_b(t) = E_b/(2\pi R_s^3) \propto t^{-38/11}$.

\begin{figure}
\centerline{
\includegraphics[width=0.45\textwidth]{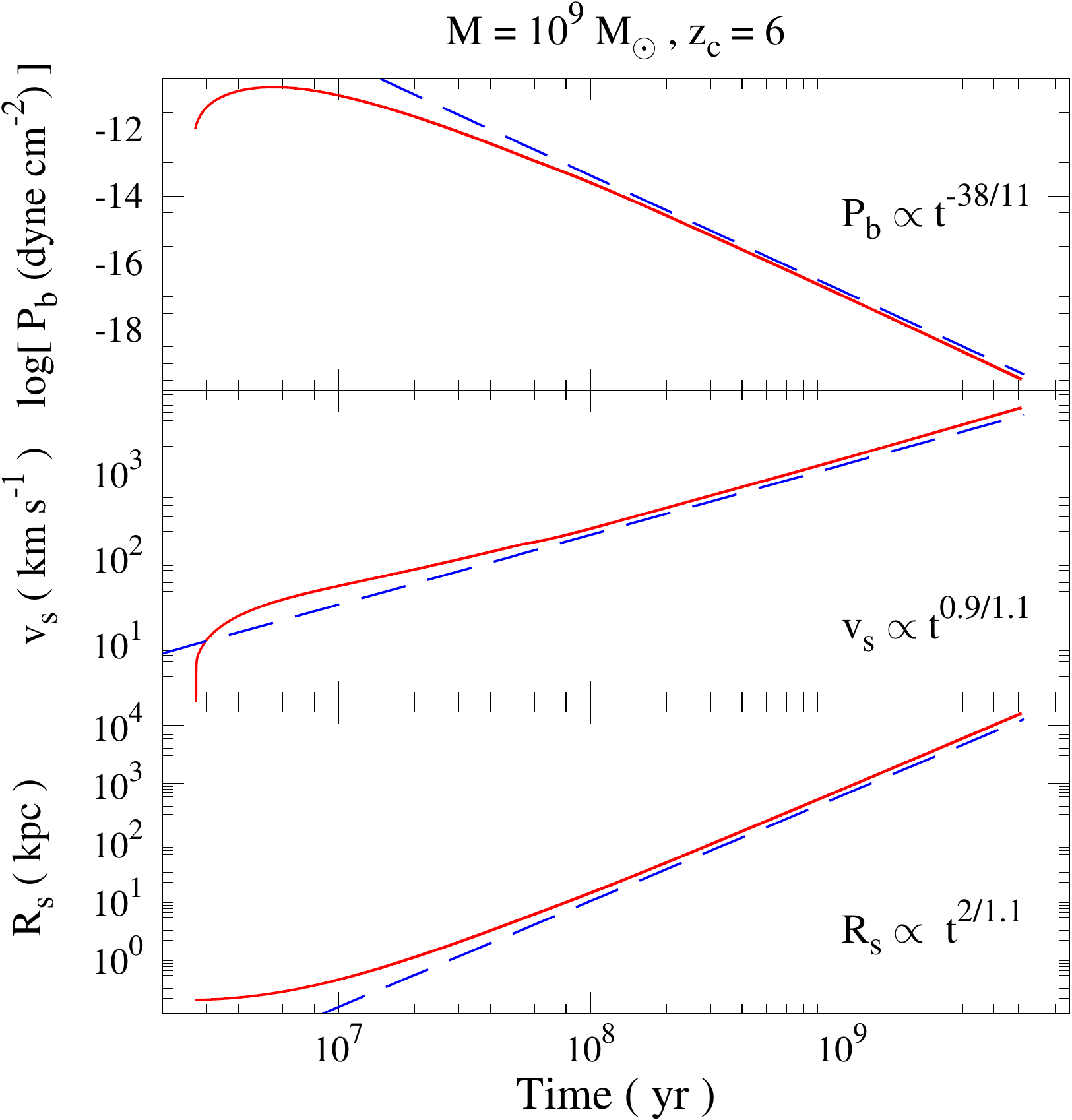}
}
\caption[]{The asymptotic self-similar behavior of properties of the
outflow. The position of outer shock ($R_s$), velocity ($v_s$)
and hot bubble pressure ($P_b$) of the outflow coming out
from a halo of mass $10^9~M_\odot$
are shown in bottom, middle and top panel respectively. The input energy
is assume to be $L_0(t)\propto t$ and the outside gas density
is $\rho_B\propto R^{-2.8}$. The corresponding
expected self-similar power-law solutions are shown by the dashed line.
The power-law relationships expected from the scale free solutions
are also given in the corresponding panels.
}
\label{fig_selfsimilar}
\end{figure}

As a simple test of some aspects of our model, 
we first show in Fig.~\ref{fig_selfsimilar}, 
the evolution of an outflow with $L_0(t) \propto t$ and $\rho_B \propto R^{-2.8}$ 
and compare it with the expected scalings of $R_s(t)$, $v_s(t)$ and $P_b(t)$ 
derived above. Here we have also used the initial conditions
as explained earlier. We see that, after a short initial period, there is indeed
excellent agreement of the computed evolution of $R_s$, $v_s$ and $P_b$, 
with the expected scaling laws derived above.

\begin{figure}
\centerline{
\includegraphics[width=0.45\textwidth]{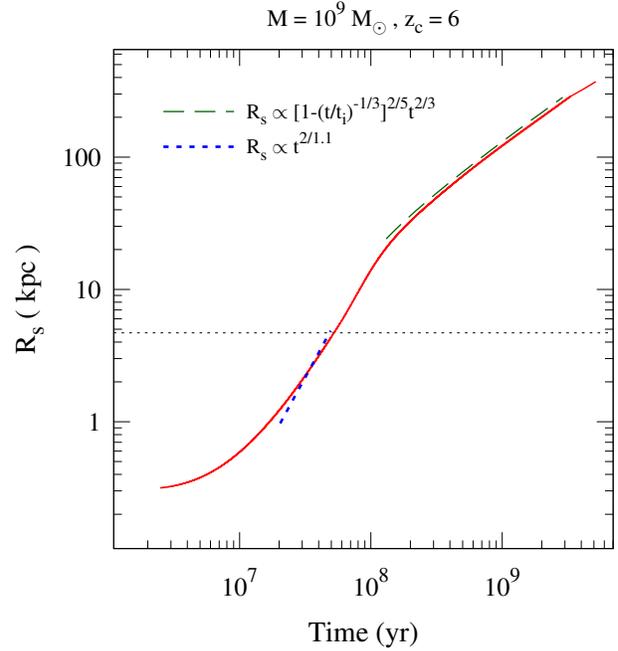}
}
\caption[]{
The actual solution for the time evolution of $R_s$ in our model.
The expected self-similar
behavior are shown for two different regimes. Within the
halo, $R_s (t) \propto t^{2/1.1}$ (short dashed line) and far away
from the halo, $R_s(t) \propto [1 - (t/t_i)^{-1/3}]^{2/5} t^{2/3}$
(long dashed line) with $t_i = 5 \times 10^7$~yrs.
}
\label{fig_fit}
\end{figure}

We also examine in Fig.~\ref{fig_fit} the actual scaling behavior
of the outflow radius, $R_s(t)$, for a halo of 
$10^9M_\odot$ considered above, now incorporating gravity,
external pressure and the effect of the inner shock.
One sees that, inside $R_{{ \rm vir}}$, $n$ is somewhat smaller than that expected
for the pure self-similar evolution law, assuming
$\rho_B\propto R^{-2.8}$ and $L_0(t)\propto t$. 
This is to be expected due to the
influence of gravity and outside pressure, which both go to slow down
the expansion of the outflow.
At distances much larger than $R_{{ \rm vir}}$, 
the gas density decreases with the
expansion of the universe approximately as $\rho_B \propto t^{-2}$
and $L_0(t)=0$. Even if we assume negligible radiative cooling of the hot bubble,
$E_b$ would decrease due to the expansion of the universe.
This case has been analyzed by Voit (1996), by a conformal
transformation of the fluid equations. Voit finds that in the
scaled variables the structure of the shocked IGM gas would follow
the usual Sedov self-similar solution. In terms of the actual time
variable, Voit predicts that for a flat matter dominated universe,
$R_s(t) \propto [1 - (t/t_i)^{-1/3}]^{2/5} t^{2/3}$.
Such an evolution is indeed close to that
obtained for the actual outflow, as can be seen from Fig.~\ref{fig_fit}.


We now examine the sensitivity of the outflow evolution to the initial
conditions that we have adopted.

\subsection {Sensitivity to the adopted initial conditions}
\begin{figure*}
\centerline{
\includegraphics[width=0.95\textwidth]{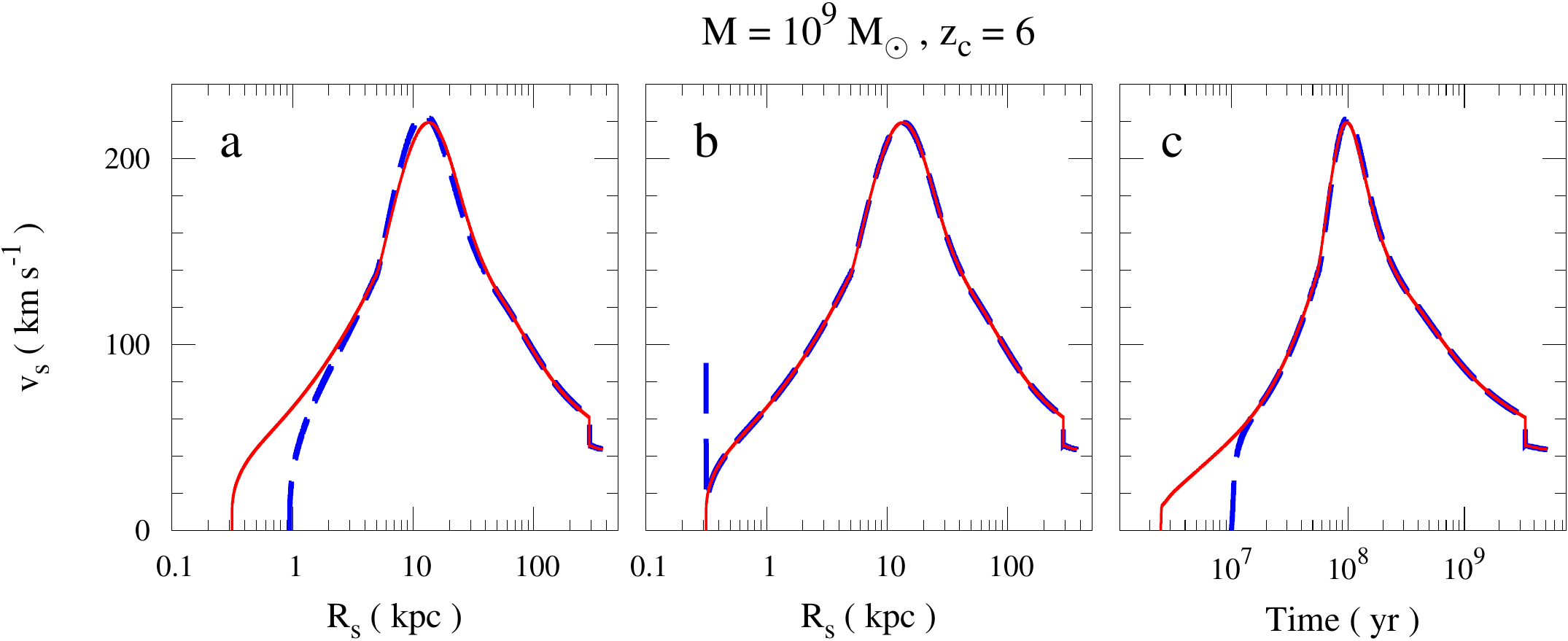}
}
\caption[]{Effect of using different initial conditions on the outflow
evolution. Panel (a) shows the effect of changing the initial radius
from $R_{{ \rm vir}}/15$ to $R_{{ \rm vir}}/5$. Panel (b) shows the
effect of changing the initial velocity from $0$
to $3v_c$. Panel (c) shows the effect of changing initial starting time.
In all the panels the results for our fiducial model are shown by the
solid lines and the evolution of the modified models are shown
by dashed lines.
}
\label{fig_ins_cond}
\end{figure*}

Let us begin with the sensitivity of the outflow solution to
the adopted initial radius of the supper bubble. In panel (a) of
Fig.~\ref{fig_ins_cond}, we show the shock velocity $v_s = \dot{R_s}$ 
as a function of the shock radius $R_s(t)$, for a $10^{9}~M_\odot$ halo collapsing
at $z_c=6$ discussed above. The solid line is for $R_{i}=R_{{ \rm vir}}/15$
while the dashed lines is for $R_{i}=R_{{ \rm vir}}/5$ keeping all other
parameters as in the fiducial model.
It is evident from the figure that a modest change in $R_i$
has very little effect in the evolution of the outflow.

Panel (b) of Fig.~\ref{fig_ins_cond} looks at the effect of changing the initial
velocity $v_i$ from $v_{i}=0$ (solid line) to $v_{i} = 3v_c$ (dashed line)
with other parameters as in the fiducial model.
Starting the evolution of the outflow with such a larger $v_i$ has very
little effect on the time evolution of the outflow.
This is because the mass loading in the shell very rapidly decreases
the velocity to that which can be consistently maintained with the
existing bubble pressure.

Finally, panel (c) of Fig.~\ref{fig_ins_cond} shows the effect
increasing $t_i$ by a factor $5$. After an initial period the shell
velocity again latches on the fiducial model solution.
And the radius to which the wind propagates is almost independent
of $t_i$ for a reasonable range around our fiducial value.
Therefore we conclude that the final results of our models are almost
insensitive to the initial conditions, for reasonable variations around
our fiducial values.  

In the following section, we study how the model parameters
influence the the outflow properties, in particular the extent of 
the outflow ($R_s$).

\section{Dependence of outflow properties on model parameters}

The free parameters in our model are,
$f_*$, $\kappa$, IMF, $\epsilon$, $\eta$, $f_h$, and $\epsilon_w$
in addition to the background cosmological parameters.  As pointed out
before, for most part of this paper we  will use cosmological parameters 
from the WMAP 3rd year data and  $f_*=0.5$ and $\kappa = 1$ that
fits the observed galaxy luminosity functions at $3 \le z \le 6$.
We are left with IMF and 4 parameters of the model that are associated
with wind dynamics. In this section we explore the dependence of our
results on the choice of these parameters. In addition we also explore
models that have star formation in a burst mode.

\subsection{The halo mass fraction}

\begin{figure}
\centerline{
\includegraphics[width=0.45\textwidth]{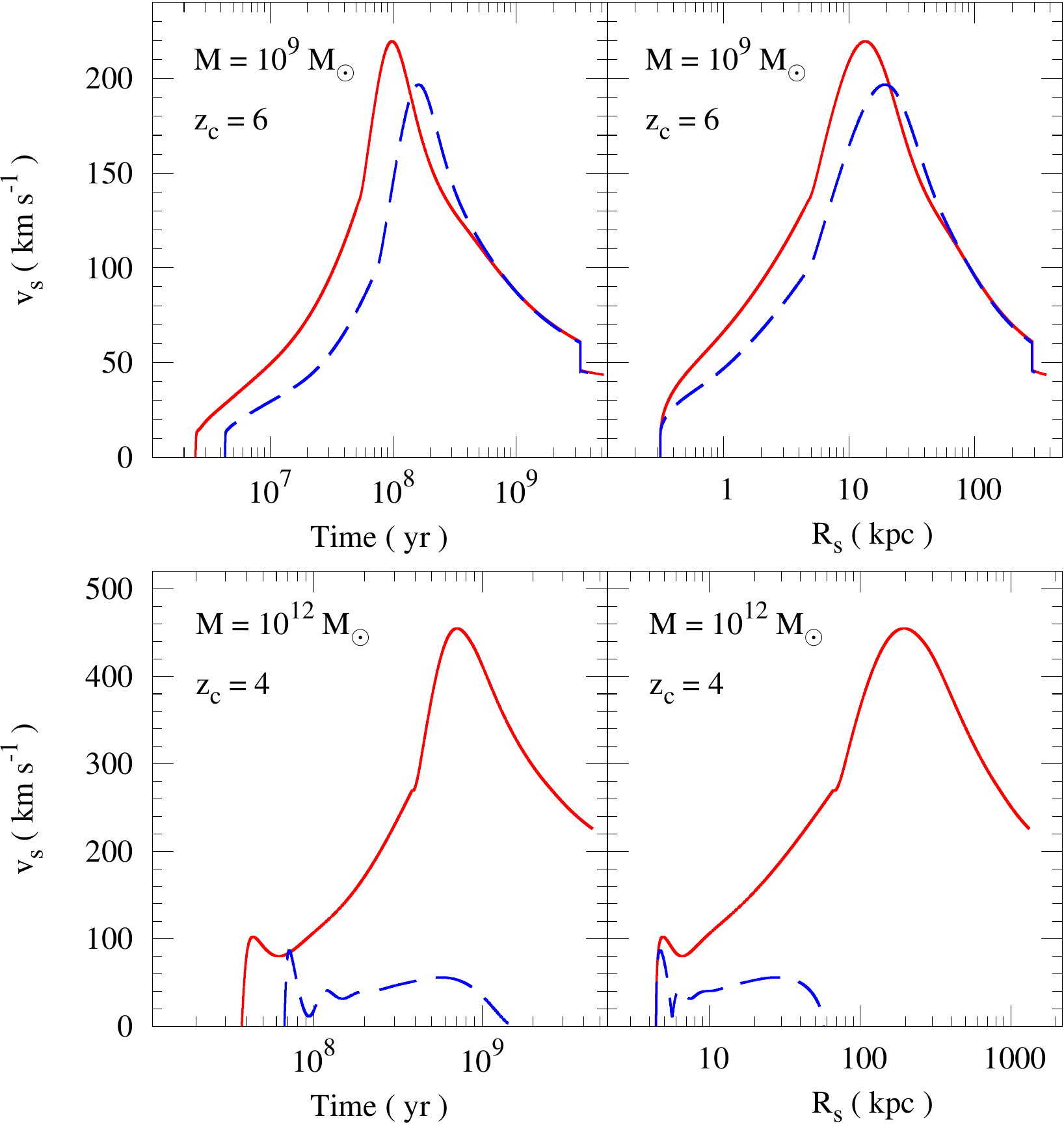}
}
\caption[]{Effect of halo mass fraction on the outflow. The top panels show the
evolution of $v_s$ as a function of time (left panel) and $R_s$ (right panel)
of the outflow coming out from a halo of mass $10^9~M_\odot$.
The solid lines are for $f_h = 0.1$ and the dashed lines
are for $f_h=0.3$. In the bottom panels we show the corresponding curves
for the outflow origination from $10^{12}~M_\odot$ halo.
}
\label{fig_halo}
\end{figure}

We first consider how the outflow properties change when we change
 the halo fraction, $f_h$. Fig.~\ref{fig_halo} shows the velocity
evolution characteristics for two halos of masses $10^9$ (top panels) and
$10^{12}~M_\odot$ (bottom panels), for both $f_h =0.1$ (solid line)
and $f_h=0.3$ (dashed lines). Higher value of $f_h$ increases the
pressure in the halo ($P_0$). Thus our initial condition that bubble pressure
equals to the halo pressure is achieved in a latter time compared to
the model with smaller $f_h$.  The mass loading of the shell when 
it propagates inside the halo is larger in the case of high $f_h$.
For the $10^9 M_\odot$ halo, this leads to a 
lower peak velocity in the case of high $f_h$ (see 
upper panels of Fig.~\ref{fig_halo}). It is however interesting to see
the final velocity and radius of the outflow remains the same in both
cases. This is mainly because when the shell travels in the IGM most
of its mass comes from the swept up IGM material and
the final velocity and radius are insensitive to $f_h$. 
The above conclusion holds provided that the initial decrease 
of $v_s$ does not lead to confinement of the outflow within $R_{{ \rm vir}}$.
Such a confinement is in fact obtained for the higher mass halo
shown in the bottom panels of Fig~\ref{fig_halo}.
For this halo of mass $10^{12}~M_\odot$, $f_h=0.1$ leads to an
outflow which escapes the halo while increasing to an $f_h=0.3$
leads to confinement of the wind.
Hence the value of $f_h$ could crucially impact upon whether one has
an outflow or not,  especially for high mass halos. 

\subsection {Mass loading from the galaxy}

\begin{figure}
\centerline{
\includegraphics[width=0.45\textwidth]{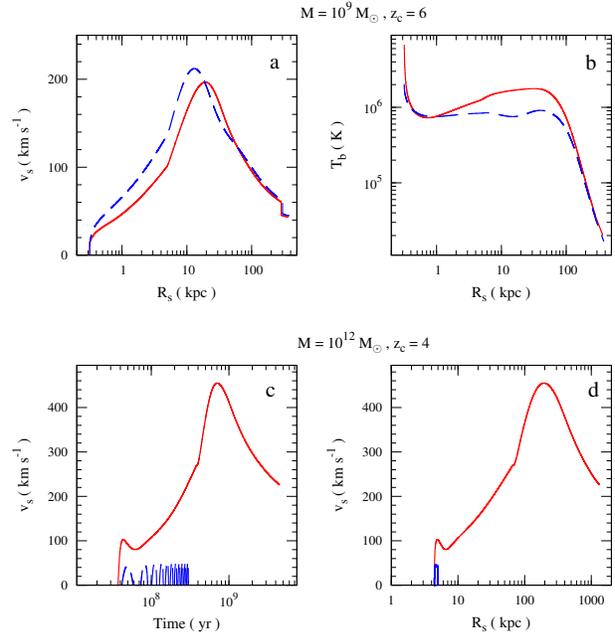}
}
\caption[]{Effect of $\eta$ on the outflows.
Panel (a) and (b) show respectively the shell velocity, $v_s$ and hot bubble
temperature, $T_b$ as a function of $R_s$ for the outflow coming out from a
$10^9~M_\odot$ halo. Panel (c) shows the time evolution of $v_s$ while panel (d)
shows the $v_s$ as a function of $R_s$ for the outflow originating from a
$10^{12}~M_\odot$ halo. In all the panels solid lines are for $\eta=0.3$ and
the dash lines are for $\eta=1.0$.
}
\label{fig_eta}
\end{figure}

Here we discuss the dependence of the outflow evolution on the mass loading factor,
$\eta$. The top panels of Fig.~\ref{fig_eta} show the results for $M=10^9 ~M_\odot$
for $\eta =0.3$ (continuous curves) and $\eta=1.0$ (dashed curves). We do
not notice much difference in the evolution of $\dot{R_s}$ however
the temperature of the bubble in the case of $\eta = 1$ is lower than that
in the case of $\eta = 0.3$. Higher values of $\eta$ reduces the initial
temperature and increases the density of the bubble when it is inside the
virial radius.  The bottom panels in Fig.~\ref{fig_eta} shows the results for 
$M=10^{12}~M_\odot$. When $\eta = 0.3$ the model produce outflow that escapes the
galaxy. However, for $\eta = 1.0$ the flow is confined. This is mainly because
in addition to the initial temperature being low the bubble gas cools much more
efficiently in the case of $\eta = 1.0$,  due to its increased density.
The other quantity that changes appreciably to the change of $\eta$ 
is the asymptotic metallicity limit. This is higher in the case of high $\eta$
(see Eq.~(\ref{eqn_metals})).
Like in the case of $f_h$, the influence of $\eta$ on the nature of the
flow seems to be strong when one considers high mass halos.

\subsection{Energy input efficiency}
\begin{figure}
\centerline{
\includegraphics[width=0.45\textwidth]{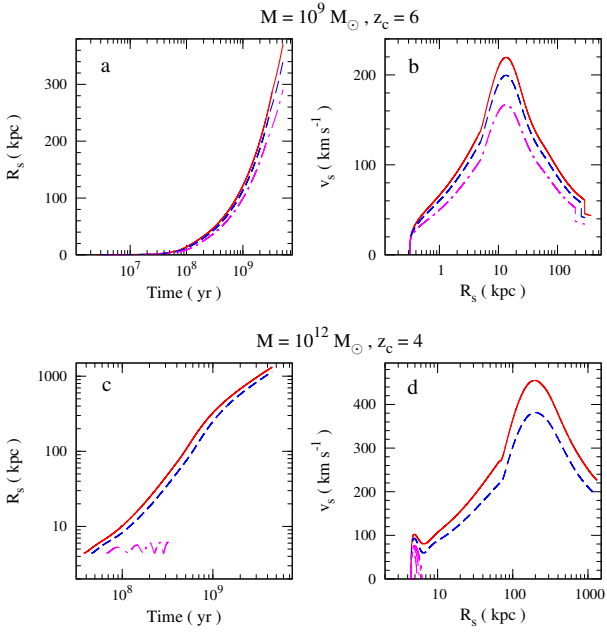}
}
\caption[]{The effect of IMF on outflows. The top panels show
$R_s$ (panel (a)) and $v_s$ (panel (b)) of an outflow
coming out from a halo of mass $10^{9}~M_\odot$ collapsed at $z_c = 6$.
The solid, dashed and dotted-dashed lines are for $1-100~M_\odot$,
$0.5-100~M_\odot$ and $0.1-100~M_\odot$ Salpeter IMF respectively.
The bottom panels show the corresponding results for the outflow
coming out from a $10^{12}~M_\odot$ halo collapsed at $z_c = 4$.
}
\label{fig_IMF}
\end{figure}
Now consider the effect of varying the efficiency of energy
input into the wind which is characterized by the factor 
$(\epsilon_w \nu f_*)$. This can arise for example in changes to the adopted 
IMF, since such a change will lead to different $\nu$.
A $1-100~M_\odot$ Salpeter IMF gives $\nu = 1/50$, while changing the
lower mass cut-off to $0.5~M_\odot$ and $0.1~M_\odot$ leads to
$\nu=1/70$ and $\nu=1/130$ respectively. In most of our models 
discussed till now we have assumed $\epsilon_w$ to be 0.1. A smaller 
value of this efficiency factor will have an effect similar to reducing the
lower mass cut-off in the IMF.
In the top panel of Fig.~\ref{fig_IMF} we have shown the results
from a halo of mass $10^{9}~M_\odot$ with the above three IMFs.
Two major effects are noticeable. The maximum velocity attained by the
outflow and the final outflow radius are both higher for higher $\nu$ 
(i.e IMF with higher values of low mass cut off), 
as expected because of the increase in  input energy.  
Note that the same effect can be obtained by lowering the $f_*$ 
(or  $\epsilon_w$) by a factor 2.6. In the case of $10^{12}~M_\odot$
(bottom pane of Fig.~\ref{fig_IMF}) the material is confined to the halo
for $\nu = 1/130$. Remember this corresponds to standard Salpeter
mass function.  In Paper I, we have argued that the redshift evolution of
UV luminosity function of galaxies can be explained by slowly evolving
low mass cut of in the IMF over the redshift range 3 to 6. Such an
evolution will have important effects on the influence of winds in
the global properties.

\subsection{Burst versus continuous made of star formation}

\begin{figure}
\centerline{
\includegraphics[width=0.45\textwidth]{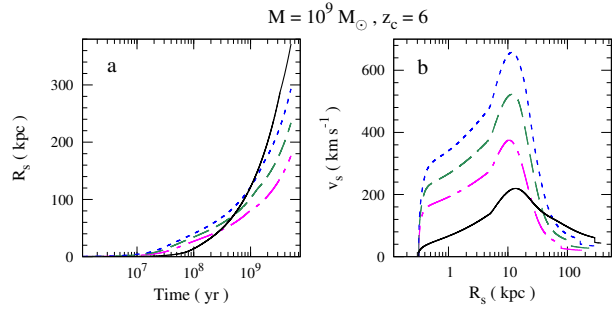}
}

\caption[]{The evolution of an outflow originating from a $10^9~M_\odot$
halo for the burst mode of star formation.
The time evolution of $R_s$ is shown in panel (a) while panel (b)
shows $v_s$ as a function of $R_s$. The short-dashed, long-dashed
and dotted-dashed lines are for $\epsilon_{SF} = 0.5$, $0.25$ 
and $0.1$ respectively. The other parameters are same as our
fiducial model. For comparison we have shown as a solid line, the evolution of the
same outflow in a continuous star formation model with $f_*=0.5$.
}
\label{fig_burst_rv}
\end{figure}

We now consider the effect of burst mode of star formation on the
wind dynamics. In Fig.~\ref{fig_burst_rv}
we show the evolution of an outflow emerging from a halo of
mass $10^9~M_\odot$ collapsed at $z_c=6$. Panel (a) of Fig.~\ref{fig_burst_rv}
shows the radius, $R_s$ of the outflow as a function of time for
$\epsilon_{SF} = 0.5$ (short-dashed line), $0.25$ (long-dashed line)
and $0.1$ (dotted-dashed line). The other parameters are same as our
fiducial model. For comparison we have shown as a solid line, the evolution of the
same outflow in a continuous star formation model with $f_*=0.5$.
It is evident from the figure that the maximum velocity achieved
by the outflow is much higher in the case of burst mode of star formation
compared to the continuous one. However the final radius of the outflow
is smaller in the burst model. Further, as one expects, both the maximum velocity
and the final radius of the outflow decrease with lower values of
$\epsilon_{SF}$. Also note that the duration of initial acceleration phase
is shorter in case of the burst model.
The smaller size to which burst models drive outflows will have
consequence on their volume filling factor (see below).

\subsection {Results for different halos}

We now consider outflows originating in halos of different mass scales.
We assume the parameters of SFR and wind models identical to the
fiducial model discussed in section~\ref{sec_wind_profile}.
For each mass we take the collapse redshift to 
be approximately the epoch when a $3\sigma$ fluctuation becomes unity.
\begin{figure}
\centerline{
\includegraphics[width=0.45\textwidth]{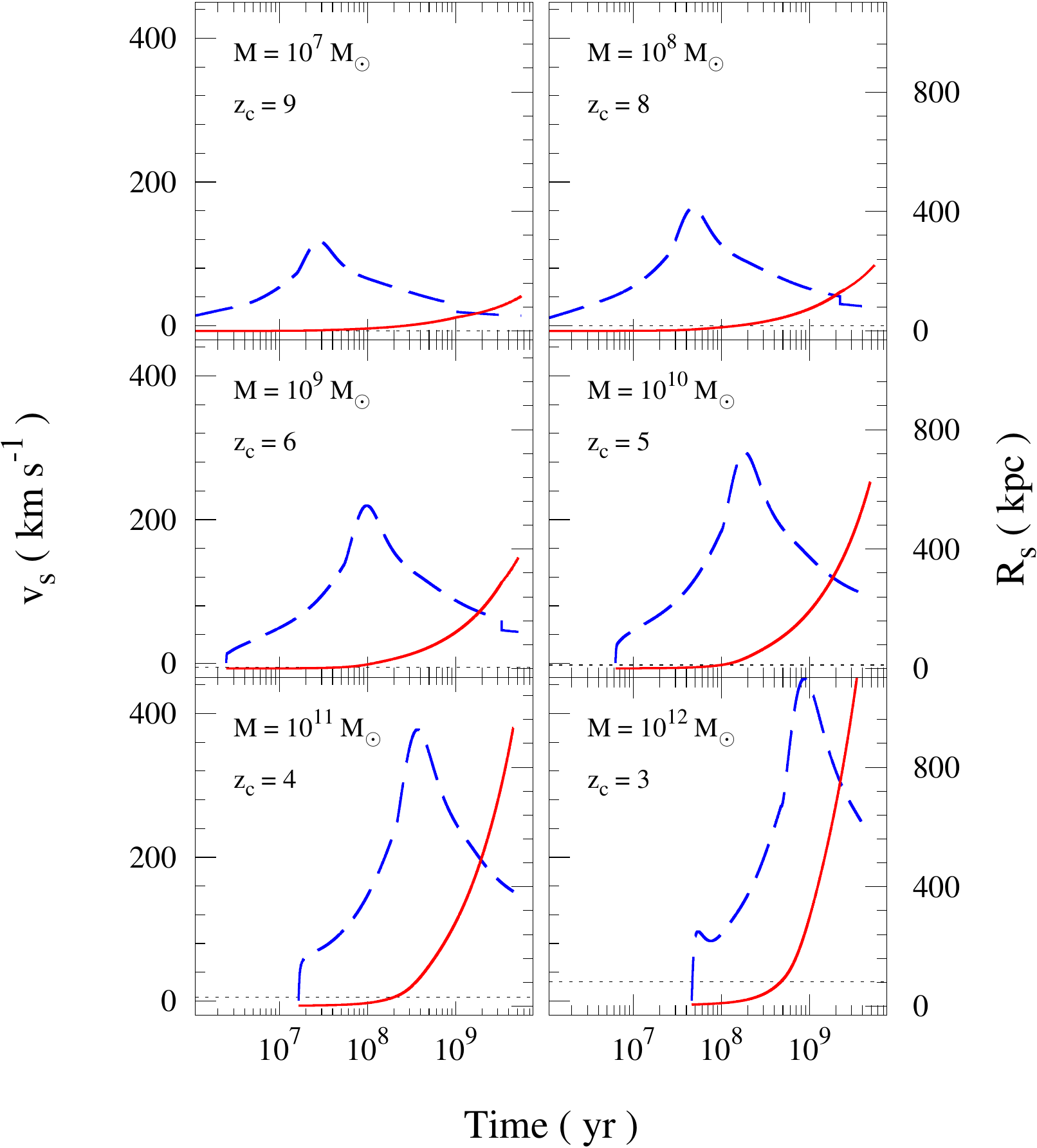}
}
\caption[]{The time evolution of the radius (solid lines) and the 
velocity (dashed lines) of the shell arising from halos of
different masses. The mass of the halo and its collapsed
redshift are given in each panel The horizontal dotted lines in
each panel represent the virial radius of these halos.
}
\label{fig_mass}
\end{figure}
In Fig.~\ref{fig_mass} we have shown the time evolution of  
the velocity ($v_s$) and radius ($R_s$) of outflows arising
from halos of different masses, collapsing at redshifts, $z_c$,
as indicated in the figure. All other model parameters are taken
to be same as the fiducial model. It is evident from the
figure that the over-all qualitative features of outflow properties
remain the same for all the relevant halo masses (for the assume set
of parameters). The final radius of the wind and also the maximum 
outflow velocity increases with the mass of the halo.
This is a manifestation of the fact that SFR hence the energy available for the 
wind is higher for higher mass object  when the bubble cooling is not 
that efficient. 
The halo gravitational force which increases with mass
could have reversed this trend, but is generally sub-dominant compared to
the pressure of the hot bubble driving the outflow, provided
the outflow escapes the halo. We also notice in the previous sections,
changing the parameters from the values used in the fiducial models
may stop the large scale outflows in the high mass halos. Basic requirement
in this case is that the bubble cooling should be fast enough. This 
can be achieved in the high mass halos by increasing $\eta$ or $f_h$
or by decreasing $\epsilon_w$.
We will fold in these properties of different mass halos to
calculate the global effects of outflows below.

\section{Rayleigh-Taylor instability and shell fragmentation}

We have seen in the last section that when the large scale
outflows are possible, the shell of gas in the outer shock in our models 
generically accelerates till just beyond the virial radius.
This arises because the input kinetic luminosity increases with time,
till $t = t_{\rm dyn}$, while at the same time the confining halo density 
decreases with radius. As mentioned earlier, acceleration of the outflow 
obtains for $n>1$ or $\alpha + \beta > 2$ (see section~\ref{sec_comp}).
In our case where $\alpha \sim 2.8$ within the virial radius, 
and $\beta \sim 1$ for $t < t_{\rm dyn}$, one gets
$n \sim 1.8$ and so the shell accelerates while it traverses the
halo. In the rest frame of the shell, this acceleration corresponds to 
a pseudo-gravitational force pointing from the dense shell
to the low density hot bubble.
Such an accelerating dense shell driven by a low density
hot bubble is subject to the Rayleigh-Taylor (RT) instability.
(Physically, it corresponds to the analogous case of 
a heavy fluid lying `on top' of a light fluid in a gravitational field).
This can lead to fingers of the shell material penetrating
the hot gas, while bubbles of the hot fluid rise into the shell,
resulting in turbulent mixing of the two fluids.
If the mixing scale becomes of order the shell thickness
the RT instability will also lead to the fragmentation
of the shell. We now examine in a simple manner,
some of the consequences which could result
from this instability.

In an expanding bubble, the evolution of the Lagrangian perturbation ($\phi$)
in the shell with co-moving wavenumber $k_h$ is given by (Pizzolato \& Soker 2006)
\bea
\ddot{\phi}+ \left( 2\f{\dot{a}}{a} + \f{\nu_e k_h^2}{a^2}
\right)\dot{\phi} - \omega ^2 \phi = 0
\label{eqn_phy}
\eea
where (cf. Chandrasekhar,1981; Padmanabhan, 2002),
\bea
\omega ^2 = \left[ \left| g \right| + \ddot{R}_s\right] \f{k_h}{a}
\f{\rho_s - \rho_b}{\rho_s \coth (k_hh/a) + \rho_b}.
\eea
Here, $g$ is the gravitational acceleration of the halo potential,
$a$ is the expansion scale factor define as $a(t)=R_s(t)/R_s(t_0)$ and
$h$ is the thickness of the shell.
We have also assumed that the hot bubble is ``thick", that is $k_hR_s/a \gg 1$.
We take the initial time $t_0$ as the time when we start
calculating the growth of $\phi$. 
We have also included above the effect of viscosity which
stabilizes the RT instability
for large $k_h$, in a heuristic manner akin to Piriz et. al. (2006).
The effective kinematic viscosity $\nu_e$ is obtained
using (cf. Piriz et. al. 2006)
$$
\nu_e = \f{2 ( \mu_b + \mu_s )}{\rho_b + \rho_s}.
$$
where $\mu_b$ and $\mu_s$ are the dynamical viscosity of the bubble material
and shell material respectively while $\rho_b$ and $\rho_s$
the corresponding densities. For a fully ionized gas the
Spitzer viscosity $\nu =\mu/\rho \approx 6.5\times 10^{22}
(T/10^6)^{2.5} (n_i/{\rm cm}^{-3})^{-1}$~cm$^{2}$~s$^{-1}$,
where $T$ is the temperature and $n_i$ the ion density of the gas.
We have already seen the shell gas can cool faster than the expansion time
while it is within the virial radius, to a temperature $T_s \sim 10^4$ K.
The density of the shell is taken to be $\rho_s = M_s^2 \rho_B$, where
the Mach number $M_s = (v_s/c_s)$, with $c_s$ the sound speed corresponding to 
a temperature $T_s$. So the above $\nu_e$ is largely determined by the bubble
temperature and the shell density (since $T_b \gg T_s$ and $\rho_s \gg \rho_b$).
\begin{figure}
\centerline{
\includegraphics[width=0.45\textwidth]{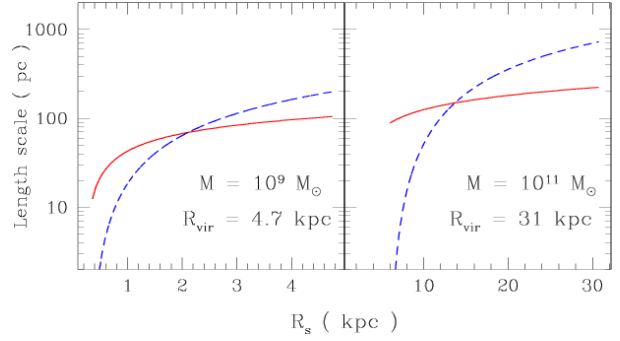}
}
\caption[]{Thickness of the shell (solid lines) and the mixing
length (dashed line) for RT instability as a function of the outer shock location, 
for $R_s \le R_{{ \rm vir}}$. Left panel is for
$10^{9}~M_\odot$ halo collapsed at $z_c = 6$ and right panel is for
$10^{11}~M_\odot$ halo collapsed at $z_c = 4$.
 }
\label{fig_length}
\end{figure}

In the absence of viscosity small scale perturbations
always grow faster than larger scale ones. However, viscosity
damps the growth of small scale perturbation. Expansion on the other hand
damps the growth of perturbations on all scales. 
The initial perturbation amplitude is unknown, but one 
does not expect the contact discontinuity to be smooth
because of the small-scale inhomogeneities in the wind fluid and
the outside halo gas. When perturbations go nonlinear, such that
the displacement of the shell-bubble interface ($R_c$ surface) becomes 
comparable to the thickness $h$ of the shell, one expects the shell to
fragment. One can estimate $h$ using $m_s = 4\pi R_s^2 h \rho_s$.
This is shown in Fig.~\ref{fig_length} for two representative halos,
one of $10^9 M_\odot$ (left panel) and for $10^{11} M_\odot$ (right panel).
For the $10^9 M_\odot$ halo the thickness grows from $h \sim 10$pc to
$h\sim 100$pc, while for a $10^{11} M_\odot$ halo, $h \sim 100-200$ pc.

\begin{figure}
\centerline{
\includegraphics[width=0.45\textwidth]{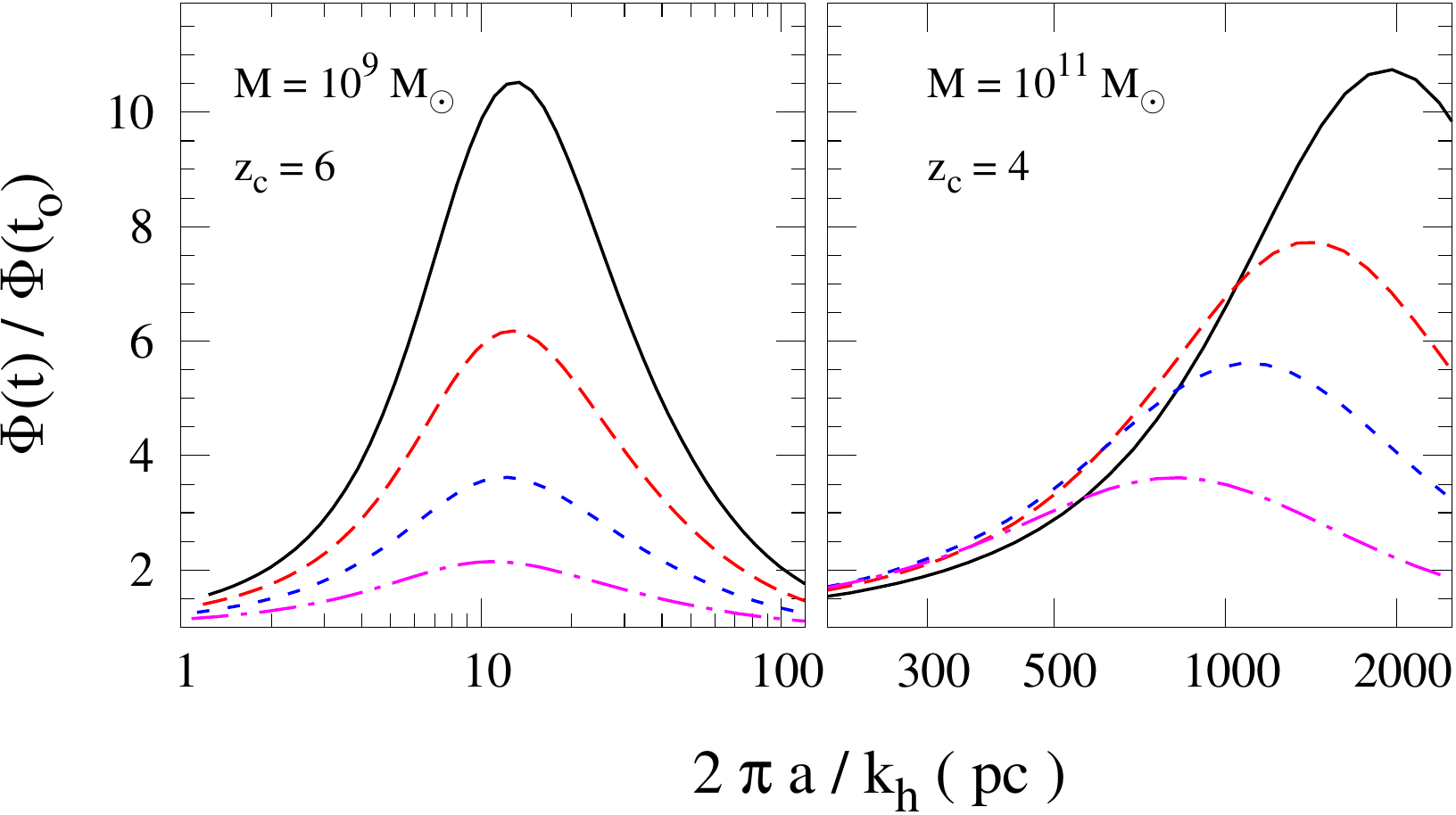}
}
\caption[]{The growth of perturbations at different physical length scale.
We show the $\phi(t)/\phi(t_0)$ at $t=10^6,~1.5\times10^6,~2\times10^6$
and $2.5\times10^6$ yrs from bottom to top for $10^9~M_\odot$ (left panel) halo.
For $10^{11}~M_\odot$ (right panel) halo we have plotted $\phi(t)/\phi(t_0)$
at $t=4\times10^7,~6\times10^7,~8\times10^7$
and $1.1\times10^8$ yrs from bottom to top.
 }
\label{fig_inst_growth}
\end{figure}

We can solve Eq.~(\ref{eqn_phy}) to find the evolution of $\phi(t)$
as a function of $k_h$. In Fig.~\ref{fig_inst_growth} we show the
resulting growth factor $G = \phi(t)/\phi(t_0)$
as a function of the proper wavelength of the perturbation, 
$\lambda_p = (2\pi/k_h) a$, for various times $t$.
We see that there can be significant growth of perturbations, by
over an order of magnitude, due to the RT instability.
For a $10^9 M_\odot$ halo, the most unstable modes have 
$\lambda_p = l_f\sim 10$ pc, while for a $10^{11} M_\odot$ have one gets a 
much larger $l_f \sim 0.7-2$ kpc.
So the shell will keep losing mass due to the RT instability with the scale
of fragments being comparable to $l_f$ if all scales have similar
fractional perturbations. There could of course be a range of
scales which become nonlinear, 
because we do not know the initial spectrum of inhomogeneities.

Further, one expects the RT fingers/bubbles to have a random terminal velocity
$v_f \sim (g_{eff} l_f)^{1/2}$ (cf. Dimonte et al 2005 and
references therein). Here $g_{eff} = \left| g \right| + \ddot{R}_s$.
Taking typical values for $g_{eff}$ and $l_{eff}$, one gets
$v_f \sim 18~ {\rm km~s^{-1}}~(g_{eff}/10^{-8}
~{\rm cm~s^{-2}})^{1/2}$ $(l_f/100~{\rm pc})^{1/2}$ 
For a large enough Reynolds number ${\rm Re}=(v_f l_f/\nu_e)$,
this contact layer between the two
fluids could also become turbulent. In this case the growth of the dominant
bubbles become self-similar, the two fluids mix in a mixing
layer growing in size as $l_{mix} \sim 0.05 g_{eff} t^2 \sim 75~{\rm pc}~
(g_{eff}/10^{-8}~{\rm cm~s^{-2}})~(t/(5\times10^7~{\rm yr}))$ (Dimonte et al 2005). 
Again if $l_{mix}$ becomes of order of the shell thickness
the shell will be fragmented with the additional effect that the shell gas
and the hot bubble gas within $l_{mix}$ also turbulently mix
with each other. We show in Fig.~\ref{fig_length} the evolution of the scale $l_{mix}$
with time. We see that $l_{mix} \sim h$ well within the virial radius,
and so the shell is likely to fragment in this case due to the RT instability.

The shell fragments resulting from the RT instability will be moving
in the hot bubble and so will suffer significant mass loss due to evaporation.
The rate of evaporation is given by (Cowie \& McKee 1977)
\bea
\dot{M}_{ev} = 4.1\times 10^{-5} \left( \f{T_b}{10^6~{\rm K}} \right)^{5/2}
\left( \f{a}{100~{\rm pc}} \right)~M_\odot~yr^{-1}
\eea
where $a$ is the radius of the fragment. This correspond to an evaporation time scale
$t_{ev} = m_c/\dot{M}_{ev}$ given by,
\bea
t_{ev} = 2.3\times 10^{8} \left(\f{n_s}{0.1 {\rm cm}^{-3}}\right)
\left( \f{T_b}{10^6~{\rm K}} \right)^{-5/2}
\left( \f{a}{100~{\rm pc}} \right)^2~{\rm yr}
\eea
where we have taken the cloud mass to be $m_c = (4\pi/3)\rho_s a^3$. 
In fact if the bubble-shell interface becomes
turbulent, such evaporation could be enhanced due to an enhanced contact
surface. For $10^9 M_\odot$ halos, where $l_f \sim 10$ pc,
one expects the shell fragments to evaporate and mix with the bubble gas
fairly rapidly, within say a few times $10^{8}$ yrs. 
But for large galactic scale higher mass halos,
$l_f$ is an order of magnitude larger, and $t_{ev}$ is likely to
be much larger.

In order to examine if the effect of the RT instability significantly changes
the further evolution of the outflow, we have considered the following simple
toy model: We assume that the shell is broken periodically due to
the RT instability during the acceleration phase. 
At each such epoch we assume the shell looses a significant amount of 
its mass. However the hot bubble is still expanding outwards
and sweeps up material from the surrounding to form a new shell. We assume the
newly created shell has initially some small fraction ($1\%$) 
of the original mass and expands with
the same velocity it had before fragmentation. 

\begin{figure*}
\centerline{
\includegraphics[width=0.95\textwidth]{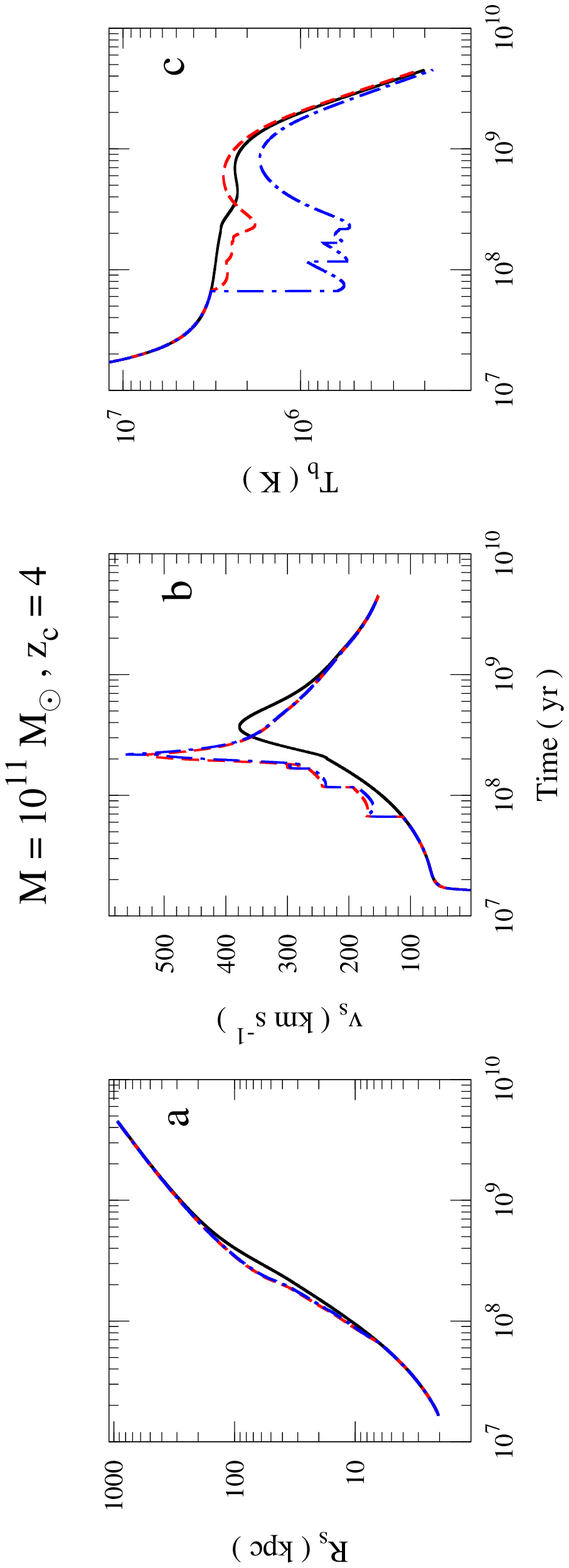}
}
\caption[]{The effect of RT instability on outflow. We show
$R_s$, $v_s$ and $T_b$ of an out flow originating from
a $10^{11}~M_\odot$ halo collapsed at $z_c=4$ in panel
(a), (b) and (c) respectively. The solid curves are for
unperturbed solution while the other two are the solution taking account
of the RT instability. The dashed lines are obtained when we consider
no evaporation of the fragments while the dotted-dashed lines are
plotted when we assume complete evaporation of the fragments
and instantaneous mixing of the fragmented mass into the hot bubble.
 }
\label{fig_inst}
\end{figure*}

The effect of the Rayleigh-Taylor instability following the
above prescription is shown in Fig.~\ref{fig_inst} for a halo 
of mass $10^{11}~M_\odot$ collapsing at $z_c=4$. 
This example would be relevant for Lyman break galaxies which are thought 
to be galactic scale objects having outflows.

In panel (a) and panel (b) of Fig.~\ref{fig_inst} we compare the
radius and velocity profile for the cases with and without the RT instability.
The unperturbed solutions are shown by solid lines while the solutions
taking account of RT instability as described above are shown by
dashed and dotted-dashed lines.
We have assumed that the shell gets fragmented periodically every 
$4\times10^7$ yrs while it is accelerating. 
At such times the shell looses
$99\%$ of its mass and continues
to sweep with the same speed as it had before fragmentation.
We show two cases: one in which we 
add the fragmented cloud mass into the hot bubble assuming that they evaporate
completely (dotted-dashed lines) and a second in which the fragmented clouds do not
add any mass into the hot bubble (dashed lines).

We see from this figure (panel (b)) that the velocity in
the accelerating stage are affected by the breaking of the shell due
to the RT instability, however the final radius
and velocity remain practically unaltered. The maximum velocity achieved 
by the wind increases in the case where we take account of the RT instability,
as the mass of the shell reduces periodically.
The breaks seen in the velocity
of the outflow mark the position when the shell fragments. 
 
The temperature of the hot bubble (panel (c) in Fig.~\ref{fig_inst})
also finally reaches the same asymptotic value. However,
depending on whether the fragments can evaporate (dotted-lines)
or not (dashed lines) the exact evolution of the temperature
can be different. The fragment/hot bubble interfaces will have a range
of temperatures from $T\sim 10^4-10^6$ K, with the metals from the 
hot fluid mixed in with the cooler shell gas. 
There could also be mixing due to secondary Kelvin-Helmoltz instabilities
which operate at the interfaces and/or the turbulence mentioned above.
These interface regions could lead to a host of absorption
lines in both the spectrum of the wind blowing galaxy and background
quasars seen through the outflow region.

From the above analysis we also see that the final radius and
velocity of the outflows do not change significantly. Hence we
will not consider the RT fragmentation while
computing the global properties of the outflow.

\section{Detection of metals}
In the semi-analytic models described  and explored in detail above,
we can see that winds can efficiently transport metals into the IGM
for a wide range of model parameters. The question we wish to ask now
is that, what are the required physical conditions in this expelled
gas that will enable us to detect them in spectroscopic observations. 
The detectability of the metals in a given ionization state depends
on collisional ionization, photoionization and recombination. Thus
one needs to incorporate the effect of these processes into our
calculations. The most easily detectable doublet transitions
from the highly ionized gas 
are C~{\sc iv}$\lambda\lambda1548,1550$, Si~{\sc iv}$\lambda\lambda$1393,1402 and 
O~{\sc vi}$\lambda\lambda$1032,1037. The respective ionization potentials are 
64.5 eV, 45.1 eV and 138.2 eV. The ionization fraction  for C~{\sc iv}, Si~{\sc iv}
and O~{\sc vi} peaks at temperatures of $10^5$~K, $6.4\times10^{4}$~K and 
$2.8\times10^{5}$~K respectively for collisional ionization
(Sutherland \& Dopita 1993).   
As the recombination time-scales are larger for mean IGM density at $z\sim3$
the gas temperature is not controlled by the ionization equilibrium and the wind
dynamics plays an important role in the gas temperature and density.  

\subsection{Ionization correction}
\begin{figure}
\centerline{%
\includegraphics[width=0.45\textwidth]{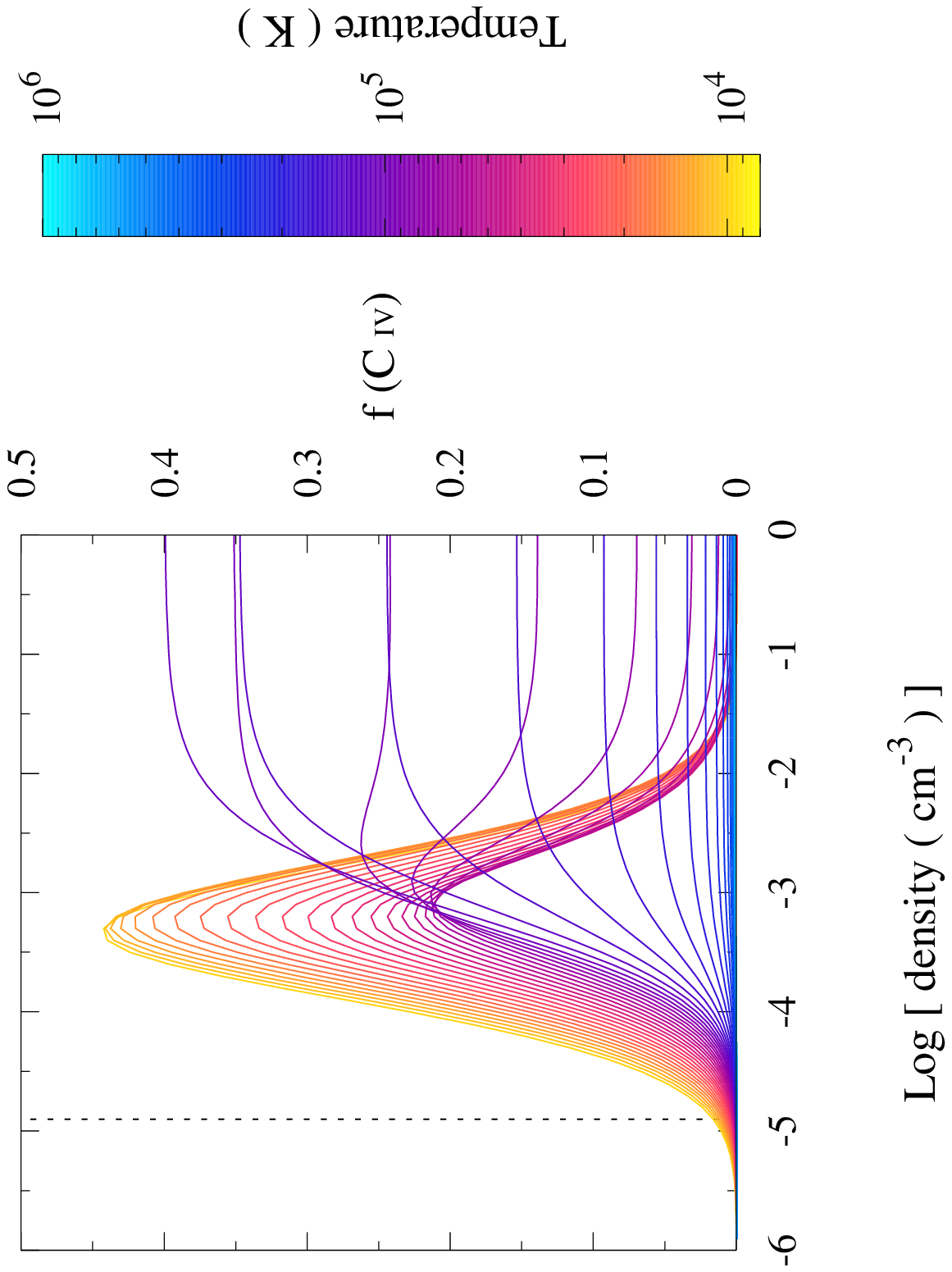}
}
\centerline{%
\includegraphics[width=0.45\textwidth]{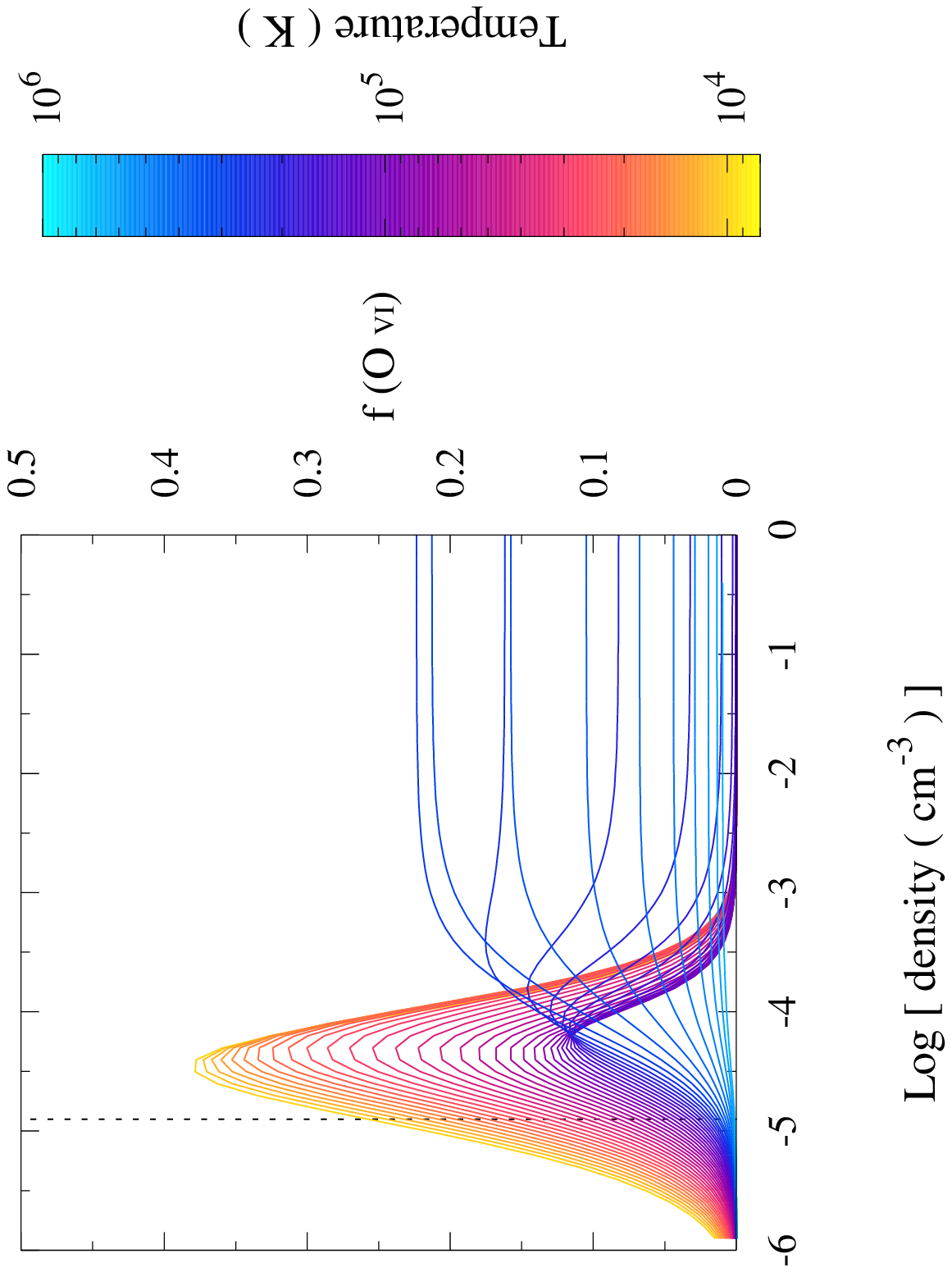}
}
\caption[]{Ionization fraction of C~{\sc iv} (top) and O~{\sc vi}
(bottom) as a function of density of the medium. Different colour
coded curves are for different temperature.
}
\label{fig_ionization}
\end{figure}
Now we calculate the ionization fraction for the three species namely C~{\sc iv},
Si~{\sc iv}  and O~{\sc vi} as a function of density and temperature of the medium.
In order to compute the ionization corrections correctly we have used the
spectral simulation code cloudy (version C06.02 - see Ferland et al. 1998
for details).
We assume the gas to be a plane parallel slab with total hydrogen column 
density of $10^{19}$ cm$^{-2}$. We have used the QSO and Lyman break galaxy
dominated meta-galactic UV background at $z=3$ computed by Haardt
and Madau (2001) as the ionizing source. The gas density is varied between
10$^{-5}$ to 1 cm$^{-3}$ and the temperature of the gas is varied between
10$^4$K to 10$^6$ K.
\begin{table}
\caption{Ionization fraction}
\begin{center}
\begin{tabular}{|c| c| c| c| c|}
\hline
 \hline
Density &species &\multicolumn{3}{c}{fractional ionization at T = }\\
	&	& $10^4$~K & $10^5$~K & $5\times 10^5$~K \\ \hline
	& C~{\sc iv}	& $1.6 \times 10^{-2}$ & $4.8 \times 10^{-4}$ & $1.0 \times 10^{-5}$ \\ 
\raisebox{1.5ex}[0cm][0cm]{$\bar{\rho}$}
        & O~{\sc vi}	& $2.6 \times 10^{-1}$ & $4.1 \times 10^{-2}$ & $3.5 \times 10^{-3}$ \\ 
        & Si~{\sc iv}   & $5.6 \times 10^{-7}$ & $3.6\times10^{-9}$ & $1.6\times10^{-8}$\\ \hline
        & C~{\sc iv}	& $9.4 \times 10^{-5}$ & $9.1 \times 10^{-7}$ & $3.6 \times 10^{-8}$ \\ 
\raisebox{1.5ex}[0cm][0cm]{$0.1\bar{\rho}$}
        & O~{\sc vi} &$1.5 \times 10^{-2}$ & $4.4 \times 10^{-4}$ & $2.1 \times 10^{-5}$ \\ 
        & Si~{\sc iv}& $2.0\times10^{-10}$ & $1.0\times10^{-12}$ & $1.8 \times 10^{-3}$\\ \hline
	& C~{\sc iv}&$2.7 \times 10^{-1}$ & $3.8 \times 10^{-2}$ & $3.3 \times 10^{-4}$ \\ 
\raisebox{1.5ex}[0cm][0cm]{$10\bar{\rho}$}
        & O~{\sc vi}	& $1.7 \times 10^{-1}$ & $6.3 \times 10^{-2}$ & $2.2 \times 10^{-2}$ \\ 
        & Si~{\sc iv}   & $6.3\times10^{-3}$ & $7.2\times10^{-5}$ & $9.7\times10^{-6}$ \\
\hline
\end{tabular}
\label{table_ionization}
\end{center}
\end{table}

The ionization fraction of C~{\sc iv} and O~{\sc vi} as a function of density 
for a range of temperatures a are shown in Fig.~\ref{fig_ionization}.
In Table.~\ref{table_ionization}, we give the fractional ionization of the
three species (C~{\sc iv}, Si~{\sc iv} and O~{\sc vi} ) at three
different temperature and densities. Here, $\bar\rho$ is
the mean IGM density at $z=3$.
We wish to point out that results will not depend too much on our choice of 
N(H) as long as N(H~{\sc i}) is optically thin at the Lyman limit.  
It is clear from the figure that at low densities and temperatures the 
effect of photoionization is very important. The ionization fraction of the two 
species are controlled by collisions (when it is independent of density)
at higher densities. The vertical line in this figure marks mean IGM
density at $z=3$. We notice that the maximum Si~{\sc iv} fraction is
$6.3\times10^{-3}$ when we consider
$T=10^4$~K, $\rho = 10\bar\rho$. For higher temperature and lower density
gas the fraction of Si~{\sc iv} is negligibly small.

From the table it is clear that when one considers the underdense region
(density $\sim0.1\bar\rho$) ionization fraction of C~{\sc iv} is less than 
$10^{-4}$ for $T\ge10^5$~K. This will make detection of C~{\sc iv}
in absorption towards high redshift bright objects virtually impossible. 
This exercise clearly states that if the metals are going to be distributed in
voids with average density less than the cosmological mean density then
it will be very difficult to detect them in C~{\sc iv} absorption.  The maximum
ionization fraction for O~{\sc vi} we get for this case is 1.5\%.
The column density of O~{\sc vi} is given by,
\begin{equation}
N(O~{VI}) = 7.4\times10^{-8} \times N(H) \left(\f{f_{O VI}}{0.01}\right)
\left(\f{Z}{0.01~Z_\odot}\right).
\label{colovi}
\end{equation}
Typically a low density region with N(H)~=~$10^{21}$~cm$^{-2}$ will produce 
detectable O~{\sc vi} absorption when the gas is at $10^{4}$~K. 
However, the detection becomes difficult if the temperature of the gas is 
higher than $10^4$~K.
Clearly when O~{\sc vi} is detected from the low density medium it will
have signatures of photoionization and not that of collisional ionization.

If the over dense regions, like that probed by the high column density
Lyman-$\alpha$ forest, are polluted by metals then it will become easy to 
detect C~{\sc iv} and O~{\sc vi} absorption from this gas if the temperatures
are close to photoionization temperatures. Again if the gas temperature
is higher than $10^5$K it will become more difficult to detect C~{\sc iv}.
Having got a rough idea of the temperature density ranges where O~{\sc iv}
and C~{\sc iv} fraction is higher in the presence of meta-galactic UV background
radiation, we now consider different cases in our models.

Spectroscopy of bright QSOs, GRBs and Lyman break galaxies 
allow us to probe the metals in the IGM through absorption line seen
in their spectra. Direct spectroscopy of galaxies probes the ongoing winds
in the galaxies where as the spectroscopy of QSOs allow us to probe
the global enrichment of the IGM and enrichment around bright galaxies.
Clearly different observations probe the nature of the wind at different
stages of its evolution.

\subsection{Metals in free wind}

As seen from  Fig.~\ref{wind_profile}, the metal enriched gas
after escaping the galactic disk will travel outwards
in the form of free wind before
encountering the inner shock at $R_1$. Metal absorption 
in the spectrum of galaxies with small radial separation could 
come from the free wind material coming out of the galaxy.
In our model the free wind material extends upto the inner shock ($R_1$)
which can go beyond virial radius of the halo.
When this gas is shocked at the inner shock the temperature
goes up and hence detection probability of the metals goes down. 
Even a naive calculation (see below) then shows that the free
wind material has sufficient column density for detection
in line absorption. 

We assume that the free wind is not very hot. This may happen
because cooling is very efficient in the ejecta of individual
SNe or due to subsequent adiabatic expansion of the thermally
driven galactic wind. We integrate the density
of the free wind from $R_i$ to $R_1$ to find out the column
density of H, assuming for illustration a constant mass outflow rate. 
Since the free wind material is expanding
asymptotically at a constant velocity $v_w$, 
the density ($\rho_w$) will scale $\rho_w(r) \propto r^{-2}$.
The normalization can be fixed 
at the initial radius $R_i$ from mass conservation to be
$\rho_w(R_i) = \dot M_w /( 4 \pi R_i^2 v_w)$.
Integrating this profile we get column density of hydrogen
\begin{equation}
N(H) = \f{\dot M_w}{4 \pi m_p v_w} \left[\f{1}{R_i} -\f{1}{R_1}\right]
\end{equation}
where $m_p$ is the proton mass. 
Taking $\dot M_w = 10 M_\odot$~yr$^{-1}$,
$v_w = 400$~km~s$^{-1}$, $R_i = 1$~kpc and assuming $R_1 >> R_i$ we get
N(H)~$\sim 2 \times 10^{20}~{\rm cm}^{-2}$. 
From Eq.~(\ref{colovi}) it is clear that O~{\sc vi} column density of this gas
will be $1.5\times10^{13}$~cm$^{-2}$ if the metallicity is 0.1 and $f_{OVI}$
is 0.001. Note that in our models ${\dot M_w}$ is a function of
time and the above exercise gives us only an order of magnitude
estimate of typical expected column density.
 
In our models high mass outflow rates are expected in high mass
galaxies as they have higher SFR.
However, the exact value of $f_{O~{\sc vi}}$ (or $f_{C~{\sc iv}}$) will depend on 
the metallicity and temperature of the free wind gas.
Ideally we expect the initial temperature of the free wind to
be equal to or less than the temperature of the superbubble that initiates the
flow. This will be typically in the range few time $10^5$~K to $10^6$~K
(see also Fig 3. of Efstathiou 2000). Simulations with single starburst in the
disk suggest the temperatures could be as high as $10^7$~K in the
inner side of the free wind (see Fig.~4 of Mac Low \& Ferrara, 1999
and Fig.~6 of Fujita et al. 2004). If the free wind is at high temperature
then detectability in the UV absorption is difficult. 
However if the temperatures are  more like few $10^5$ K then we 
expect to detect the free wind in UV absorption lines discussed above.

It is important to note that the outflows detected in the spectra of
high redshift Lyman break galaxies, starburst and dwarf galaxies 
at low redshifts also show standard low ionization interstellar absorption
in the outflowing gas (Pettini et al. 2000; Rupke et al. 2002;
Martin 2005). Presence of Na~{\sc i} absorption in the outflow 
means,  clearly some part of the outflowing gas 
is optically thick to the Lyman continuum photons originating
either from galaxy itself or from the meta-galactic UV background
radiation. Note that most of the observational papers favor the distance
of the outflowing gas to be few kpc away from the galactic disks.
If we assume that the outflowing material from the galaxy 
has properties very similar to that in ISM then we can envisage a 
situation in which the outflowing material has mixture of hot, warm 
and cold media (Heckman et al. 2000). 
In such a case  a free wind material can produce 
absorption lines covering wide range of ionization states (see also
Murray et al. 2007).

\subsection{Metals in the bubble  and shell}

Next we consider the bubble gas after the free wind enters the 
inner shock.  As discussed before, the free wind material and the
($1-\epsilon$) times the swept up material are in the bubble.
The bubble density is usually low compared to the ambient medium.
When the shell is well within the virial radius the bubble density 
is contributed by both the leftover gas from the swept up material
(or evaporated material from the shell) and the shocked wind gas.
In our models this is roughly in the range $10^{-2}$ to 
$10^{-3}$~cm$^{-3}$. However, the gas temperature is usually higher than 
$10^6$~K.  At these temperatures the gas ionization will be dominated
by collisions and both C~{\sc iv} and O~{\sc vi} will not be the 
dominant ionization state of the respective atoms.

It is clear that for a uniform bubble material it will be difficult
to detect the outflowing gas in the UV absorption.
It is also important to realize that when the outflowing gas is near the
virial radius SFR in the galaxy is also near its peak. Thus the study of
absorption lines in bright galaxies will correspond to the situation in
which the shock front is either inside the virial radius of the galaxy
or just about leaving it. Thus, if the UV absorption lines seen 
in the spectra of galaxies originate from the bubble material then
we needs inhomogeneities in the bubble gas as well. This could
either be from (i) the shocked wind material that already has 
clumps as we discussed above (ii) or the broken shell fragments
from the RT instability. In the former case the high density ISM
gas will already have metals mixed into it. This will
enable one to detect even low ionization material. However, in the
latter case the broken shell material may predominantly have swept up
material that may not be highly enriched.  High ionization species
can also originate from the conductive interface between hot  metal rich
bubble material and cold shell material (as discussed in the case of
Weaver et al. 1997). From our discussion on RT instability in 
section 6, it is clear that such fragments survive for a longer time-scale
against evaporation in high mass halos.

When the shock front leaves the virial radius, the average bubble density
is ($1-\epsilon$) times the IGM density. For $\epsilon = 0.9$, then
typically,  the bubble is underdense by a factor 10 compared to the 
uniform IGM. From our earlier discussion on individual halos we note
that the bubble temperature in this stage of evolution is decided by
adiabatic cooling and is some where in the range 10$^6$K to 10$^5$ K.
From Table~\ref{table_ionization} it is clear that it will be difficult to
detect both C~{\sc iv} and O~{\sc iv} in absorption leave alone detecting
the low ionization lines seen in the outflow. If UV absorption is produced
from the bubble gas we need inhomogeneities with higher density and
lower temperature (as discussed above) compared to the mean bubble
density and temperature.

On the other hand the shell density is high enough and temperature
is low enough (when the shock is well within the virial radius)
so that the metals can be detected in absorption.
The shell material is dominated by the swept
up mass from halo gas/IGM. If the IGM is not pre-enriched by the metals
then one would not detect the shell in metals line.

\subsection{Effect of pre-enrichment}

As discussed in the introduction, C~{\sc iv}, Si~{\sc iv} and
O~{\sc iv} absorption lines detected in the spectra of high redshift 
QSOs probe the metal enrichment history of the universe.
The discussions presented above suggests that when the metals
are expelled from the galaxies they usually displace the material
in the ambient medium into a thin shell and spread the metals
into a low density high temperature bubble. In a simple uniform
density model without inhomogeneities it  will be difficult to
detect these shells through C~{\sc iv} and O~{\sc vi} absorption
lines.

Note that when an outflow is traversing through a medium which
is already enriched by earlier generation of star formation the
swept-up shell material can be detected in absorption lines.
For example the column density of carbon at the virial radius
of the halo is
\bea
N(C) & \sim & 5\times 10^{12} \epsilon \left(\f{f_h}{0.1}\right)
\left(\f{1+z}{4}\right)^2 \times \nonumber \\
& & ~~~~~~~~~
\left(\f{M}{10^{10}~M_\odot}\right)^{1/3} \left(\f{Z}{10^{-3}~Z_\odot}\right)
~{\rm cm}^{-2}.
\eea
Here we have assume that the halo gas is pre-enriched with a
metallicity of $10^{-3}~Z_\odot$.
When the shell is within the virial radius of the halo its
density varies between $10^{-2}$ to $10^{-3}$~cm$^{-3}$ and the
temperature is $\sim 10^4$~K. The C~{\sc iv} fraction will be close to $0.3$
and the shell will produce detectable C~{\sc iv} absorption.
However when the outflow is in IGM,
the density of the shell material is very low and
the temperature is more than $10^5$~K. This makes it difficult to detect
the shell in C~{\sc iv} even though the column density of carbon is
still of similar order compared to the case when the outflow
is inside the halo.

Interestingly our knowledge of metals in the
Lyman-$\alpha$ clouds comes from observations of high
density regions with over density in excess of 6 or so.  One way of
incorporating metal into these over dense regions without 
altering the statistical properties of the Lyman-$\alpha$ forest
distribution is to inject material into the IGM very early 
(while a $z=3$ over dense region was very close to the mean
density at an earlier epoch) and allow the metal mixing through
the evolution of gravitational perturbations that produce 
Lyman-$\alpha$ forest absorption lines. In this scenario it is
quiet possible that the metal absorption lines we see in the 
high density Lyman-$\alpha$ forest without the signature of 
collisional excitations could have come out of very early generation
of low mass galaxies. Even though our models do not capture the non-linear
evolution of the IGM, in the following section when we discuss the 
global properties of the outflows we will 
show that the IGM could be pre-enriched with 
metals.  

\section{Global consequences of outflows}

From calculating the evolution of a suite of individual outflow
models, we can study several global properties of the wind affected
regions. This is mainly to understand the effect of the outflows on the IGM.
One simple quantity is the porosity $Q(z)$,
defined by adding up the outflow volumes around all the sources at any redshift:
\bea
Q(z) = \int \limits_{M_{\rm low}}^{\infty} \de M \int \limits_z ^{\infty}
\de z' \f{\de^2 N(M,z,z')}{\de z'\de M} \f{4}{3}\pi \left[R_S(1+z)\right]^3.
\eea
Here, $\de^2 N(M,z,z')/\de z'~\de M$ comes from Eq.~(\ref{eqnmPS}) and $R_s$
comes from solving for the outflow dynamics. The lower limit $M_{\rm low}$
is determined from the cooling criteria and the effects of reionization
feedback, as described in section~\ref{sfr}. 
For $Q < 1$ the porosity gives the probability that a randomly selected point 
in the universe at $z$ lies within an outflow region. 
For $Q \sim 1$, it is more useful to define the associated filling factor
of the outflow regions, which if outflows are randomly distributed is given by, 
$F=1-\exp [-Q(z)]$. These considerations ignore source clustering,
which can be important for rare sources and lead to a smaller $F$.
Note that $Q$ can exceed unity while $F(z) < 1$ always.

We can also define porosity weighted averages of various physical quantities associated
with the outflows, and their probability distribution functions (PDFs).
For any physical quantity say $X$, its porosity weighted average is defined as
\bea
\langle X\rangle &=& Q^{-1} \int \limits_{M_{\rm low}}^{\infty} \de M 
\int \limits_z ^{\infty} \de z'~ \f{\de^2 N(M,z,z')}{\de z'~\de M}\times \nonumber \\
& & ~~~~~~~~~~~~~~~~~~~~~~~~~~~\f{4}{3}\pi 
\left[R_S(1+z) \right]^3 X.
\label{eqnavg}
\eea
The cumulative PDF can be obtained by replacing $X$ in Eq.~(\ref{eqnavg})
by the Heaviside Theta function $\Theta(X(M,z,z') - X_0)$ and the differential PDF 
can be obtained by differentiating this respect to $X_0$.
We compute below such porosity weighted averages and the associated PDFs 
of various physical properties related to the outflows.

One of the main motivations to study the SNe driven galactic outflows
is to understand the metal pollution of the IGM. 
To compute this, we calculate the mass of metals coming out of the galaxy
through an outflow as described in Appendix~\ref{metal_evolution}.
We divide the total mass of ejected metals in a fiducial volume of the universe
by the average baryonic mass in this volume 
and refer to this as the average global metallicity, $\bar{Z}$ of the IGM.

Note that the global properties of the outflows
will depend on the parameters describing star formation,
reionization and cosmology. As mentioned in section~\ref{sfr}, we will
consider both `atomic cooling' and `molecular
cooling' models of Paper I, which are consistent with the observed
high redshift UV luminosity functions of galaxies and 
observed constraints on reionization.

\subsection{Atomic cooling Model}

First, consider the
model parameters for individual halos as discussed in section
\ref{sec_wind_profile} [i.e we take $f_* = 0.5$, $\kappa = 1.0$,
$f_h = 0.1$, $\epsilon=0.9$, $\epsilon_{\rm w} = 0.1$, $\eta = 0.3$, 
$\nu^{-1}=50~M_\odot$ with cosmological parameters from the WMAP 3rd yr data].
We refer to this as our fiducial model A.
\begin{figure}
\centerline{
\includegraphics[width=0.45\textwidth]{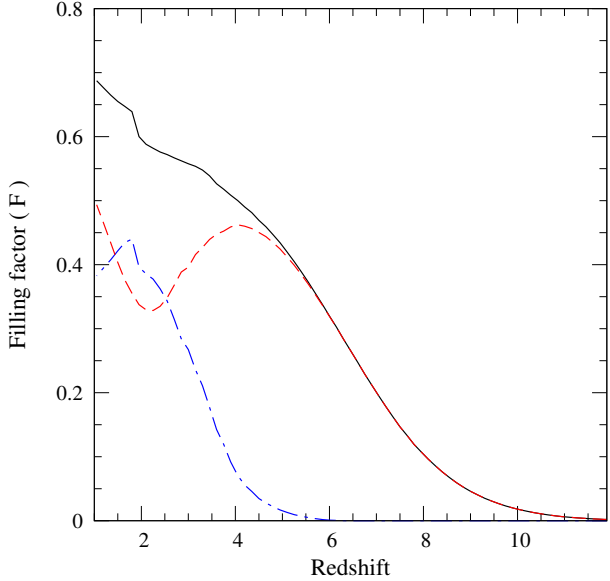}
}
\caption[]{The volume filling factor as a function of redshift
for `atomic cooling' model A.
Total filling factor of the universe is shown by the solid line.
The dotted-dashed line shows the contribution coming from the outflows
which are frozen into the hubble flow where as
the dashed curve shows contribution by the rest of the
outflows which are still moving at a supersonic velocities.
}
\label{fig_Q}
\end{figure}

\subsubsection{The volume filling factor}

In Fig.~\ref{fig_Q} we show the volume filling
factor $F(z)$ for model A, as a function of redshift. 
The solid line shows the net volume filling factor.
We can also split this up into contributions from outflows
which have already frozen into the Hubble flow, say $F_H(z)$ 
and those which have not (and moving still at a supersonic velocity), 
denoted by $F_A(z)$. The dash-dotted and dashed lines shows
respectively $F_H$ and $F_A$. At high redshifts ($z\gtrsim 6$)
the volume filling factor is dominated by $F_A$; i.e. the outflows
which have not frozen into the hubble flow.
However below redshifts of about 3, the contribution
from the hubble frozen outflows starts to dominate.
It is clear from the Fig.~\ref{fig_Q}
more than $30\%$ of the universe is affected by the
outflows even at $z\sim 6$ and this increases to $\sim 60\%$
by $z \sim 2$ for these model parameters. 

\begin{figure}
\centerline{
\includegraphics[width=0.45\textwidth]{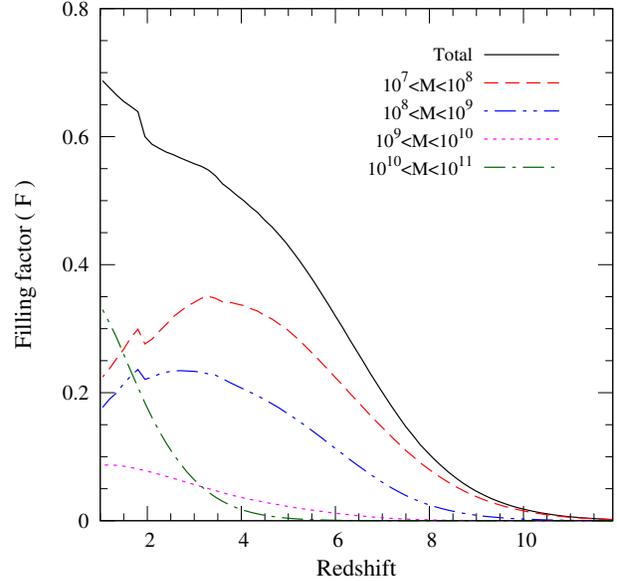}
}
\caption{Contribution of halos with different mass ranges
to the filling factor of IGM. Curves are drawn for mass range
$10^{7}-10^{8}~M_\odot$ (dashed), $10^{8}-10^{9}~M_\odot$ (dot-dot-dashed),
$10^{9}-10^{10}~M_\odot$ (dotted) and $10^{10}-10^{11}~M_\odot$
(dash-dotted). The total filling factor is also shown by
solid line.
}
\label{fig_wind_mass}
\end{figure}

It is also important to know the mass range of halos that
contribute significantly to the volume filling factor at different
epochs. Fig.~\ref{fig_wind_mass}
gives the contribution to the volume filling factor by halos in different
mass ranges. At redshifts $z\gtrsim 2$ the filling factor
is dominated by galaxies with mass range $10^{7}-10^{9}~M_\odot$
while higher mass halos start to dominate at lower redshifts. 
Earlier work by \cite{madau_ferrara_rees} in fact examined the
effect of outflows from galaxies with a typical halo mass of $10^8 h^{-1} M_\odot$.
We also find that such halos are important contributors to $F$.
However our work includes halos of all mass ranges that are allowed
by our cooling criteria, and 
we find that this leads to a significant increase in the volume
filling factor of outflows to give $F \sim 0.7$ by $z \sim 1$.
It is also interesting to note that the galaxies 
contributing to the observed high-$z$ UV luminosity functions
have typically masses $M \ge 10^9 M_\odot$ (see Fig.~4 in Paper I).
Therefore galaxies which dominate the volume filling factor
are not yet detected directly. 

\subsubsection{Porosity weighted averages and PDFs}
\begin{figure}
\centerline{
\includegraphics[width=0.45\textwidth]{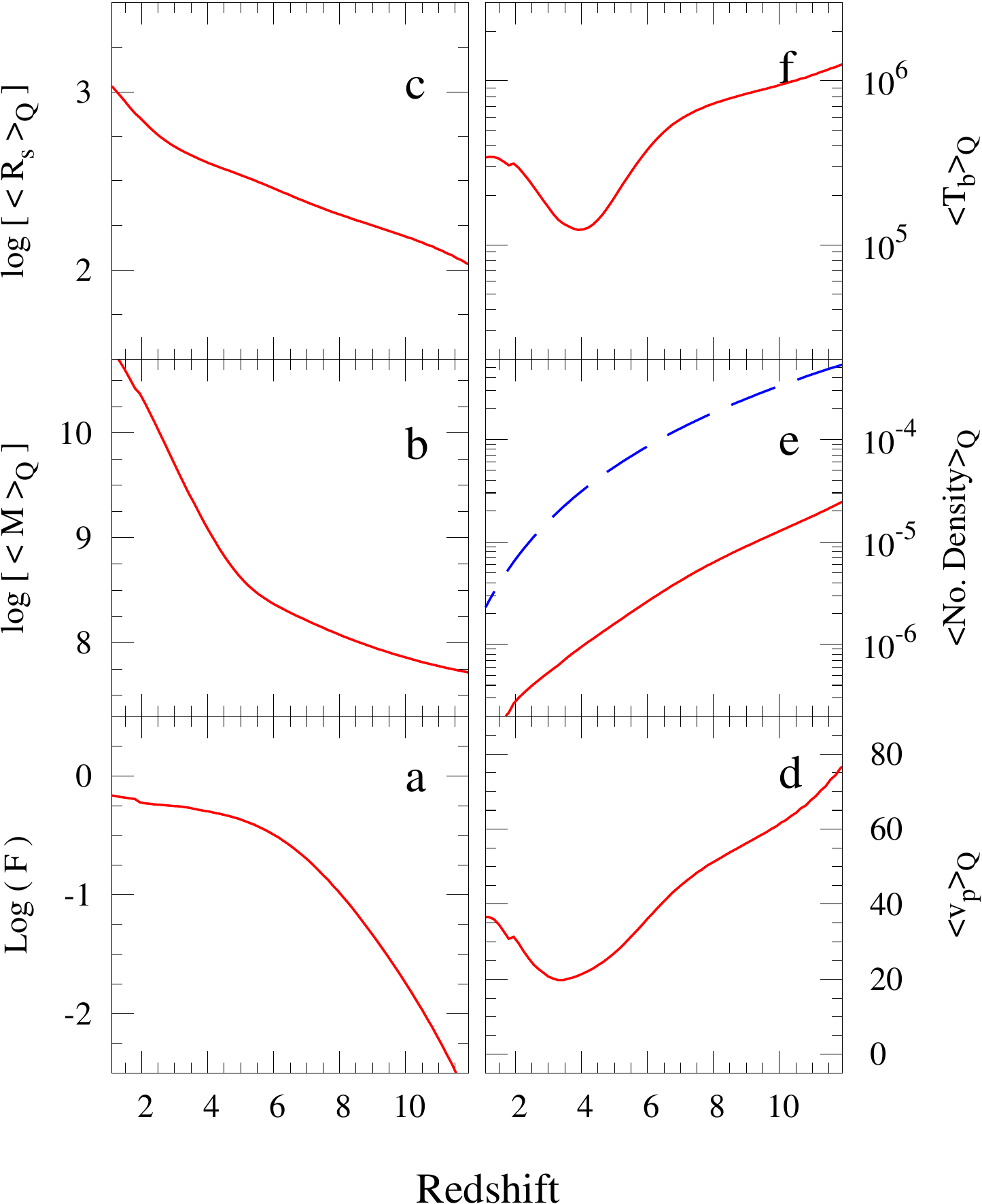}
}
\caption[]{The global properties of outflows.
Panel (a) shows the volume filling factor $F$.
The porosity weighted average dark matter mass (in $M_\odot$),
comoving radius (in kpc) and peculiar velocity (in km~s$^{-1}$) of the outflows
are shown in panel (b), (c) and (d) respectively.
The porosity weighted density (in cm$^{-3}$) and temperature (in K)
of the hot bubble are shown in panel (e) and (f). In panel (e)
we also show the mean baryonic density of IGM by dashed line.
}
\label{wind1}
\end{figure}

\begin{figure}
\centerline{
\includegraphics[width=0.45\textwidth]{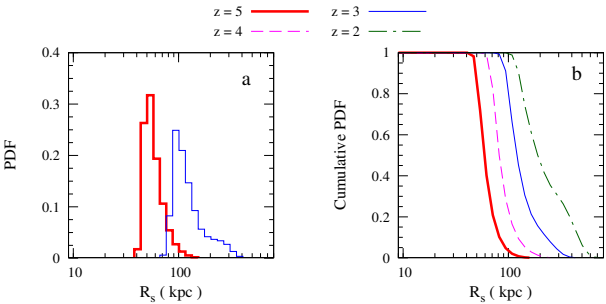}
}
\caption[]{The differential (panel a) and cumulative (panel b) PDFs of the
outflow radius ($R_s$) for model A.
}
\label{fig_radius_pdf}
\end{figure}

The porosity weighted averages of several physical properties
of outflows are shown in Fig.~\ref{wind1}.
Panel (a) shows the volume filling factor, while panel (b) gives
the porosity averaged dark matter mass of the halo from
which the outflows are originating. Both these aspects of
outflows have been discussed above.
Panel (c) shows the porosity averaged comoving radius of the 
outflow in units of kpc. This increases from $\sim 100 $ kpc at
$z \sim 10$ to $\sim 500$ kpc at $z \sim 3 $ upto $1$~Mpc by $z =1$.
More detailed information is available  
in panel (a) and (b) of Fig.~\ref{fig_radius_pdf}, where we
show respectively, the differential and cumulative PDFs of
the outflow radius (proper), for $z=2$, $3$, $4$ and $5$.
For clarity, we have shown the differential PDFs only for $z = 5$ and $3$.
It is clear from the figure that the median proper radius of the bubble is
$60$, $80$, $110$ and $200$~kpc, respectively, for  $z = 5$, $4$, $3$ and $2$.
These scales are perhaps just below the scales in the matter
power spectrum probed by the Lyman $\alpha$ forest at $z \sim 3$
(Croft et al. 1998). The outflows also fill a significant fraction
of the volume, with $F \sim 0.6$ by $z \sim 2$.
However the outflows are believed to propagate more into
the voids than along filaments (Theuns et al. 2002), and so may not 
perturb the overdense regions from which the Lyman $\alpha$ forest 
absorption originates.  We discuss this issue further below.

In panel (d) of Fig.~\ref{wind1} we show the porosity
weighted peculiar velocity of outflows. This velocity is $\sim 75 $ km s$^{-1}$
at high redshifts $z \sim 12$, reflecting the fact that
most of the outflows are very young at this stage.
As time increases the average peculiar velocity decreases to
about $20$ km s$^{-1}$ at $z \sim3-4$,  as considerable 
volume is filled by the flows, originating from early generation of
galaxies, that are either Hubble frozen or moving with much lower 
peculiar velocities. However, at $z<3$ the porosity weighted 
peculiar velocity increases due to an increase in the contribution
of outflows from high mass halos.
Note that as the volume filling factor increases, more and more
outflows will interact and the peculiar velocity of the
outflows can then lead to supersonic turbulence in overlapping 
regions. Such turbulence can also lead to amplification of
magnetic fields in the IGM, by the operation of the fluctuation dynamo
(cf. Zeldovich, Ruzmaikin \& Sokoloff 1990; Brandenburg \& Subramanian 2005).

The solid and dashed curves in panel (e) of Fig.~\ref{wind1} show
respectively, the porosity weighted number density of the hot bubble
and the mean baryonic density of the IGM. 
It is clear that the bubble density is always
less than the mean IGM density at the same epoch by a factor
$\sim 10$. This factor is determined by the value of the entrainment
parameter, $\epsilon$, since the bubble mass is eventually dominated
by the mass swept from the IGM and not 
by the mass coming from the galaxy.  As discussed in the previous
section if the hot bubbles have low density then the metals in these
bubbles will not be detected in the UV spectroscopy of the bright 
objects. As the fresh outflow most probably propagates into the voids
the metals detected in the high column density Lyman-$\alpha$ 
absorption lines have to come from the early generation of outflow
that pre-enriched the IGM.

\begin{figure}
\centerline{
\includegraphics[width=0.45\textwidth]{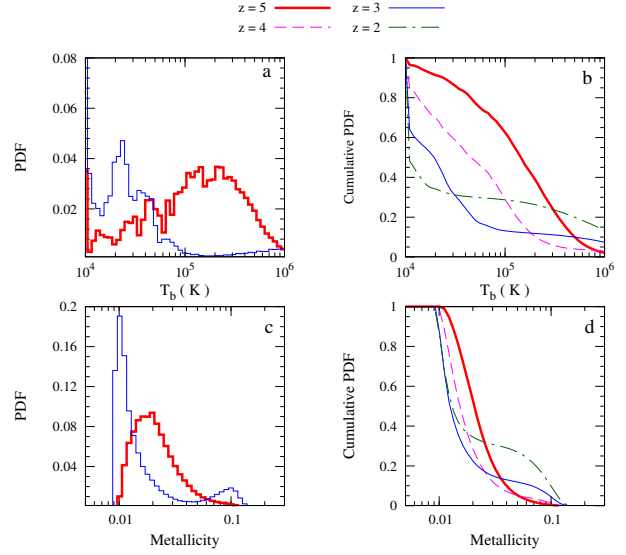}
}
\caption{Probability distribution function (PDF) of
temperature and density of the hot bubble. In panel (a)
we show the PDF of bubble temperatures. In panel (b) we show the
cumulative distribution of bubble temperature.
In panel (c) and (d) we show the similar distribution of bubble density.}
\label{fig_wind_T_Z}
\end{figure}

The average bubble temperature is shown in panel (f) of Fig.~\ref{wind1}.
It decreases from about $10^6$~K at $z\sim 10$
to $\sim 10^5$~K at $z \sim 3-4$. More detailed information is available  
in panel (a) and (b) of Fig.~\ref{fig_wind_T_Z}, where we
show respectively, the differential and cumulative PDFs of
the bubble temperature, for $z=2,3,4$ and $5$.
For visual clarity, the differential PDF is given only for $z=3$ and $z=5$.
At $z=5$ , more than 70\% of the bubbles are at
a temperature higher than $10^5$~K where as by $z=3$ this
fraction decreases to less than 15\%. This is mainly due
to the adiabatic expansion of the bubbles. The void regions
these bubbles fill will nevertheless be at a higher temperature than
the photoionized IGM. At even lower 
redshifts when halos with $M \gtrsim 10^9 M_\odot$ start to contribute
significantly to the volume filling factor, 
the fraction of bubbles at $T_b \gtrsim 10^5$~K
increases, although more than 80$\%$ of the volume filled by 
the bubbles still have $T_b \lesssim 3 \times 10^4$~K (see the
cumulative PDF). The photoionized IGM is expected to have temperatures
in the range $2-4\times10^{4}$~K in the redshift range $2\le z\le 4$
(Schaye et al. 2000). The model discussed above has 55\% and 20\%
of the volume being filled by the gas with $T\ge5\times10^4$~K
at $z = 4$ and $3$ respectively.  Influence of such a gas to the observed
properties of the Lyman-$\alpha$ forest is  negligible if the hot gas 
predominantly percolates into the low density voids. However,
as pointed out by Theuns, Mo \& Schaye (2001), if the gas
around massive galaxies are uniformly heated to high temperatures
their effect will be felt in the high column density end of the Lyman-$\alpha$
forest.

\begin{figure}
\centerline{
\includegraphics[width=0.45\textwidth]{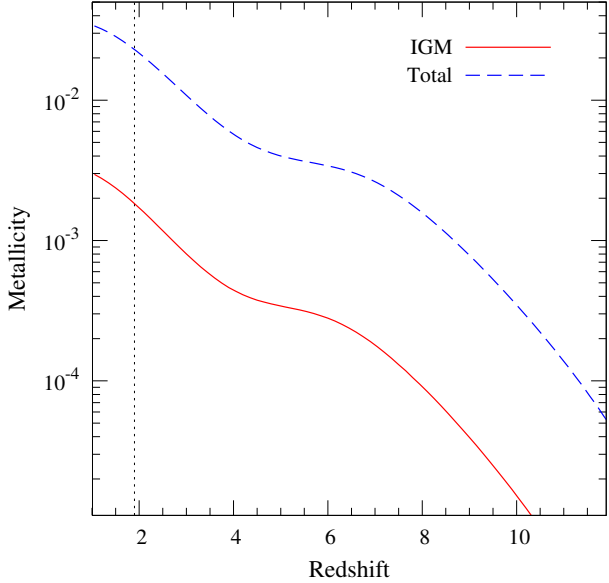}
}
\caption{The global average metallicity evolution of the IGM (solid)
 and galaxies (dashed) as a function of redshift. The Vertical dotted line 
corresponds to the redshift where $Q=1$.
}
\label{fig_metalA}
\end{figure}

In panels (c) and (d) of Fig.~\ref{fig_wind_T_Z} we also show the 
corresponding PDFs of bubble metallicity. 
Most of the bubbles have metallicities between $0.01-0.1~Z_\odot$,
with the differential PDF peaked around the lower value.
The average global metallicity evolution is also of interest.
We show this in Fig.~\ref{fig_metalA} for our fiducial model A. 
The epoch when the porosity becomes unity is marked by a vertical
dotted line. One can see that the globally averaged
metallicity gradually builds up from about $\bar{Z} \sim 10^{-5}~Z_\odot$ at
$z\sim 10$ to $\bar{Z} \sim 2\times10^{-3}~Z_\odot$ at $z \sim 2$, by which
time the porosity has just exceeded unity.
Further, any halo collapsing after the epoch where $Q \sim 1$, should have
at least this amount of metals present in its ISM even before
it starts forming stars. We will also compare below, the global
metallicity evolution among different models.
Fig.~\ref{fig_metalA} also show the average of all metals produced in the galaxies.
It is clear that for the model parameters we have chosen here considerable 
percent of the metals that are produced remains in the halos.
Thus accurate estimation of metal budget at high-$z$ can be used to constrain
the models. It has been argued that only $\sim60\%$ of the metals
produced can be accounted for by the metals in the high redshift galaxies
and $\ge$ 40\% of the global metals produced have to be accounted for by
the metals spread outside the bright galaxies (see Bouche et al. (2007)
for recent metal budget). This means we may need to expel more metals
than what is the case with our fiducial model considered here.

\subsection{Reionization feedback}

\begin{figure}
\centerline{
\includegraphics[width=0.45\textwidth]{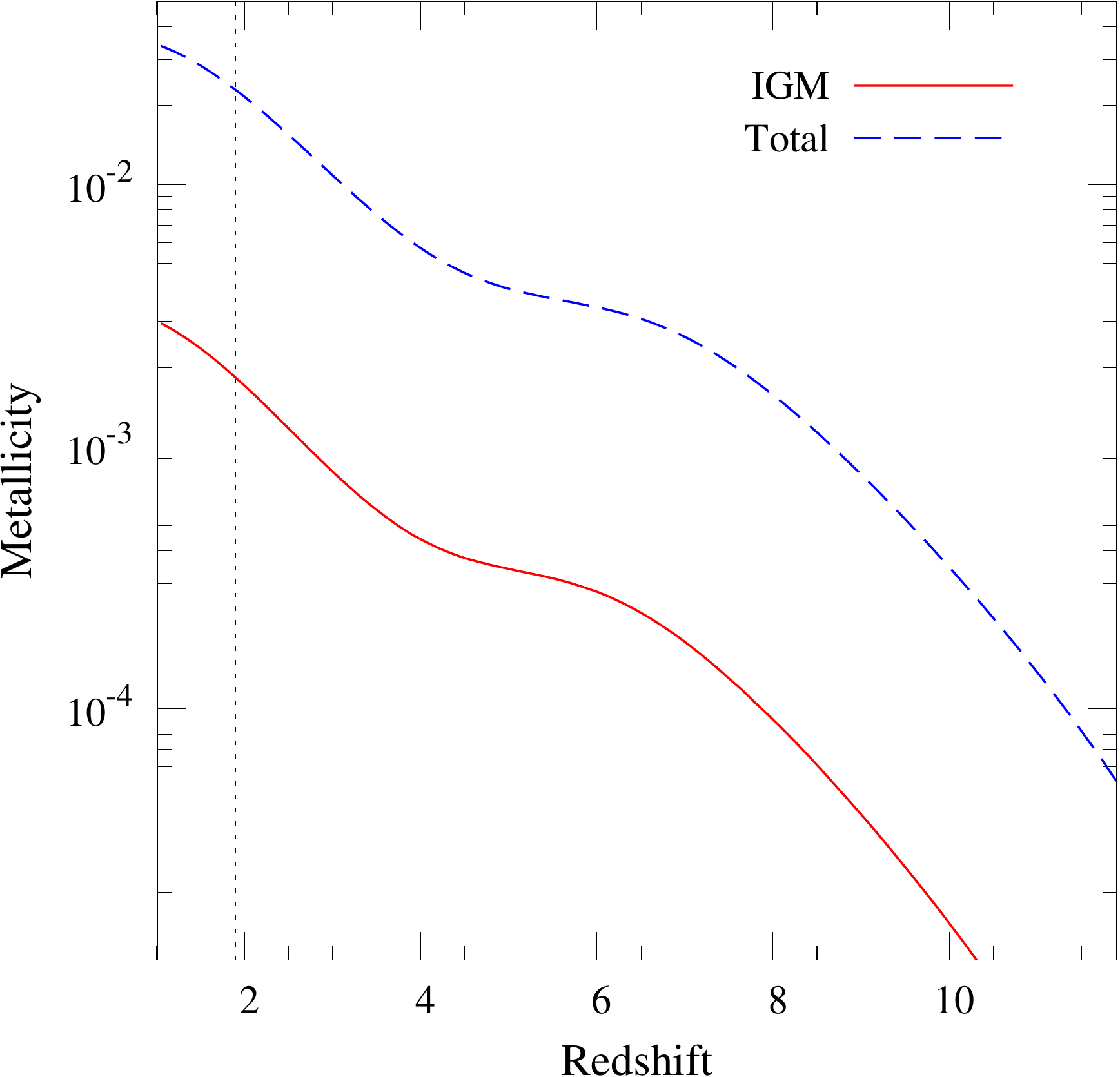}
}
\caption{Effect of reionization on the volume filling factor.
The volume filling factor for model A is shown by solid line.
The dashed and dotted-dashed lines correspond to a step reionization
model with $z_{re} = 6$ and $z_{re} = 11$ respectively keeping all other
model parameters identical to that of model A.
}
\label{fig_wind_reion}
\end{figure}

In this paper we have calculated the global outflow properties
taking into account self-consistently the feed back due to
the reionization history. This is an important improvement as most
of the earlier works on galactic outflows do not take into account the 
radiative feedback in the star formation models.
In Fig.~\ref{fig_wind_reion}  we highlight this effect by 
comparing the result of our self-consistent model with two toy
models with sudden reionization.
The solid line corresponds to the evolution of $F(z)$ in Model A
calculated with the self-consistent ionization history as in
Paper I  (See their Table~2 and Fig.~2). In this model the 
reionization occurs at $z_{re}=7$ and the universe gets 50\%
ionized at $z\sim8.5$.

The dashed line assumes that the universe underwent an
abrupt reionization at $z_{re}=6$, while the dash-dotted line assumes
$z_{re}=11$. We see that early abrupt reionization at $z_{re} =11$ 
leads to a significant fall in the volume filling factor, especially
at $z \gtrsim 2$. This is because the smaller mass halos which contribute
dominantly to volume filling the IGM are suppressed from forming stars
due to early reionization. Below this redshift, higher mass halos
start to contribute leading to a rise in $F(z)$. On the other hand
having an abrupt reionization at a latter redshift $z_{re}=6$, leads
to 90\% of the volume being filled with outflow material by $z=3$.
Therefore a determination of the volume filling factor due to outflows
is a sensitive probe of the ionization history. This point has also
been independently made recently by Pieri et al. (2007).

\subsection{Global properties for a range of model parameters}

\begin{table*}
\begin{center}
\caption{Model parameters for our different models.}
\begin{tabular}{c c c c c c c c c c}
\hline
model	& $f_*$ & $\kappa$ & $f_{esc}$ & $\sigma_8$ & $n_s$ & $\eta$
 & $\epsilon_w$ & $Z~(Z_\odot)$ & Remarks \\ \hline
A & 0.50 & 1.00 & 0.10 & 0.75 &	0.95 &	0.30 &	0.10 & 	0.01 & standard \\

B & 0.25 & 0.50 & 0.10 & 0.75 & 0.95 &	0.30 &	0.10 & 	0.01 &	change in $f_*$ \& $\kappa$ wrt A\\

C & 0.25 & 0.50 & 0.10 & 0.75 &	0.95 &	2.00 &	0.10 &  0.01 &	change in $\eta$ wrt B \\

D & 0.50 & 1.00 & 0.10 & 0.75 & 0.95 &  0.30 & 	0.02 & 	0.01 &	change in $\epsilon_w$ wrt A \\

E & 0.25 & 1.00 & 0.10 & 0.85 &	1.00 &	0.30 &	0.10 & 	0.01 &	change in $f_*$ \& cosmology wrt A \\

F & 0.50 & 1.00 & 0.10 & 0.75 &	0.95 &	0.30 &	0.10 &  0.03 &	change in $Z$ wrt A \\

\hline
\end{tabular}
\label{table_difrn_models}
\end{center}
\end{table*}
\begin{figure*}
\centerline{
\includegraphics[height=0.95\textwidth,angle=-90]{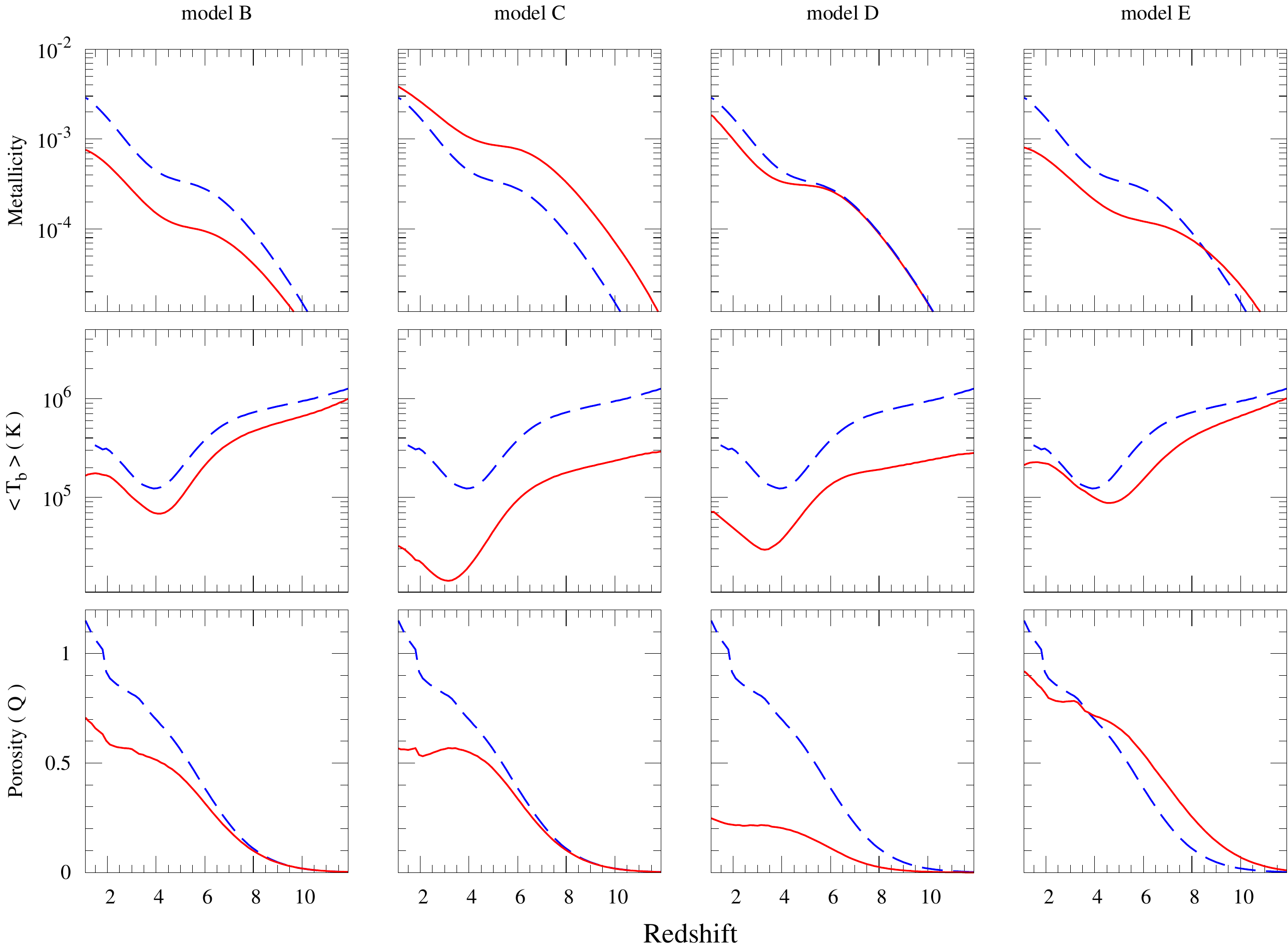}
}
\caption{Porosity (Q), Q weighted average temperature and
IGM metallicity as a function of
redshift are shown in bottom, middle and top panels respectively
for some of the models in Table~\ref{table_difrn_models}. 
The dashed curve in all the panels are the
results for our model A. This is provided for comparison.
}
\label{fig_wind_all_models}
\end{figure*}

In Table~\ref{table_difrn_models} we have listed a few models where
we vary some parameters around that adopted in the fiducial model A.
We wish to investigate the effect of this change on the global
outflow properties. These models are all consistent with the observed
high-$z$ UV luminosity functions below $z =6$. We also have taken
the self-consistent reionization feedback as in Paper I.
In Fig.~\ref{fig_wind_all_models} we show the global
properties of the outflows for models listed in Table~\ref{table_difrn_models}.
In this figure we only concentrate on {porosity (Q)}
(bottom panels), porosity weighted average temperature of the
hot bubble (middle panels) and the global metallicity of the
IGM (top panels). The other physical quantities, which are shown in
Fig.~\ref{wind1}, do not vary significantly while changing the model parameters.

Model B examines the effect of adopting a lower $f_*=0.25$ 
compared to model A. In order to still fit the UV luminosity functions of 
high redshift Lyman break galaxies, one then needs to take a lower value of 
$\kappa=0.5$ (Paper I). The porosity weighted mean temperature
of the bubble for model B is slightly lower
than that of  model A. However, both the porosity and 
the average global metallicity $\bar{Z}$
of the IGM are smaller than in model A.
The porosity in fact never reaches unity in this case.
The smaller values of $\bar{Z}$ and $Q$ are the  manifestation
of the simple fact that total star formation in model B is
half of model A. As discussed before the parameters of this
SFR model may correspond
to a model with strong negative feedback from the SNe.

In model C we explore the effect of increasing the the mass loading factor 
to $\eta=2$. Note that we can not do this in model A because
it assumes $f_*=0.5$, and such a large $\eta$ would require more baryons
to be expelled than the galaxy originally contained.
Hence in model C we assume all parameters as in model B, except for $\eta$.
Two major differences arise due to a larger $\eta$ now, compared to model A 
or model B. The average temperature of the bubble is significantly lower,
and also the average global metallicity is higher. In section 5.2 (see also
Fig.~\ref{fig_eta}), we have shown that increase in $\eta$ 
reduces the bubble temperature, prevents the escape of winds
from massive halos and increases the asymptotic metallicity.
The temperature is lower because in the initial stage of the
outflow evolution, an increased mass loading (by a factor 2/0.3), 
leads to a significantly reduced temperature, and this in turn 
reduces the porosity averaged temperature. The increase in average 
global metallicity is also due to the increase in $\eta$.
From Eq.~(\ref{eqn_metals}) one 
can compute the mass of metals coming out of the galaxy. 
For $\eta = 2$, it is factor of $8$ higher than for $\eta = 0.3$, 
assuming the same $f_*=0.25$.  The porosity curve
does not show an increase at $z\le 2$ (as seen in model B) due to the
fact that the winds from the high mass halos do not escape efficiently
due to higher cooling rate in the halos.

We explore the possibility of having a lower wind efficiency of $\epsilon_w=0.02$
in model D. All other parameters are as in model A.
In section~5.3 (see Fig.~\ref{fig_IMF}) we have shown the decreasing the 
energy efficiency (i.e changing either of IMF or $\epsilon_w$) reduces the 
radius outflow radius. If the change in the efficiency factor is more than $2.5$
we see that winds will not escape from the high mass halos. As the 
adopted change in $\epsilon_w$ for model A is factor $5$ of that used in D we 
do not expect all the halos that have outflows in A to have one in D as well.
As expected both the porosity and the bubble temperature are lowered.
On the other hand it is interesting to note that the average global
metallicity does not change for $z \gtrsim 6$. This is because every halo which
had an outflow in model A still has one in model D at these high redshifts.
And the mass of metals ejected remains the same with redshift. However,
at $z \lesssim 6$ the outflows do not escape from some higher mass halos
due to lower $\epsilon_w$ and hence reducing the mass of ejected metals
at these epochs. In any case it is interesting to note that
the models with $\epsilon_w \le0.02$ do not achieve $Q = 1$.

In model E we change the cosmological parameters i.e. $\sigma_8$ and $n_s$.
We take $\sigma_8=0.85$ and $n_s=1.0$, more in tune with the WMAP 1st year data. 
We self-consistently calculate the SFR  and reionization for this model. 
To fit the observed {luminosity function} we need $f_* = 0.25$ with $\kappa = 1$.
The reionization occurs at $z_{re} = 8.4$ with an electron optical depth of
$\tau_e \sim 0.096$. In this model, the porosity is initially (at $z \gtrsim 3.5$) 
higher than for model A, primarily because of the larger abundance of halos
due to the higher $\sigma_8$ and $n_s$. But at lower redshifts
the porosity decreases below that in model A, perhaps due to the lower $f_*$.
The porosity weighted temperature and metallicity are close to
that seen in model B confirming the above reasoning.

We have also explored the effect of changing the bubble metallicity in
calculating the cooling of the bubble gas (model F). This has negligible 
effect on the global properties of the outflows.

The most interesting outcome of the exercise presented in this section
is that $Q \ge 1$ only for our fiducial model. Thus if star formation activities
are sustained only in the atomic cooled halos then the parameters of the 
models should be close to that of our fiducial model in order to completely
fill the IGM with metals. Overall the alternative models lead to a
lower porosity and volume filling factor, as well as lower bubble temperatures. 
The average global metallicity can be larger than in model A if
the mass loading factor $\eta$ is higher, but for the other alternatives
$\bar{Z}$ is generally lower. 

Presence of radiative feedback makes it difficult to predict the 
trend of various quantities discussed above when we add molecular
cooled halos in our models. This needs to be explored in a self-consistent
way. This is what we do in the following sub-section.

\subsection{Models including Molecular Cooled halos}

In Paper I we considered a number of models where the lower mass
cut-off of a halo which can host star formation is decided by
assuming efficient molecular cooling ($T_{\rm vir} > 300$~K).
Such small mass halos are not detectable directly. 
But their influence can be felt indirectly
via their ionizing efficiency. Several of the molecular cooling
models of Paper I are also consistent with the available constraints 
on reionization. Outflows from such molecular cooled halos could 
also affect the properties of the IGM in important ways.
Here we study the outflow properties of two such models, namely
Model M$_2$ and M$_3$ ( see Table 3 of Paper I). In model M$_2$
we assume that a fraction $f_*=0.1$ of the baryons in a molecular cooled
halo is turned into stars having a normal salpeter IMF.
This is in addition to the star formation in the atomic cooled
halos, where the parameters for star formation are as in model A.
Such a model not only fits the high-$z$ UV luminosity function,
but also has an electron scattering optical depth 
$\tau_e \sim 0.105$, consistent with WMAP 3rd year data.
The model leads to a complex ionization history, with the final reionization
at $z_{re}=5.9$ (Fig. 2 of Paper I). 

\begin{figure}
\centerline{
\includegraphics[width=0.45\textwidth]{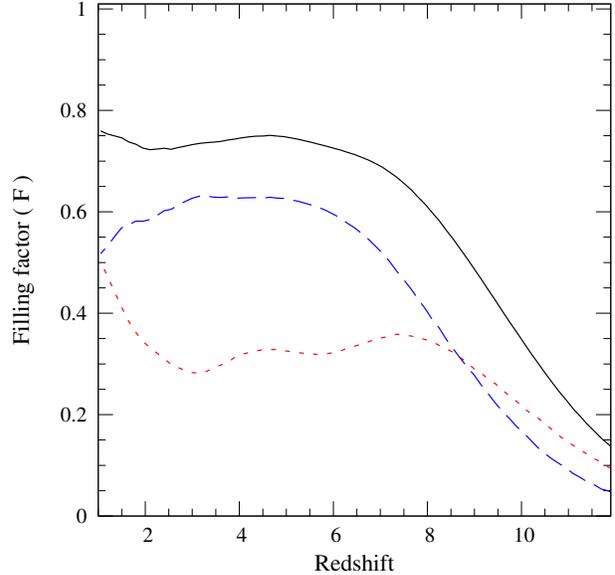}
}
\caption[]{ The volume filling factor as a function of redshift
for molecular cooled model (model M$_2$ of Paper~I).
The solid line shows the total filling factor of the universe.
The dashed curve is the filling factor by the outflows
which have frozen into the hubble flow of the IGM where as
the dotted curve is the filling fraction by the rest of the
outflows. 
}
\label{fig_Q_m}
\end{figure}
In Fig.~\ref{fig_Q_m} we have shown the predicted volume filling
factor of outflows obtained in this  model. Comparing it with Fig.~\ref{fig_Q}
of model A, we see that star formation in the molecular cooled halos leads to 
significant volume filling of the universe, with a porosity $Q(z)=1$,
at even at $z\sim 8$. In fact, the hubble frozen outflows dominate below 
this redshift and the $Q$ contributed by such outflows 
itself becomes of order unity below $z\sim 6$.
Subsequent structure formation below $z \sim 6$ can lead to a
significant fraction of the metal enriched outflowing material from the 
molecular cooled halos being incorporated into
the mildly overdense regions of the IGM (relevant for the Lyman-$\alpha$
forest lines), and lead to a metallicity floor of the IGM.

\begin{figure}
\centerline{
\includegraphics[width=0.45\textwidth]{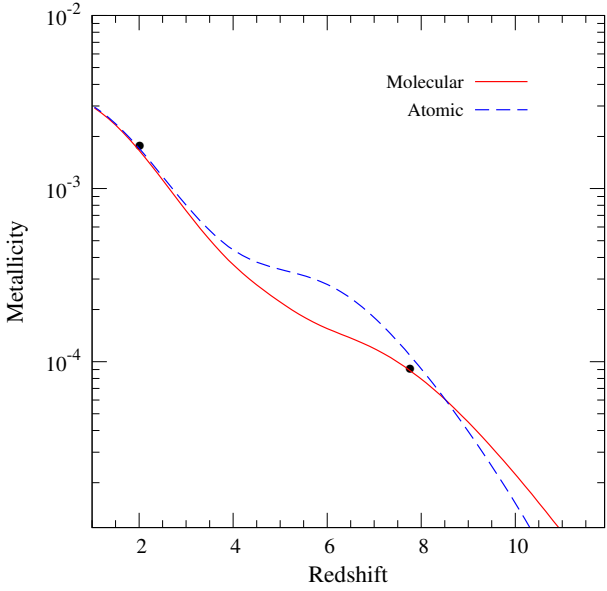}
}
\caption[]{
Global average IGM metallicity as a function of redshift. The solid line is for
molecular cooled model and dashed line is for atomic cool model, model A.
The points indicate when the porosity, $Q$, becomes unity.
}
\label{fig_metallicity}
\end{figure}

In Fig.~\ref{fig_metallicity} we show $\bar{Z}$ evolution
in the molecular cooled model (solid line) and compare it
with atomic cooled model (model A: dashed line). It is evident 
that at any particular epoch the average global metallicity
produced by these models do not differ much.
However, as pointed out above, the porosity of outflows is larger,
at an earlier redshift, in the molecular cooled model.  
In this model, any non-linear or mildly non-linear structures
that are formed  after  $z\sim 8$ will have an initial metallicity
of $Z\ge10^{-4}~Z_\odot$ even before the new stars are formed.
Thus a normal mode of star formation in such small mass halos is a very effective
means of spreading metals  in to the mildly non-linear regimes at low redshifts
that are traced by the Lyman-$\alpha$ forest.
Note that one can increase the metallicity of the IGM at $z\sim8$ by
changing our model parameters.
For example, we get $\bar{Z}\sim10^{-3.5} Z_\odot$
when we use $\eta =2$ and $f_* =0.1$ in the molecular cooled halos.
Note that this is very close to the minimum metallicity suggested by 
observations of Songaila (2001). Increasing $f_*$ to $0.2$ will increase
the metallicity in the outflow by a factor of $4.6$.
However, if we wish to preserve the ionization history 
then the UV escape fraction $f_e$ should be lowered to keep $f_*f_e$ conserved.
Note the metallicity of the IGM can be increased if we relax the
condition of uniform mixing (in Appendix~\ref{metal_evolution}) and allow
the wind to have higher metallicity than the average ISM.

\begin{figure}
\centerline{
\includegraphics[width=0.45\textwidth]{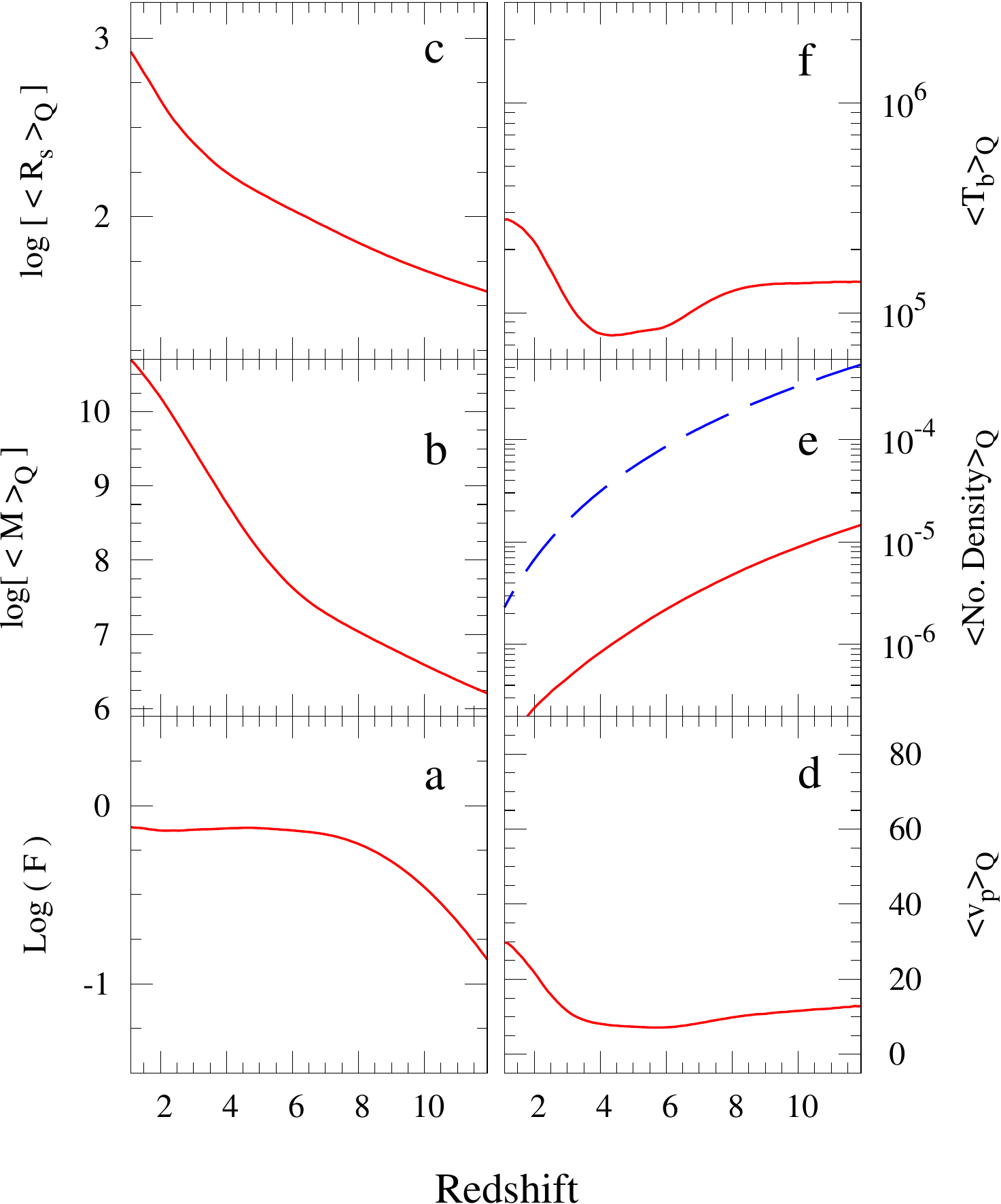}
}
\caption[]{The global properties of outflows for molecular
cooling model (M$_2$ of Paper~I). 
}
\label{fig_wind_mol}
\end{figure}

\begin{figure}
\centerline{
\includegraphics[width=0.45\textwidth]{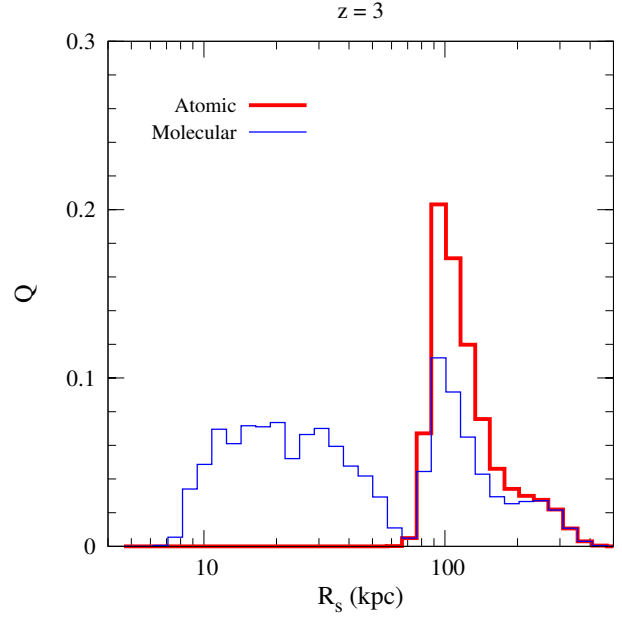}
}
\caption[]{Porosity contributed by outflows of different sizes.
The thick line shows the distribution for atomic cooling model A
where as the thin line is for molecular cooled model M$_2$.
}
\label{fig_radius_pdf_AM}
\end{figure}

To compare the predictions of this molecular cooling model (M$_2$)
with the atomic cooling model A, in more detail,
we show the corresponding porosity averaged physical characteristics
of the outflows in Fig.~\ref{fig_wind_mol}. 
The panels in this figure correspond to the same quantities as in 
Fig.~\ref{wind1}.
Comparing Fig.~\ref{fig_wind_mol} with Fig.~\ref{wind1},
we see that in the case of the molecular cooling model M$_2$,
on the average, the contribution to the filling factor comes from
very low mass halos, { with mass $\sim 10^7~M_\odot$ at $z=8$
to $\sim 10^9~M_\odot$ at $z=3$}. For such 
low mass halos, the outflow radius is small
on average. But their number density is high enough that
they can significantly pollute the IGM.
The porosity averaged comoving size of the outflows $\sim 300$~kpc at $z=3$,
is somewhat smaller than for model A.
This can also be seen from Fig.~\ref{fig_radius_pdf_AM} where the
distribution of the bubble radius in molecular cooled
model M$_2$ is compared with the atomic cooled model A. The suppression
of the large size bubbles ($\sim 100$~kpc) in M$_2$ is due to
a larger radiative feedback than in model A.
We also note that the median bubble size in the model is $\le 40$ kpc (proper)
for $z>3$. This is also a factor two lower than that of model A.
The average density of the hot bubble
does not show any significant difference between atomic
and molecular cooling models as the density of the bubble
is mainly govern by entrainment parameter, $\epsilon$.
One of the major differences is the lower
porosity averaged bubble temperature.
Since star formation and hence the total number of SNe
is less in smaller mass halos the hot bubble ends up 
with a lower temperature. Also the outflows produced
by these smaller halos lead to smaller peak velocities.
And the resulting porosity averaged peculiar velocity 
for the molecular cooled model is much smaller at
$z \gtrsim 3$, compared to that of atomic cooling models.
Low values of porosity weighted  temperature, radius and 
the peculiar velocity coupled with $Q > 1$ for the
bubble makes this model more favorable to pollute the IGM
without disturbing the observable properties of the Lyman-$\alpha$
forest.

Many earlier works related to molecular cooled halos, 
assume a top heavy mode of primordial star formation
in molecular cooled halos.  For a top heavy IMF one may 
expect high mass SNe to dominate increasing the mass 
of ejected metals. It would be interesting to examine if these 
features can lead to a larger metallicity floor or a larger 
volume filling factor. To examine the effect of such a top heavy 
mode,  we consider the model M$_3$ in Paper I. This model assumes
that molecular cooled halos form stars in the mass range
$50-500~M_\odot$ in a salpeter IMF. In this model the
reionization occurs at $z_{re}$ = 11.6 with $\tau_e =0.155$
slightly higher than that constrained by the WMAP 3rd
year data.
 
\begin{figure}
\centerline{
\includegraphics[width=0.45\textwidth]{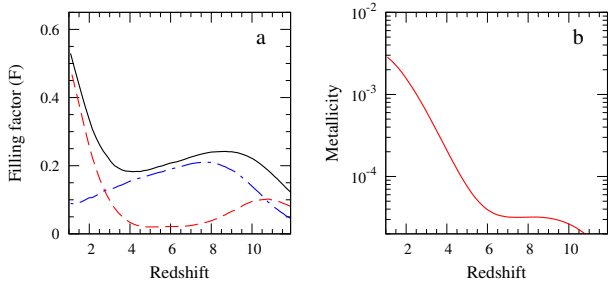}
}
\caption[]{The volume filling factor and global average
metallicity predicted by `top heavy' molecular cooled model.
Panel (a) shows the volume filling factor where as panel (b)
shows the average global metallicity of the IGM. In panel (a)
we also show the break up $F_A$ (dashed line) and $F_H$
(dotted-dashed line). 
}
\label{fig_mol_top}
\end{figure}

We follow Furlanetto and Loeb (2003) and assume that in this case, 
one SNe explodes for every $460~M_\odot$ of star formed,
with an energy output of $10^{52}$~ergs, and ejecting $4~M_\odot$
of carbon.  We show in panel (a) and panel (b) of Fig.~\ref{fig_mol_top}, 
the evolution of the volume filling factor and average global metallicity
respectively, for this model.
It is clear that even at high redshifts $z\sim10$, outflows fill
a significant fraction $F\sim 0.25$ of the volume. However the volume
filling factor hovers around this value even till $z\sim 3$,
and only after this epoch $F$ continues to increase.
This is because the universe is reionized at a very
early epoch ($z_{re}=11.6$) in this model. Subsequently, 
reionization feedback leads to a strong suppression of star formation 
in dwarf galaxies below this high redshift. 

This suppression also affects the average global metallicity.
Even though each SNe in this model is producing $40$ times more carbon
than for model $M_2$, for a given amount of star formation the 
excess metallicity in M$_3$  only by a factor 4. However, M$_3$
produces 17 time more UV ionizing photos compared to that of M$_2$.
At higher redshifts M$_3$ has slightly higher metallicity. However at low redshifts
higher mass unsuppressed `atomic cooled' halos dominate 
in the global properties and hence the predicted average IGM metallicity 
for this model is similar to other models discussed earlier.
The volume filling factor at low redshifts is moderately smaller than 
in model $M_2$.

Hence from above results we can conclude that the inclusion
of star formation in molecular cooled halos will help in enriching
the IGM at higher redshifts, and lead to a metallicity floor,
provided the molecular cooled halos are less efficient in 
reionizing the universe (for example, if stars form still with a normal IMF
or the escape fraction of the UV photons are very low).

\section{Discussion and Conclusions}

We have examined in some detail here semi-analytic models of 
galactic outflows and their consequences for the intergalactic medium.
Our models are constrained by available observations of both star formation
and reionization. The nature of star formation is one of the key
elements which controls the energetics of galactic outflows and
also their metal content. At the same time, it also decides the
reionization history which is an important input for the radiative
feedback that suppresses star formation in low mass halos.
We improve on earlier semi-analytical modeling of galactic outflows
in several important ways. 
Galaxies form stars continuously with the duration and fraction of
baryons going into stars being constrained by the observed high-$z$
UV luminosity functions (see Paper I for details). We take into
account of existing constraints 
on reionization and most importantly, the implied radiative feed back
on the star formation in a self-consistent manner.
We adopt the WMAP 3rd year cosmological parameters.
We model galactic outflows in a manner similar to
stellar wind blown bubbles (cf. Weaver et al. 1977),
following the dynamics of both the outer shock $R_s$ and also 
a possible inner reverse shock at $R_1$.
This can naturally incorporate possible smooth transitions from 
pressure to momentum driven outflows. 
We use the modified PS formalism of \cite{sasaki94} formalism to
calculate the formation rate of dark matter halos and the global
consequences of outflows, instead of taking just the time derivative
of the PS function, which does not account for the destruction rate 
of halos.

Outflows generically accelerate initially due to the 
increasing energy input from the galaxy and decreasing halo density profile.
The outer shock velocity reaches a peak value which increases with halo mass,
typically ranging from $\sim 100-400$ km s$^{-1}$ for halo masses 
$\sim 10^7-10^{12} M_\odot$ respectively. This phase lasts for a dynamical 
time-scale after which the outflow decelerates, till it becomes subsonic 
and freezes to the Hubble flow. The hot bubble of shocked wind material, 
has initial temperatures $\gtrsim 10^6$ K
but subsequently cools due to adiabatic expansion.
If there is significant mass loading from the galaxy,
it can also cool radiatively to transit to a momentum
driven flow.  However, for most model parameters here, this does 
not occur. 
The swept up shell gas typically cools efficiently while
the outflow traverses the halo. During the acceleration phase
the thin shell is also unstable to R-T instability 
leading to shell fragmentation (see also Ferrara \& Ricotti (2006)) without
significantly altering the final outflow radius. 
Such shell fragmentation could enhance the detectability
of metal lines from outflows by providing evaporating interfaces where 
the hot metal-enriched bubble gas mixes with cool, dense shell material.

During its evolution, individual outflows can travel
well beyond the virial radius of the host halo to proper
distances $\sim 200-1000$ kpc for the above halo mass range.
The inner shock at $R_1$ keeps up with the outer shock while
the galaxy is actively forming stars, with typical values of $R_1/R_s \sim 0.4-0.6$.
This is very similar to the wind structure seen in the simulations of
Fujita et al. (2004). By the time the outflow becomes subsonic and
freezes onto the Hubble flow, both the hot bubble and shell temperatures
are $\sim 10^4$~K (determined now by photoheating), and the shell
is likely to fragment and mix with the bubble and IGM gas.

We explored in some detail the dependence of outflow properties
on the assumed initial conditions and various model parameters. 
We show that the initial conditions play very little role in deciding
the nature of the outflows.  We also check this by comparing our model
predictions with scale-free solutions.

We find that outflows can generically escape from the 
low mass halos ($ M \lesssim 10^9 M_\odot$), 
that dominantly contribute to the volume filling of the IGM.
For galactic scale halos, as expected, having higher halo density $f_h$, higher
wind mass loading $\eta$ or lower energy input efficiency  
($\epsilon_w\nu f_*$) makes it more difficult for outflows to escape.
A burst mode of star formation generically leads to a smaller outflow radius,
even for the same values of other parameters. As the outflow properties
of low mass halos are less sensitive to above mentioned parameters the
nature and efficiency of star formation in these objects decide the 
feedback due to galactic winds.

The detection of metals from the outflows and in the IGM is one of
the crucial issues in our paper. The gas phase metallicity in the
ISM and the wind are self-consistently computed for a given star formation
rate and IMF assuming instantaneous uniform mixing in the ISM. The detection
of the expelled metals either in different stages of the outflow or in the IGM
depends crucially on the ionization state of the gas.  Using photoionization
calculations performed using Cloudy we have shown that the metals in 
free wind and low density bubble will be very difficult to detect through
standard UV  absorption lines of C~{\sc iv} and Si~{\sc iv} and O~{\sc vi}.
We need clumped high density gas either coming from the ISM in the form
of free wind or from the R-T instabilities to detect these absorption lines.
In particular some of
these high ionization species can be detected in the conductive interfaces
between the cold clumps and hot bubble material. We show that the 
metals in the underdense regions ($\rho\sim 0.1\bar{\rho}$) will be very
difficult to detect if the temperatures are higher than 10$^5$ K. Thus
fresh outflow from galaxies that enter the IGM will be very difficult to
detect.  C~{\sc iv} and O~{\sc vi} absorption lines are easily detectable
if they originate from overdense regions ($\rho>10\bar{\rho}$) that are
typically probed by high column density Lyman-$\alpha$ absorption
line with $T\sim10^4$~K. This is possible if IGM is already filled with 
pre-enriched gas by the time these over densities were order unity
fluctuations (say $z\ge 8$).

One of the important issues is to understand is how outflows
impact on the physical properties of the intergalactic medium. 
We therefore computed the volume of the IGM affected by outflows, 
the porosity weighted averages and PDFs of several important outflow 
characteristics, for a number of atomic and molecular cooling models.

For our fiducial atomic cooling model A, more than $30\%$ of the universe 
is affected by the outflows even at $z\sim 6$ and this increases to 
$\sim 60\%$ by $z \sim 2$. Galaxies with mass range $10^{7}-10^{9}~M_\odot$ 
dominantly contribute to the volume filling factor; higher masses only dominate
at low redshifts. This is consistent with the suggestion of 
Madau, Ferrara and Rees (2001), 
of the dominant influence of halos of $\sim 10^8 M_\odot$ in
filling the universe with outflows; our work however 
includes halos of all mass ranges. Further, these galaxies
which dominate in filling the IGM are not yet detected directly in the 
high-$z$ UV luminosity functions (see Fig.~4. in Paper I),
The porosity averaged outflow comoving radius, peculiar velocity and
bubble temperature, evolve from $\sim 100 $~kpc, $\sim 60 $~km~s$^{-1}$
and $10^6$~K, respectively, at $z \sim 10$ to $\sim 500$~kpc
$20$~km~s$^{-1}$ and $\sim 10^5$~K at $z \sim 3 $.
The median value of the outflow radius is $60 $~kpc (proper)
at $z =5$ which increases to $100$~kpc at $z=3$.
Whereas at $z=5$ , more than 60\% of the volume filled by bubbles are at
a temperature higher than $10^5$~K this fraction decreases to less than
15\% by $z=3$. This is mainly due to the adiabatic expansion of the bubbles.
The void regions these bubbles fill will nevertheless be at a
higher temperature than the photoionized IGM. 
Most of the bubbles have metallicities between $0.01-0.1~Z_\odot$,
with the differential PDF peaked around the lower value.
The average global metallicity evolution is also of interest.
For our fiducial model, this gradually builds up from about 
$\bar{Z} \sim 10^{-5}~Z_\odot$ at
$z\sim 10$ to $\bar{Z} \sim 2 \times 10^{-3}~Z_\odot$ at $z \sim 2$, by which
time the porosity has just exceeded unity.

We have examined several other atomic cooling models which are all
consistent with the constraints on star formation obtained in paper I.
These models also lead to significant filling of the IGM at $z\sim3$ with metals
(with $-2.5\gtrsim [Z/Z_\odot]\gtrsim -3.7$), the actual extent depending on
the efficiency of winds, the initial mass function (IMF) and the fractional
mass that goes through star formation and cosmological parameters. 
The reionization history has a significant effect on the volume
filling factor, due to radiative feedback (see also Pieri et al. 2007).
Further, a large fraction of outflows at $z\sim3$ are supersonic,
hot ($T\ge 10^5$~K) and have low density, making metal lines difficult to detect.
These models may also result in significant perturbations in the IGM gas on scales
probed by the Lyman-$\alpha$ forest. 

On the other hand, we find that models including star formation in
molecular cooled halos with a normal mode of star formation
(or a lower UV escape fraction) can potentially volume fill
the universe at $z\ge 8$ without drastic dynamic effects
on the IGM, thereby setting up a possible metallicity floor
($-4.0\le [Z/Z_\odot]\le-3.6$). In fact, the hubble frozen
outflows dominate below this redshift and the $Q$ contributed by such outflows
itself becomes of order unity below $z\sim 6$. On the average, very low
mass halos, with mass $\sim 10^6-10^7~M_\odot$ at $z=8$ to
$\sim 10^8-10^9~M_\odot$ at $z=3$ dominantly contribute to the volume
filling factor. The porosity averaged comoving radius of the outflows
is less than 100 kpc at redshifts where the $Q\sim 1$. The bubble
temperature and peculiar velocities are also smaller than for the
fiducial atomic cooling model. The above features make this model
ideal to spread metals into the regions which will
subsequently collapse to form the Lyman alpha forest regions, without unduly
disturbing these regions dynamically.
To some extent this scenario is the extrapolation of the scenario of
Madau, Ferrara and Rees (2001) which was applied to outflows from
dwarf galaxies, to even lower mass halos.
Interestingly, molecular cooled halos with a ``top-heavy'' mode of star
formation are not very successful in  establishing the metallicity floor
because of the additional radiative feedback, that they induce.

As we discussed above we use a functional form for SFR (constrained
by the observations of high-$z$ UV luminosity functions) without doing
self-consistent calculations. However, to get reliable results from the
self-consistent calculations one needs to be very clear about various
physical processes that are involved. This is reflected in the fact that
two recent simulations addressing this issue conclude differently.
Scannapieco et al. (2006) find that supernova feedback and the resulting outflows 
decrease the fraction of baryons which are turned into stars, by factor
ranging from 2-4 as one changes the mass of the object, and also a related
more rapid fall in the star formation rate with time. However,
Koboyashi et al. (2007) find somewhat different results. They note that the 
two effects of supernovae, the increased metal line cooling due to the chemical 
feedback and the increased heating due to the energy feedback both have opposite 
effects on the star formation rate in the galaxy. And indeed these two effects seem 
to cancel to produce no net effect due to supernovae feedback on the
star formation rate (see their Fig.~2 and the discussion).
Therefore it is not entirely clear from these works the extent to which one
needs to change $f_*$ and $\kappa$. 
If SNe produce strong feedback effects as found by Scannapieco et al. (2006)
the star formation rate will fall sharply as a function of time.
This corresponds to our models with low $\kappa$ that requires low value
of $f_*$ in order to reproduce the UV luminosity functions.

Future improvement of our work would involve replacing Eq.~(\ref{eqnsf}) with
a model of star formation in a multiphase ISM including
various heating and cooling processes,
possible effects of dark halo clustering
on the outflow properties  and importantly setting up our semi-analytical 
model in the framework of a large-scale structure simulation.

\section*{acknowledgements}
SS thanks CSIR, India for the grant award
No. 9/545(23)/2003-EMR-I. SS also thanks
Andrea Ferrara, Simon White and Cecilia Scannapieco for useful discussions.


\appendix

\section{Metallicity evolution}
\label{metal_evolution}

We calculate the metallicity of the bubble material as follows
(also see Binney \& Tremaine 1994). Suppose at any given time
$\delta M_s$ is the amount of mass goes into star formation.
Let's assume at that instance the metallicity of the IGM is $Z$.
The amount of gas mass lost from the ISM is $\delta M_g=-(1+\eta)\delta M_s$
where we take the mass loss rate due to wind is $\eta \delta M_s$
and we neglect the time delay between the star formation and SNe
explosion. If one takes that $p$ is the amount of heavy metals
ejected per solar mass of star formed then the change of metal mass
in the ISM is given by $\delta M_h = [p-Z(1+\eta)]\delta M_s$.
On the other hand the increase of metals in the wind material
is $\delta m_h = Z\eta\delta M_s$. Now $Z=M_h/M_g$. Differentiating
this and substituting for $\delta M_g$ and $\delta M_h$ from above,
one can get $\delta Z = (p/M_g)\delta M_s$. Taking the initial
gas mass in the ISM as $M_0$, we get $\de Z/\de M_s = p/(M_0-(1+\eta)M_s)$.
Integrating this with the boundary condition that $M_h=0$ when
$M_s=0$ we get
\bea
Z=-\left(\f{p}{1+\eta}\right)\ln\left[1-(1+\eta)\f{M_s}{M_0}\right]
\eea
Using this relation we obtain the metal mass in the wind material is
\bea
m_h&=&\f{\eta p}{(1+\eta)^2}M_0\left[(1+\eta)\f{M_s}{M_0}\right. \nonumber\\
&& \left.+ \left(1-(1+\eta)\f{M_s}{M_0}\right)\ln\left(1-(1+\eta)
\f{M_s}{M_0}\right)\right]
\eea
In the asymptotic limit when all the star formation is over i.e.
$M_s/M_0=f_*$ we get
\bea
m_h&=&\f{\eta p}{(1+\eta)^2}M_0\left[(1+\eta)f_*\right. \nonumber\\
&& \left.+\left(1-(1+\eta)f_*\right)\ln\left(1-(1+\eta)f_*\right)\right]
\label{eqn_metals}
\eea
To give a rough idea of numbers involved, we take $\eta = 0.3$,
$f_*=0.5$ and $p=0.1/50$ (one SNe will form per $50~M_\odot$ of star
formation and $0.1~M_\odot$ carbon will produce from each SNe).
In the asymptotic limit this will give the metallicity
of the wind material as $0.2 Z_\odot$. For $\eta = 1.0$ and
$f_*=0.25$ the metallicity is $0.1 Z_\odot$. However, in realistic
situation this gas is also going to mix with some fraction of halo/IGM
material reducing the metallicity of the hot bubble.
Another thing to note is that if the ISM of the galaxy is already
enriched with metals then the amount of heavy elements transfered
to the hot bubble from the wind is
\bea
m_h&=&\f{\eta p}{(1+\eta)^2}M_0\left[(1+\eta)\f{M_s}{M_0}+
\left(1-(1+\eta)\f{M_s}{M_0}\right)\right. \nonumber \\
&&\left.\ln\left(1-(1+\eta)\f{M_s}{M_0}\right)\right]
+Z_0\eta M_s
\eea
where $Z_0$ is the metallicity of the ISM when the galaxy is formed.

\section{Structure of the shell in adiabatic regime}
\label{sec_shell_struc}
\begin{figure}
\centerline{
\includegraphics[width=0.45\textwidth]{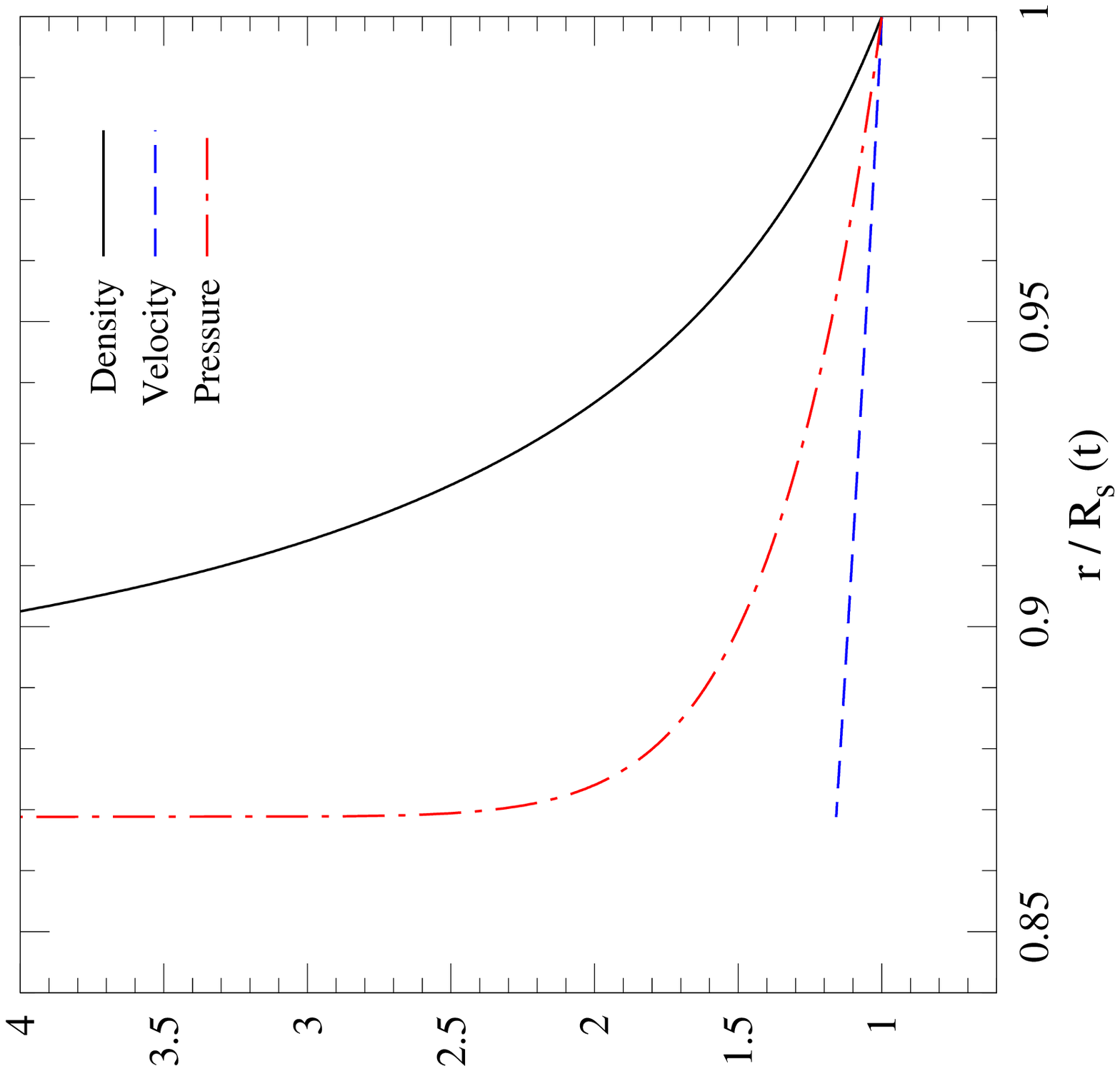}
}
\caption[]{The shell structure for our model parameters
when outflow is traversing inside the halo. We show the
dimensionless scaled density, velocity and pressure as solid,
dashed and dash-dotted line respectively. We show the
profile upto the contact discontinuity.
}
\label{fig_shell}
\end{figure}

We consider here the inner structure of the swept up shell.
This is important for detection of the shell material and
for determining its cooling efficiency.
Inside the halo, the density falls approximately as $\rho_B\propto
r^{-2.8}$ and $L(t)\propto t$ which leads to $R_s(t)\propto t^{2/1.1}$.
In this limit one can obtain a self-similar solution for the
shell structure following Weaver et al (1977) and Koo \& McKee (1992).
We solve the continuity equation, momentum conservation
equation and energy conservation equation assuming
a self-similar solution (c.f. Eq.~(2)-(8) in Weaver et al.
(1977) and Appendix B of Koo \& McKee (1992)). The boundary
conditions are obtained from the shock jump conditions assuming
a strong shock. Fig.~\ref{fig_shell} gives the self similar
scaled structure of the shell.
We show as solid, dashed and dash-dotted lines, respectively, 
the dimensionless scaled density $\tilde{\rho}$, scaled velocity
$\tilde{v}$ and scaled pressure $\tilde{P}$ as a function
of similarity variable $\lambda = r/R_s$.
The dimensionless scaled parameters
are defined as $\rho(r) = \rho_1(t)\tilde{\rho}(\lambda)$,
$v(r) = v_1(t)\tilde{v}(\lambda)$ and $P(r) = P_1(t)\tilde{P}(\lambda)$
where the $\rho_1(t)$, $v_1(t)$ and $P_1(t)$ are the post shock values
given by the usual Rankine-Hugoniot jump conditions.
We show the density, velocity and pressure
upto the contact discontinuity,  whose location is
obtained from the criteria
given in Koo \& McKee (1992). From the figure it is clear that
when the outflow is traversing through the halo the shell
can indeed be approximated as a thin shell.
It is only $\sim 10\%$ of the
radius. Within the shell the density can be very high.
This help in cooling the shell material and which in turn
will enhance the density further, making the shell much thinner.
\end{document}